%% file: natural_orbital_thesis_draft_170402.tex
\documentclass[final,numrefs,sort&compress,noinfo]{nddiss2e}

\usepackage[pdftex]{graphicx}
\usepackage{amsmath}
\usepackage{multirow}
\usepackage{longtable}
\usepackage{enumerate}
\usepackage{verbatim}
\usepackage{pdfpages}
\usepackage{mathtools}
\usepackage{mathrsfs}	
\usepackage{placeins}
\usepackage{isotope}
\usepackage{epstopdf}

\begin{document}

\setcounter{topnumber}{99}
\setcounter{bottomnumber}{99}
\setcounter{totalnumber}{99}
\renewcommand{\topfraction}{.99}
\renewcommand{\bottomfraction}{0.99}
\renewcommand{\textfraction}{0.01}
\setcounter{dbltopnumber}{99}
\renewcommand{\dbltopfraction}{.99}



\newcommand{\Nmax}{N_{\mathrm{max}}}
\newcommand{\Ncut}{N_{\mathrm{cut}}}
\newcommand{\Ntot}{N_{\mathrm{tot}}}
\newcommand{\hw}{\hbar\omega}
\newcommand{\mnc}{m_{\mathrm{N}} c^{2}}

\newcommand{\Gj}[6]{ \begin{Bmatrix}
  #1 & #2 & #3 \\
  #4 & #5 & #6
 \end{Bmatrix}}


\newcommand{\plotspath}{plots}
\newcommand{\bibpath}{bibliography}

\frontmatter 

\title{NATURAL ORBITALS FOR THE NO-CORE CONFIGURATION INTERACTION APPROACH}
\author{Chrysovalantis Constantinou}
\work{Dissertation} 
\degaward{Doctor of Philosophy} 
\advisor{Mark A. Caprio}
\department{Physics}
 
\maketitle
%
%

\copyrightholder{Chrysovalantis Constantinou} 
\copyrightyear{2017}           
\makecopyright

\begin{abstract}
\textit{Ab initio} calculations face the challenge of describing a complex multiscale quantum many-body system. The nuclear wave function has both strong short-range correlations and long-range contributions.

Natural orbitals provide a means of adapting the single-particle basis for \textit{ab initio} no-core configuration interaction (NCCI) calculations to better match the many-body wave function. Natural orbitals are obtained by diagonalizing the one-body density matrix from a calculation using an initial single-particle reference basis, such as the traditional harmonic oscillator basis. The natural orbital basis builds in contributions from high-lying oscillator shells, thus accelerating convergence of wave functions, energies, and other observables.

The convergence of the ground and excited state energies, radii, and electromagnetic observables of $\isotope{He}$, $\isotope{Li}$, and $\isotope{Be}$ isotopes calculated using natural orbitals in \textit{ab initio} NCCI calculations is discussed. It is found that electromagnetic observables involving the $M1$ operator fully converge, while the calculated energies, radii, and observables involving the $E2$ operator converge significantly faster with the natural orbital basis than with the harmonic oscillator basis. The use of infrared (IR) extrapolation schemes with the natural orbital calculations is also explored.
 
\end{abstract}

\renewcommand{\dedicationname}{\mbox{}}

\begin{dedication}

To my cousin Ioannis Constantinou who was fighting a hard battle when this dissertation was under way.

\end{dedication}

\tableofcontents
\listoffigures


%
%



\begin{acknowledge}

This is perhaps the easiest to write part of the thesis. However, it is not easy to include all the people who contributed in the materialization of this work into a written document. Please forgive me if you do not find your name listed here.

I have to start with my advisor Professor Mark Caprio to whom I am grateful for all the time he devoted to me while I was a graduate student at the University of Notre Dame. My interest in theoretical nuclear physics started in my undergraduate studies when, after learning about the shell model of the nucleus, I became curious to understand how one can derive the magic numbers starting from the Schr\"{o}dinger equation. However, it was Mark's lectures on nuclear physics the fall semester of $2010$ that ``lured'' me into the subject. Mark demonstrated a great amount of patience in helping me develop (often from scratch) the coding, writing, and mathematical skills needed in modern theoretical nuclear physics research. Of course the process is not complete however Mark pointed out all the steps, readings, and projects I can work on to further sharpen my skills.

I want to thank the members of my Ph.D. committee, Professors Stefan Frauendorf, Kathie Newman, and Ani Aprahamian for reading the manuscript and offering suggestions for improvements. When I came to Notre Dame in $2009$, Professor Newman was the director of graduate studies and the instructor of the graduate level classes electromagnetism $\mathrm{I}$ and $\mathrm{II}$. I want to thank her for her initial mentorship. Professor Frauendorf was the instructor of the classes on nuclear physics and nuclear reaction theory for which I served as a teaching assistant. I want to thank him for the physics discussions we had and for giving me the opportunity to deliver the lectures for some of his classes.

The completion of this thesis would not have been possible without the computer codes provided by Professors Pieter Maris and James Vary of the nuclear theory group at the Iowa State University. I want to thank them for their help, support, and encouragement.

I also want to thank Dr. Guillaume Hupin who participated in the initial discussions about the implementation of natural orbitals, and Mitch McNanna who was the first in our group to implement natural orbitals in their one dimensional version.

Throughout my studies at Notre Dame I benefited from my interaction with various faculty members who taught me physics or provided guidance at various stages of my graduate studies. I want to especially thank Professors Maxime Brodeur and Tan Ann for helping me understand the methods used to study the structure of $\isotope[6]{He}$, $\isotope[7]{Li}$, and $\isotope[7]{Be}$ experimentally. Moreover, I want to thank Professor Christopher Kolda for being a good teacher and mentor.

It is difficult to list all the people at the University of Notre Dame who graciously offered me support when I needed it. I want to thank Shari Herman for all her help with the administrative part of my studies (and her emotional support). I want to thank my officemates Dr. Weichuan Li and (soon to be Dr.) Anna McCoy for standing me. I also want to thank the Greeks at Notre Dame, Antonis Anastasopoulos, Dr. Ioannis Gidaris, Alexandros Lamprou, Dr. Antonios Kontos, and Professor Alex Taflanidis for their support.

Two great teachers in Cyprus and Greece played an important role in my education. Andreas Constantinou (to whom I am not related however he does have my father's name and surname), taught me physics in high school and helped me develop problem solving techniques which I still use. Professor Michael Kokkoris was my undergraduate advisor at the National Technical University of Athens and the man who is responsible for me becoming a nuclear physicist after he invited me to participate in experiments conducted at the National Center for Scientific Research ``Demokritos''.

Nothing would be possible without the constant guidance and support of my parents Andreas and Maria Constantinou who are always there when I need emotional, financial, or any other form of support. I want to thank them for everything.

Professor Stavros Constantinou and Dr. Georgette Constantinou provided a home away from home here in the US. They were very supportive throughout my studies and I am very grateful for that.

When this work was finalized, I was already a post-doc at Yale University under Professor Francesco Iachello whom I want to thank for his understanding, guidance, and support.

Support for this work was provided by the U.S. Department of Energy, Office of Science, under Award Numbers DE-FG02-95ER-40934, DESC0008485 (SciDAC/NUCLEI), DE-FG02-87ER40371, DE-FG02-91ER-40608, and the University of Notre Dame. Computational resources were provided by the University of Notre Dame Center for Research Computing and the National Energy Research Scientific Computing Center (NERSC), a U.S. Department of Energy, Office of Science, user facility supported under Contract DE-AC02-05CH11231.

\end{acknowledge}


\mainmatter

\include{chapters/chapter1/chapter1_draft_170402}

%
%

\include{chapters/chapter2/chapter2_draft_170402}

%
%

\include{chapters/chapter3/chapter3_draft_170402}

%
%

\include{chapters/chapter4/chapter4_draft_170402}

%
%

\include{chapters/chapter5/chapter5_draft_170402}

%
%

\include{chapters/chapter6/chapter6_draft_170402}

%
%



%
%


 \backmatter
 \bibliography{\bibpath /books,\bibpath /data,\bibpath /expt,\bibpath /master,\bibpath /mc,\bibpath /misc,\bibpath /proc,\bibpath /theory}   
%
%
%
%

\end{document}

%% file: chapters/chapter1/chapter1_draft_170402.tex
\chapter{\textit{AB INITIO} METHODS IN NUCLEAR STRUCTURE THEORY}
\label{chap-chap1}

This work introduces natural orbitals for nuclear no-core configuration interaction (NCCI) calculations. The NCCI approach is an \textit{ab initio} method (i.e., from first principles) used for the description of the structure of light nuclei ($A \lesssim 24$)~\cite{navratil2000:12c-ab-initio,navratil2000:12c-ncsm,barrett2013:ncsm}. This introductory chapter attempts to provide an overview of \textit{ab initio} methods used in nuclear physics, discuss various (realistic) internucleon interactions used with \textit{ab initio} methods, and discuss the challenges \textit{ab initio} methods face, with an emphasis on the NCCI approach which is used in this work.

The atomic nucleus is a strongly-correlated, self-bound, quantum many-body system. Its building blocks are protons and neutrons which are themselves complex relativistic many-body systems consisting of quarks and gluons. In the low energy regime of nuclear physics the underlying quantum chromodynamics (QCD) degrees of freedom are not excited. To obtain information about the structure of the atomic nucleus one needs to solve the many-body nuclear Hamiltonian using a method appropriate to the nucleus under study.

\textit{Ab initio} methods strive to describe the nucleus using internucleon interactions as the starting point. Due to the large model spaces used, these methods require the computational power provided by supercomputers~\cite{vary2009:ncsm-mfdn-scidac09}. The main differences between \textit{ab initio} and traditional methods include the use of realistic internucleon interactions instead of phenomenological interactions and the equal treatment of all the nucleons in the problem. For example, in the NCCI approach all the nucleons interact via realistic internucleon interactions which are usually constructed by fitting experimental data (as briefly described below). The NCCI model space is constructed by distributing all the nucleons over a selected set of single-particle orbitals (usually a truncation scheme is used which restricts the possible configurations of the nucleons among these orbitals). In contrast, in the traditional shell model a phenomenological mean field created by all the nucleons is taken as the potential which confines the nucleons. The model space is constructed by distributing the valence nucleons over single-particle orbitals residing outside the ``frozen'' single-particle orbitals occupied by the core nucleons (the core is usually a doubly magic nucleus). The core does not explicitly interact with the valence nucleons. To account for configuration mixing a residual interaction is included in the shell model Hamiltonian~\cite{suhonen2007:nucleons-nucleus,talmi1993:shell-ibm}, and the solution is then obtained by diagonalizing the Hamiltonian in the selected model space.

Internucleon interactions can be derived in a number of ways. For example, chiral effective field theory ($\chi$EFT)~\cite{weinbergxeft:1990,weinbergxeft:1991} is used to derive internucleon interactions from first principles. The Idaho N$^3$LO~\cite{entem2003:chiral-nn-potl} nucleon-nucleon ($NN$) interaction is an example of an interaction derived using this approach. In chiral effective field theory the QCD Lagrangian is written as an infinite series of terms with increasing number of derivatives and/or nucleon fields. Applying this Lagrangian to $NN$ scattering generates an infinite number of Feynman diagrams. Expanding the nucleon potential in terms of $(Q/\Lambda_{\chi})^{\nu}$, where $Q$ denotes a momentum or pion mass, $\Lambda_{\chi} \approx 1$ GeV is the chiral symmetry breaking scale, and $\nu \geq 0$ makes the problem tractable. For a given $\nu$ the number of terms is finite and calculable. The N$^3$LO interaction is a fourth order ($\nu = 4$) interaction, including charge dependence, which reproduces scattering data below $290$ MeV lab energy with accuracy comparable to the one of phenomenological high-precision potentials~\cite{entem2003:chiral-nn-potl}.

Alternatively, internucleon interactions can be derived by fitting an interaction to experimental nucleon-nucleon scattering data and energies of a few bound states of light nuclei. An example of an interaction derived using this approach is the JISP$16$ internucleon interaction~\cite{shirokov2007:nn-jisp16} which we will use in this work. The derivation of the JISP$16$ interaction is based on the $J$-matrix inverse scattering method and the NCCI approach~\cite{shirokov2004:nn-jisp}. The interaction is derived with two goals in mind. Namely, to minimize the need for three-body ($NNN$) interactions (which we often need to condider in order to reproduce experimental data in the expense of adding computational complexity) and to achieve faster convergence in small model spaces. The starting point is the charge independent JISP$6$~\cite{shirokov2005:nn-jisp6} interaction which provides an excellent description of the properties of the deuteron as well as $NN$ scattering data. Using phase equivalent transformations, the JISP$6$ scattering phase shifts are modified to provide a good description of nuclei with $A \leq 16$.

Once an internucleon interaction is selected, one can proceed to solve the nuclear many-body problem using one of the many available \textit{ab initio} methods. For example, in the NCCI approach, which we study in this work, the nuclear Hamiltonian is cast into an eigenvalue problem in terms of a many-body basis. The resulting Hamiltonian is then diagonalized using the Lanczos algorithm, which yields the first few low-lying eigenvalues and eigenvectors of the Hamiltonian~\cite{lanczos:method}.

In the coupled-cluster approach, a set of filled single-particle orbitals defines a reference state, and the many-body problem is expanded in terms of particle-hole excitations among these orbitals \cite{coester1958:coupled-cluster,gour2006:coupled-cluster-near-16o,hagen2007:coupled-cluster-benchmark,hagen2009:coupled-cluster-com,hagen2010:coupled-cluster}. In both the coupled-cluster and the NCCI approaches the goal is to obtain converged results which do not depend on the model space or the single-particle basis length parameter (which is ususally the characteristic length of the oscillator functions discussed below).

In the Green's function Monte Carlo approach, the Green's function for a Hamiltonian without a potential term is used to cast the problem of finding the many-body wave function into an integral equation which is then solved iteratively. The goal is to then obtain the wave function in a small number of iterations~\cite{pieper2004:gfmc-a6-8,wiringa2000:gfmc-a8,pervin2007:qmc-matrix-elements-a6-7}. 

Using Jacobi coordinates along with a set of hyperspherical harmonic functions~\cite{barnea-hyperspherical-functions:1998}, one can solve the problem directly using antisymmetrized many-body basis states~\cite{navratil-ncsm-jacobi-coordinates:2000,navratil2009:ncsm}. However, Jacobi coordinates are limited to nuclei with  mass number $A \lesssim 6$ because the antisymmetrization of the many-body states becomes cumbersome as the mass number increases.

Let us now focus on the NCCI approach. To work with the NCCI approach we must first choose a single-particle basis and subsequently construct a many-body basis using an appropriate truncation for the many-body basis. In this work we use the $\Nmax$ truncation scheme for reasons related to the removal of spurious center-of-mass states as described in Chapter~\ref{chap-chap2}. According to the $\Nmax$ truncation scheme, only many-body states having a total number of oscillator quanta $N = \sum_{i=1}^{A} N_{i} \leq N_{0} + \Nmax$, where $N_{0}$ is the total number of quanta in the configuration where the nucleons occupy the lowest allowed single-particle states, and $N_{i}$ is the oscillator quantum of a single-particle state are allowed in the many-body basis (here, it is implied that the harmonic oscillator single-particle basis is used). The nuclear Hamiltonian is then cast into a square matrix the diagonalization of which yields the energies and the many-body wave functions of the nucleus. The calculated results depend on both the $\Nmax$ truncation of the many-body basis and the characteristic length of the underlying single-particle basis. In the case of the harmonic oscillator basis, the characteristic length is the familiar oscillator length $b\equiv (\hbar c)/\sqrt{(m_{\mathrm{N}} c^{2}) (\hbar \omega)}$,  where $m_{\mathrm{N}} c^{2} = 938.92$ MeV is the mass of the nucleon, taken as the average between the mass of the proton and the mass of the neutron, and $\hw$ is the oscillator parameter. Hence, when the harmonic oscillator single-particle basis is used the calculated results depend on $\Nmax$ and $\hw$.

Convergence of a calculated observable is signaled by an independence of the calculated result from both the $\Nmax$ truncation of the many-body basis and the oscillator parameter $\hw$. Due to the variational principle, the ground state eigenvalue approaches the ``true'' eigenvalue, of the untruncated many-body problem, as $\Nmax$ increases, i.e., as the truncated space approaches the full space. For each $\Nmax$ truncation, calculations are performed for various $\hw$ parameters to check for the convergence of the eigenvalue in terms of $\hw$. The ground state eigenvalue has a minimum at one of these $\hw$ values (which typically shifts for each $\Nmax$ truncation) called the variational minimum of the calculation.

\begin{figure}[t]
\begin{center}
\includegraphics[width=0.7 \columnwidth]{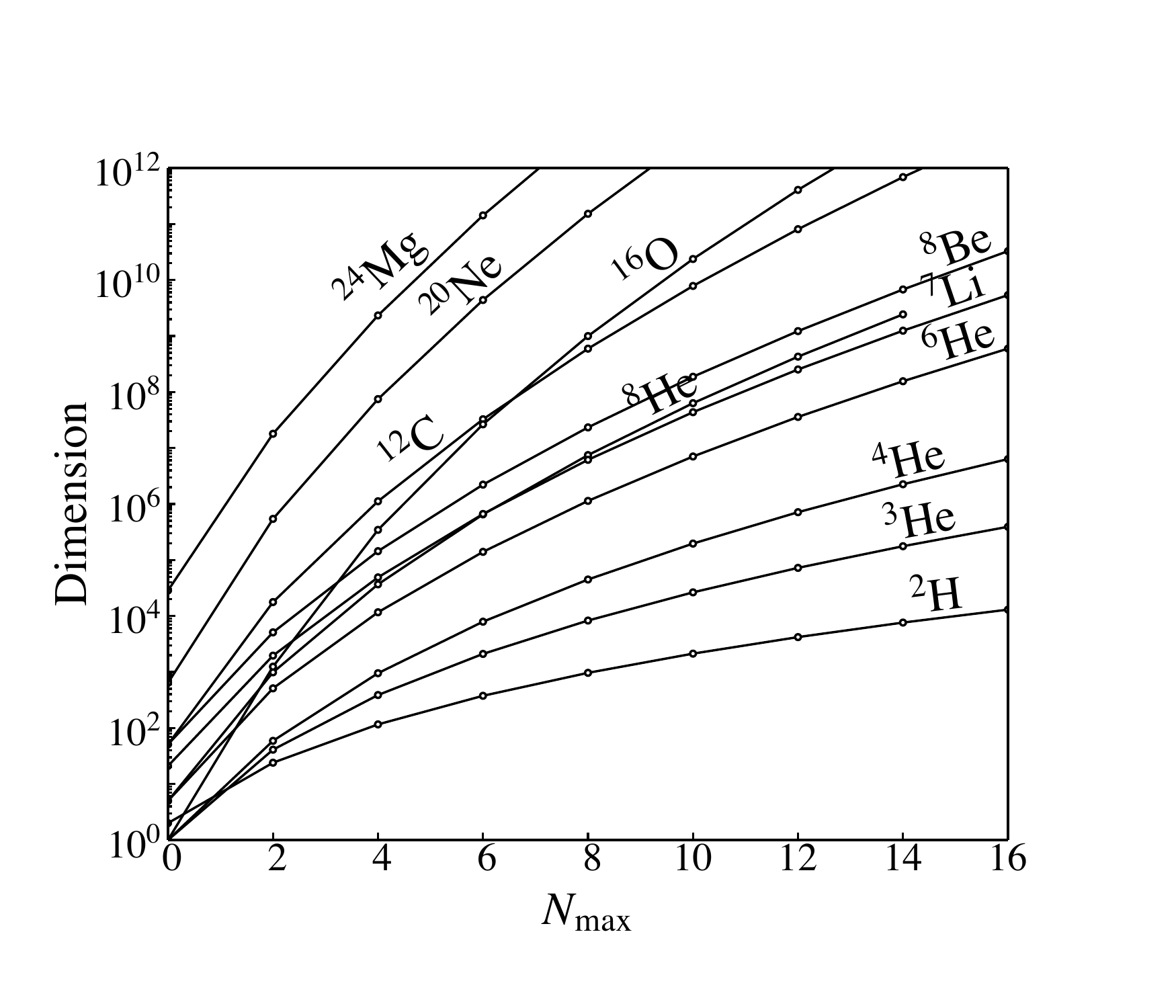}
\caption[The dimension of the no-core configuration interaction (NCCI) approach many-body basis as a function of $\Nmax$ for selected nuclei.]{The dimension of the NCCI many-body basis as a function of the $\Nmax$ truncation of the basis for selected nuclei. The dimensions shown are for spaces with zero angular momentum projection ($M = 0$) and natural parity (see Chapter~\ref{chap-chap2}).}
\label{fig-chap1-NNCI-dimension}
\end{center}
\end{figure}

The convergence problem arises from the fact that the traditionally used harmonic oscillator basis (which as explained in this thesis is not the best suited for NCCI calculations) does not lead to fast convergence in terms of the $\Nmax$ truncation. The size of the many-body basis grows rapidly with $\Nmax$ and the mass number of the nucleus under study. Therefore, reaching convergence at low $\Nmax$ truncations is highly desirable. In practice, calculations are limited by computational power to matrices with dimension $\sim 10^{10}$. In Figure~\ref{fig-chap1-NNCI-dimension}, the dimension of the NCCI many-body basis as a function of the $\Nmax$ truncation is plotted for selected nuclei with mass number $A \leq 24$ and for spaces with zero angular momentum projection ($M = 0$) and positive parity (for a detailed description of the many-body basis see Chapter~\ref{chap-chap2}).

\begin{figure}[t]
\begin{center}
\includegraphics[width=1.\textwidth]{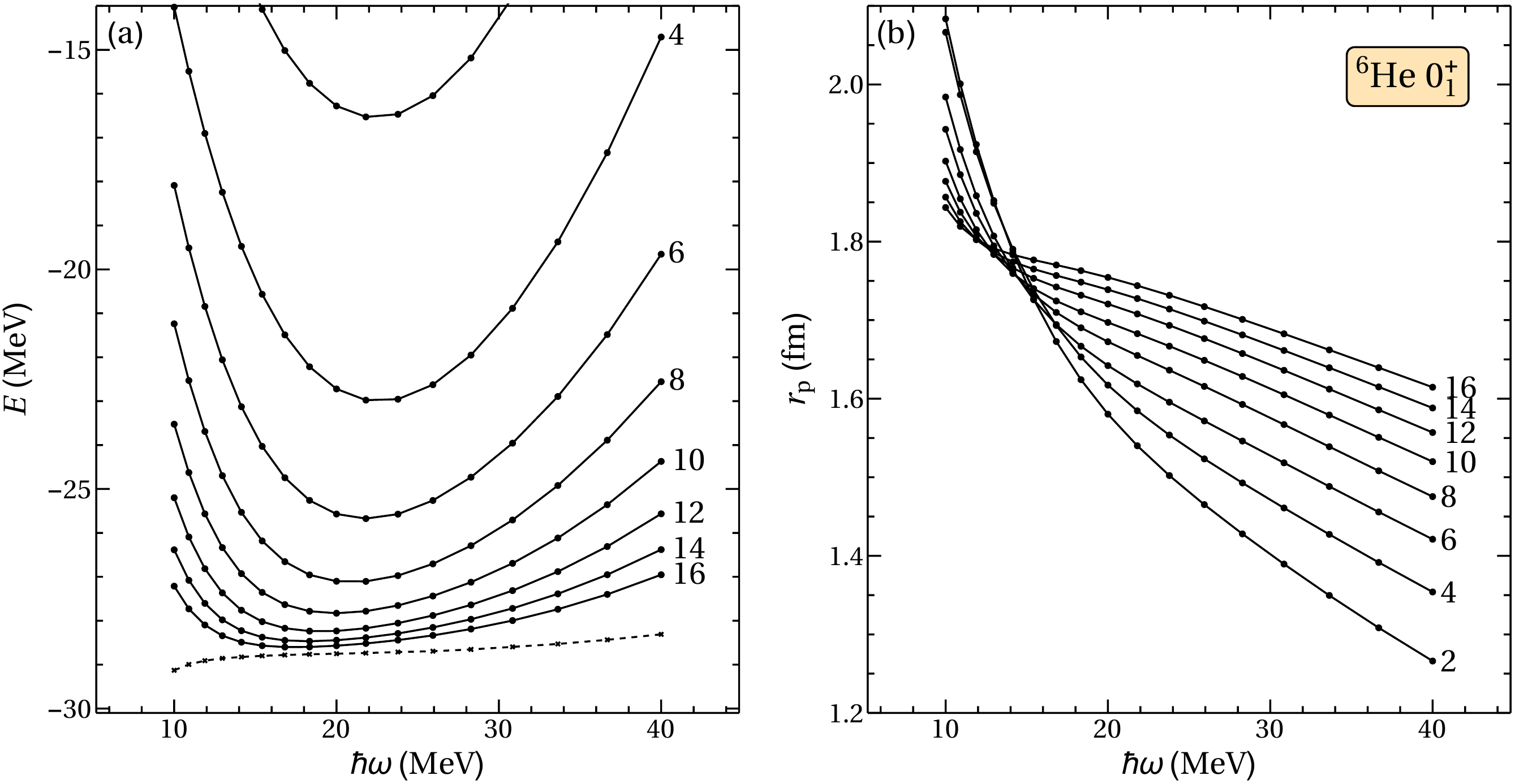}
\caption[The calculated ground state energy and proton radius in the $0^{+}$ ground state of {\protect $\isotope[6]{He}$}, obtained using the harmonic oscillator basis, the Coulomb interaction between protons, and the JISP$16$ $NN$ interaction. The calculated energies are extrapolated to the full space using an exponential extrapolation.]{The calculated ground state energy (a) and proton radius in the $0^{+}$ ground state (b) of $\isotope[6]{He}$, obtained in an NCCI calculation using the harmonic oscillator basis, the Coulomb interaction between protons, and the JISP$16$ $NN$ interaction. The $\Nmax$ truncation of the many-body basis is shown next to each successive curve. The dotted lines connect extrapolated results obtained using a three-point exponential extrapolation of the calculated ground state energy results (as described in the text).}
\label{fig-chap1-helium-6-e-rp}
\end{center}
\end{figure}

Without going into details, the results of an example NCCI calculation for the ground state energy and proton radius in the $0^{+}$ ground state of $\isotope[6]{He}$ are shown in Fig.~\ref{fig-chap1-helium-6-e-rp}, to illustrate the convergence properties of an NCCI calculation. The calculations were performed using the JISP$16$ $NN$ interaction, $\Nmax$ truncations up to $16$ (in steps of $2$ as shown next to each curve), and $\hw$ parameters in the range $10$-$40$ MeV (as shown in the horizontal axis).

As we mentioned above, full convergence of a calculated observable is signalled by an independence of the calculated results from both $\Nmax$ and $\hw$. Here, in Fig.~\ref{fig-chap1-helium-6-e-rp}(a) the calculated energy is approaching convergence around the variational minimum of the $\Nmax = 16$ curve. However, the calculated curves still depend on both $\hw$ and $\Nmax$ (i.e., full convergence is not reached). 

In order for the NCCI approach to acquire predictive power, the calculated results must converge. Ideally we want to achieve full convergence. However, even incompletely converged results may be extrapolated to the full $\Nmax \rightarrow \infty$ space using basis extrapolation methods~\cite{bogner2008:ncsm-converg-2N, maris2009:ncfc,cockrell2012:li-ncfc, coon2012:nscm-ho-regulator, furnstahl2012:ho-extrapolation, more2013:ir-extrapolation, furnstahl2014:ir-expansion, wendt2015:ir-extrapolations, furnstahl2015:ir-extrapolations, odell-papenbrock-ir-extrapolation-quadruple:2016}.

Here, we demonstrate a simple exponential extrapolation scheme which accounts for the incomplete convergence of the calculated energies. Results obtained at the three highest $\Nmax$ truncations (here $\Nmax = 12,\,14,\,16$) and the same $\hw$ are extrapolated to the full $\Nmax \rightarrow \infty$ space using the prescription $E(\Nmax) = E_{\infty} + A e^{-b \Nmax}$~\cite{maris2009:ncfc}. (The parameters $E_{\infty}$, $A$, and $b$ are fitted to the truncated results thus an estimate for the ground state energy $E \rightarrow E_{\infty}$ as $\Nmax \rightarrow \infty$ is obtained). The extrapolated results are shown with dotted lines in Fig.~\ref{fig-chap1-helium-6-e-rp}. They still depend on $\hw$, however the dependence is not as strong as the dependence of the calculated results (on $\hw$) at $\Nmax = 16$ for example. Despite the approximate $\hw$ independence of the extrapolated results, only a rough estimate of the converged result can be made based on the extrapolated results shown in Fig.~\ref{fig-chap1-helium-6-e-rp}. Obtaining reliable predictions using extrapolation methods is a matter of ongoing research. In Chapter~\ref{chap-chap2}, we will discuss the infrared extrapolation method which attempts to extrapolate NCCI calculations by putting $\Nmax$ and $\hw$ on an equal footing. Specifically, the approach is based on the premise that the truncated many-body basis actively imposes a cutoff on the ability of the basis to describe the many-body wave function.

The dependence of the calculated results on $\hw$ and $\Nmax$ is more severe for the calculated proton radius [Fig.~\ref{fig-chap1-helium-6-e-rp}(b)]. Convergence for this observable in terms of $\Nmax$ is so slow that no reliable quantitative estimate for the proton radius can be made based on Fig.~\ref{fig-chap1-helium-6-e-rp}(b).

Incomplete convergence might result from a poor description of the many-body wave function. The many-body wave function is expected to fall off exponentially ($\sim e^{-b r}$) at large distances. However, the harmonic oscillator basis used to describe the many-body wave function falls off as a Gaussian ($\sim e^{-b r^{2}}$). In Ref.~\cite{caprio2012:csbasis}, the Laguerre basis, which carries the correct exponential asymptotics, was introduced for the NCCI approach to overcome the slow convergence of observables in terms of $\Nmax$. However, it was found that the convergence of calculated energies and radii in the Laguerre basis is also slow~\cite{caprio2012:csbasis,  caprio2014:cshalo}.

In this work we introduce natural orbitals~\cite{loewdin1955:natural-orbitals-part1} in our attempt to overcome the slow convergence of observables. Natural orbitals are custom tailored orbitals adapted to the specific many-body problem (i.e., nucleus). They are obtained by diagonalizing a one-body density matrix deduced from an initial many-body calculation in a reference single-particle basis (in this work the initial calculations are performed using the harmonic oscillator single-particle basis). We will demonstrate that using natural orbitals reduces the need for high-lying single-particle orbitals (see Chapters~\ref{chap-chap3},~\ref{chap-chap4},~\ref{chap-chap5}), hence leading to faster convergence of observables in truncated spaces.

We begin with an overview of the no-core configuration interaction approach (Chapter~\ref{chap-chap2}). Natural orbitals for NCCI calculations are then derived and used in example calculations for $\isotope[3,4]{He}$ (Chapter~\ref{chap-chap3}). We then use natural orbitals to study the ground state properties of the halo nuclei $\isotope[6,8]{He}$ (Chapter~\ref{chap-chap4}) and the mirror nuclei $\isotope[7]{Li}$ and $\isotope[7]{Be}$ (Chapter~\ref{chap-chap5}). Finally, we offer some suggestions for further development of natural orbitals in future applications (Chapter~\ref{chap-chap6}).

%% file: chapters/chapter2/chapter2_draft_170402.tex
\chapter{INTRODUCTION TO THE NO-CORE CONFIGURATION INTERACTION APPROACH}
\label{chap-chap2}

\section{Overview}
\label{sec-chap2-overview}

In this chapter a broad overview of quantum many-body theory is presented focusing on the tools necessary to work with the no-core configuration interaction (NCCI) approach. The NCCI approach utilizes second quantization to cast the nuclear Hamiltonian into a square matrix in terms of a many-body basis. Here we describe all the steps needed to build and diagonalize the Hamiltonian matrix in some detail. Moreover, example calculations are presented and extrapolated using the infrared extrapolation method.

The general nuclear Hamiltonian for a system of $A$ nucleons interacting via two and/or three-body interactions $V(\mathbf{r}_{i},\mathbf{r}_{j})$ and $V(\mathbf{r}_{i},\mathbf{r}_{j},\mathbf{r}_{k})$ respectively is given by
\begin{equation}
H = \frac{1}{2m_{\mathrm{N}}} \sum\limits_{i = 1}^{A} p_{i}^{2} + \sum_{i < j}^{A} V(\mathbf{r}_{i},\mathbf{r}_{j}) + \sum_{i < j < k}^{A} V_{\mathrm{NNN}}(\mathbf{r}_{i},\mathbf{r}_{j},\mathbf{r}_{k}).
\label{eqn-chap2-overview-full-hamiltonian}
\end{equation}
Although three nucleon interactions are needed to accurately reproduce experimental data~\citep{pieperqmc:2008}, two-body interactions  provide a simpler context in which to focus on the main goal of this thesis, which is to introduce an alternative single-particle basis (other than the traditional harmonic oscillator basis) for NCCI calculations. Thus for the purpose of this thesis, we omit three-body terms from the Hamiltonian (\ref{eqn-chap2-overview-full-hamiltonian}).

Moreover, for reasons that have to do with the center-of-mass degree of freedom (described in Sec.~\ref{sec-chap2-com-factorization}) in the NCCI approach we do not work with the Hamiltonian (\ref{eqn-chap2-overview-full-hamiltonian}). Instead, we work with the translationally invariant intrinsic Hamiltonian \citep{vanhees:1983,lipkin:1958}
\begin{equation}
H_{\mathrm{in}} = T_{\mathrm{rel}} + V = \sum_{i \neq j}^{A} \frac{(\mathbf{p}_{i}-\mathbf{p}_{j})^{2}}{4 A m_{\mathrm{N}}} 
+ \sum_{i < j}^{A} V_{\mathrm{NN}}(\mathbf{r}_{i},\mathbf{r}_{j}) + \sum_{i < j}^{Z} V_{\mathrm{C}}(\mathbf{r}_{i},\mathbf{r}_{j}) ,
\label{eqn-chap2-overview-hamiltonian}
\end{equation}
where $m_{\mathrm{N}} c^{2} = 938.92$ MeV is the average nucleon mass, $V_{\mathrm{NN}}$ is the internucleon interaction, and $V_{\mathrm{C}}$ is the Coulomb interaction.

We start with a brief summary of second quantization (Sec.~\ref{sec-chap2-second-quantization}) which is then used to derive the two-body matrix elements of the NCCI Hamiltonian (Sec.~\ref{sec-chap2-tbme}). Consequently, we discuss the harmonic oscillator single-particle radial functions (Sec.~\ref{sec-chap2-harmonic-oscillator-radial-functions}) and the nuclear many-body basis (Sec.~\ref{sec-chap2-many-body-basis}). The complete factorization of the center-of-mass and intrinsic portions of the many-body wave function (Sec.~\ref{sec-chap2-com-factorization}) and a description of the procedure followed to perform many-body calculations using the NCCI approach (Sec.~\ref{sec-chap2-many-body-calculations}) follow. The one-body density matrix and its properties are reviewed (Sec.~\ref{sec-chap2-one-body-density-matrix}). Example calculations for the nuclei $\isotope[3]{He}$ and $\isotope[4]{He}$ are presented, and the calculated results are extrapolated using the infrared extrapolation method (Sec.~\ref{sec-chap2-example-calculations}). We conclude this chapter with an overview of the steps needed to perform an NCCI calculation using a general single-particle basis (Sec.~\ref{sec-chap2-general-basis}).

\section{Second quantization}
\label{sec-chap2-second-quantization}

Second quantization is used to describe systems of many identical particles. In what follows the notational conventions of Negele are used~\cite{negele1988:many-particle}.

The Hilbert space of $A$ distinguishable particles $ \mathcal{H}^{A}$ is spanned by the tensor product of $A$ distinguishable particles occupying single-particle states which belong to the single-particle Hilbert space $\mathcal{H}$ 
\begin{equation}
\mathcal{H}^{A} = \mathcal{H} \otimes \mathcal{H} \otimes \ldots \otimes \mathcal{H}.
\label{eqn-chap2-hilbert-space-product}
\end{equation}
If $\left\{ | a \rangle \right\}$ is an orthonormal basis for $\mathcal{H}$, an orthonormal basis for $\mathcal{H}^{A}$ can be constructed as the set of tensor product states
\begin{equation}
| a_{1} \ldots a_{A} ) \equiv |a_{1} \rangle \otimes |a_{2} \rangle \otimes \ldots \otimes |a_{A} \rangle,
\label{eqn-chap2-many-body-basis-product}
\end{equation}
where the curved parenthesis on the left hand side of (\ref{eqn-chap2-many-body-basis-product}) denotes that the many-body state is a simple product of single-particle states (i.e., no assumptions are made about the symmetry of the many-body state).

The space of $A$ indistinguishable fermions $\mathcal{F}^{A}$ is spanned by fully antisymmetric many-body states constructed using the many-body states of $\mathcal{H}^{A}$  
\begin{equation}
| a_{1} \ldots a_{A} \rangle = \frac{1}{\sqrt{A!}} \sum_{P} (-1)^{P} | a_{P_{1}} \rangle \otimes | a_{P_{2}} \rangle \otimes \ldots \otimes | a_{P_{A}} \rangle.
\label{eqn-chap2-many-body-basis-antisymmetric}
\end{equation} 
The sum over $P$ is a sum over all possible permutations $(P_{1},\,P_{2},\ldots,P_{A})$ of the set $(a_{1},\,a_{2},\,\ldots, a_{A})$. Moreover, $(-1)^{P}$ is determined by counting the number of transpositions of two elements which brings a permutation to its original form $(a_{1},\,a_{2},\,\ldots, a_{A})$.

The Fock space $F$ for fermions is the (Hilbert) direct sum of the fermion spaces $\mathcal{F}^{0}$, $\mathcal{F}^{1}$, $\mathcal{F}^{2}$, $\ldots$, $\mathcal{F}^{A}$, $\ldots$ for zero, one, two, etc particles
\begin{equation}
F = \mathcal{F}^{0} \oplus \mathcal{F}^{1} \oplus \mathcal{F}^{2} \oplus \ldots \oplus \mathcal{F}^{A} \oplus \ldots.
\label{eqn-chap2-fock-space}
\end{equation}
Here $\mathcal{F}^{0}$ is the space of no particles, $\mathcal{F}^{1}$ is identical to $\mathcal{H}$, and $\mathcal{F}^{A}$ with $A \geq 2$ is the space spanned by many-body states given by (\ref{eqn-chap2-many-body-basis-antisymmetric}).

Creation and annihilation operators generate the entire Fock space $F$ by adding (removing) single-particle states to (from) many-body states. The fermionic creation operator $c_{a}^{\dagger}$ adds a particle in a single-particle state $| a \rangle$ to the many-body state $| a_{1} \ldots a_{A} \rangle$  
\begin{equation}
c_{a}^{\dagger} | a_{1} a_{2} \ldots a_{A} \rangle = | a a_{1} a_{2} \ldots a_{A} \rangle,
\label{eqn-chap2-creation-operator}
\end{equation}
where the definition implies that the particle is added to the left of the initial many-body state. Now, because of the antisymmetry requirement for fermions we cannot act with the $c_{a}^{\dagger}$ operator on the state $| a a_{1} a_{2} \ldots a_{A} \rangle$ [and thus add the single-particle state $| a \rangle$ twice]. This state is equal to zero by definition
\begin{equation}
c_{a}^{\dagger} |a a_{1} a_{2} \ldots a_{A} \rangle = 0.
\label{eqn-chap2-creation-operator-2}
\end{equation}
Similarly an annihilation operator $c_{a}$ removes a particle occupying the single-particle state $|a\rangle$ from a many-body state
\begin{equation}
c_{a} |a a_{1} \ldots a_{A} \rangle = | a_{1} \ldots a_{A} \rangle.
\label{eqn-chap2-annihilation-operator}
\end{equation}
For completeness, note that in general $c_{a} | a_{1} a_{2} \ldots a \ldots a_{A} \rangle = (-1)^{P} | a_{1} a_{2} \ldots a_{A} \rangle$ where $P$ is the number of transpositions of two elements needed to bring the particle in the single-particle state $| a \rangle$ to the front of the many-body state. For example $c_{a_{2}} | a_{1} a_{2} a_{3} \rangle = - |a_{1} a_{3} \rangle$. If $c_{a}$ acts on a state $| a_{1} \ldots a_{A} \rangle$, i.e., a state which has no particle in the single-particle state $| a \rangle$, then the action of $c_{a}$ on the many-body state $| a_{1} \ldots a_{A} \rangle$ yields zero
\begin{equation}
c_{a} |a_{1} \ldots a_{A} \rangle = 0.
\label{eqn-chap2-annihilation-operator-2}
\end{equation}
Following from~(\ref{eqn-chap2-creation-operator})-(\ref{eqn-chap2-annihilation-operator-2}), we conclude that the creation and annihilation operators obey the fermionic anticommutation relation
\begin{equation}
\lbrace c_{a} , c_{b}^{\dagger} \rbrace = \delta_{ab}.
\label{eqn-chap2-anti-commutation}
\end{equation}

One-body operators are operators which can be written as a sum of single-particle operators acting on one particle at a time. To be able to calculate the matrix elements of one-body operators with respect to many body states like (\ref{eqn-chap2-many-body-basis-antisymmetric}), it is useful to express them in terms of creation and annihilation operators. To arrive at this expression let us first use the many-body states of $\mathcal{H}^{A}$ to calculate the matrix elements of a one-body operator $T = \sum_{i=1}^{A} t_{i}$. For the diagonal matrix elements (i.e., between the same many-body states) we have
\begin{multline}
( a_{1}a_{2} \ldots a_{A} | T | a_{1} a_{2} \ldots a_{A} ) = \\  
\langle a_{1} | t_{1} | a_{1} \rangle \langle a_{2} | a_{2} \rangle \ldots \langle a_{A} | a_{A} \rangle + \ldots + \langle a_{1} | a_{1} \rangle \ldots \langle a_{A-1} | a_{A-1} \rangle  \langle a_{A} | t_{A} | a_{A} \rangle = \\
\langle a_{1} | t_{1} | a_{1} \rangle + \ldots + \langle a_{A} | t_{A} | a_{A} \rangle,
\label{eqn-chap2-one-body-operator-matrix-elements-identical-states}
\end{multline}
where we used the fact that the single-particle states are orthonormal hence $\langle a_{i} | a_{j} \rangle = \delta_{a_{i}a_{j}}$. Let us now evaluate the non-diagonal matrix elements of $T$ starting with the matrix elements between the many-body states $ | a_{1} a_{2} \ldots a_{A} )$ and $ | a_{1} a_{2}' \ldots a_{A} )$, where $| a_{2} \rangle \neq | a_{2}' \rangle$. We have
\begin{multline}
( a_{1} a_{2} \ldots a_{A} | T | a_{1} a_{2}' \ldots a_{A} ) = \\
\langle a_{1} | t_{1} | a_{1} \rangle \langle a_{2} | a_{2}' \rangle \ldots \langle a_{A} | a_{A} \rangle + \langle a_{1} | a_{1} \rangle \langle a_{2} | t_{2} | a_{2}' \rangle \langle a_{3} | a_{3} \rangle \ldots \langle a_{A} | a_{A} \rangle + \\
+ \ldots + \langle a_{1} | a_{1} \rangle \ldots \langle a_{A-1} | a_{A-1} \rangle  \langle a_{A} | t_{A} | a_{A} \rangle = \langle a_{2} | t_{2} | a_{2}' \rangle,
\label{eqn-chap2-one-body-operator-matrix-elements-one-sp-state-different}
\end{multline}
since $\langle a_{2} | a_{2}' \rangle = 0 $. Finally, the matrix elements of $T$ between the many-body states $| a_{1} a_{2} \ldots a_{A} )$ and $| a_{1}' a_{2}' \ldots a_{A} )$, where $| a_{1} \rangle \neq | a_{1}' \rangle$ and $| a_{2} \rangle \neq | a_{2}' \rangle$, are
\begin{multline}
( a_{1} a_{2} \ldots a_{A} | T | a_{1}' a_{2}' \ldots a_{A} ) = \\
\langle a_{1} | t_{1} | a_{1}' \rangle \langle a_{2} | a_{2}' \rangle \ldots \langle a_{A} | a_{A} \rangle + \ldots + \langle a_{1} | a_{1}' \rangle \ldots \langle a_{A} | t_{A} | a_{A} \rangle = 0.
\label{eqn-chap2-one-body-operator-matrix-elements-two-sp-state-different}
\end{multline}
Thus, the matrix elements of the one-body operator $T$ with respect to many-body states of $\mathcal{H}^{A}$ are fully defined by the sum over single-particle matrix elements of $\mathcal{H}$. If we now define the single-particle vacuum state $| - \rangle$ (i.e., a non-occupied single-particle state), then $c_{a}^{\dagger} | - \rangle = | a \rangle$. Using this definition and (\ref{eqn-chap2-one-body-operator-matrix-elements-identical-states})-(\ref{eqn-chap2-one-body-operator-matrix-elements-two-sp-state-different}) we can write a one-body operator as
\begin{equation}
T = \sum_{ab} t_{ab} c_{a}^{\dagger} c_{b},
\label{eqn-chap2-one-body-operator-second-quantization}
\end{equation}
where $t_{ab} = \langle a | t | b \rangle$.

Similarly, the matrix elements of a two-body operator $V = \frac{1}{2} \sum_{i \neq j} u_{ij}$ with respect to the many-body states of $ \mathcal{H}^{A}$ are fully defined by the matrix elements $u_{abcd} = (ab | u | cd )$ calculated using the states of $\mathcal{H}^{2}$. Thus we can write $V$ as
\begin{equation}
V = \frac{1}{2} \sum_{abcd} u_{abcd} c_{a}^{\dagger} c_{b}^{\dagger} c_{d} c_{c}.
\label{eqn-chap2-interaction-second-quantized}
\end{equation}
However, it is more convenient to write the two-body operator as
\begin{equation}
V = \frac{1}{4} \sum_{abcd} \bar{u}_{abcd} c_{a}^{\dagger} c_{b}^{\dagger} c_{d} c_{c},
\label{eqn-chap2-interaction-second-quantized-antisymmetrized}
\end{equation}
where $\bar{u}_{abcd} = u_{abcd} - u_{abdc} = \langle ab | u | cd \rangle$ are normalized and antisymmetrized two-body matrix elements.

\section{Two-body matrix elements for the NCCI Hamiltonian}
\label{sec-chap2-tbme}

In the context of the NCCI approach, the most commonly used single-particle states are the harmonic oscillator single-particle states (see Sec.~\ref{sec-chap2-harmonic-oscillator-radial-functions}). The harmonic oscillator states are labeled by the quantum numbers $| \alpha \rangle \equiv |n_{a}l_{a}j_{a}m_{a} \rangle$, where $n_{a}$ is the harmonic oscillator radial quantum number, $l_{a}$ is the angular momentum, $j_{a}$ is the total angular momentum, and $m_{a}$ is the $z$ projection of the total angular momentum. Using the harmonic oscillator single-particle states we can construct fully antisymmetric many-body basis states for the NCCI approach
\begin{equation}
| (n_{a}l_{a}j_{a}m_{a}) (n_{b}l_{b}j_{b}m_{b}) \ldots (n_{A}l_{A}j_{A}m_{A}) \rangle.
\label{eqn-chap2-many-body-state-harmonic-oscillator}
\end{equation} 
Many-body states such as (\ref{eqn-chap2-many-body-state-harmonic-oscillator}) have a total $z$ projection of total angular momentum $M=\sum_{i=1}^{A} m_{i}$ and a total number of oscillator quanta $N_{\mathrm{tot}}=\sum_{i=1}^{A} (2n_{i} + l_{i}) = \sum_{i=1}^{A} N_{i}$. The harmonic oscillator single-particle states $| \alpha \rangle$ are given as the tensor product 
\begin{equation}
| \alpha \rangle = \left[| R_{n_{a} l_{a}} Y_{l_{a}} \rangle \times |\frac{1}{2} \rangle \right]_{j_{a}m_{a}},
\label{eqn-oscillator-orbitals-second-quantization}
\end{equation}
where $R_{n_{a}l_{a}}$ is a harmonic oscillator radial function, $Y_{l_{a}}$ is a spherical harmonic, and $|\frac{1}{2} \rangle$ is a spinor. The subscripts $j_{a}$ and $m_{a}$ indicate that $| R_{n_{a} l_{a}} Y_{l_{a}} \rangle$ (a spherical tensor of rank $l_{a}$) and the spinor $| \frac{1}{2} \rangle$ are coupled to total angular momentum $j_{a}$ and $z$ projection of total angular momentum $m_{a}$~\cite{rose1957:am}.

The Hilbert space of the NCCI Hamiltonian is spanned by the many-body states (\ref{eqn-chap2-many-body-state-harmonic-oscillator}) which provide a basis for the representation of the Hamiltonian as a square matrix. To cast the Hamiltonian into a square matrix we must first calculate the two-body matrix elements of the relative kinetic energy and two-body $NN$ interaction operators. The two-body matrix elements are calculated with respect to two-particle states, coupled to total angular momentum $J$. For distinguishable particles (e.g., one proton and one neutron states), angular momentum coupled two-particle states are defined as 
\begin{equation}
| a b ; J ) = \sum_{m_{a}m_{b}} \langle j_{a} m_{a} j_{b} m_{b} | J M \rangle |a m_{a} \rangle | b m_{b} \rangle,
\label{eqn-chap2-two-particle-states-1}
\end{equation}
where $| a m_{a} \rangle = | n_{a} l_{a} j_{a} m_{a} \rangle$, $\langle j_{a} m_{a} j_{b} m_{b} | J M \rangle $ is a Clebsch-Gordan coefficient, and the parenthesis implies that the two-particle state is not antisymmetrized. For two identical fermions we can obtain fully antisymmetrized states using the states $| a b ; J )$. We have
\begin{equation}
| a b ; J M \rangle_{\mathrm{AS}} = \frac{1}{\sqrt{2}} \left[ | a b ; J M ) - (-1)^{J - j_{a} - j_{b}} | b a ; J M \rangle \right].
\label{eqn-chap2-two-particle-states-2}
\end{equation}
These states have the symmetry property $| a b ; J M \rangle = - (-1)^{J -j_{a}-j_{b}} | b a ; J M \rangle$ which implies that if the states $| a \rangle$ and $| b \rangle$ are identical, only two-body states with even $J$ are allowed. The states (\ref{eqn-chap2-two-particle-states-2}) are antisymmetrized but not strictly normalized. An extra factor of $1 / \sqrt{2}$ is required for normalization in the special case when the two particles occupy the same single-particle orbital $| n_{a} l_{a} j_{a} \rangle$. Hence, the appropriate two-particle states are
\begin{equation}
| a b ; J M \rangle_{\mathrm{NAS}} = (1 + \delta_{ab})^{-1/2} | a b ; J M \rangle_{\mathrm{AS}}.
\label{eqn-chap2-two-particle-states-3}
\end{equation}
Both the normalized antisymmetrized states and the antisymmetrized states are used to calculate the two-body matrix elements $\langle c d ; J | H_{\mathrm{in}} | a b ; J \rangle$ of the (scalar) intrinsic Hamiltonian (\ref{eqn-chap2-overview-hamiltonian}). The change of basis relation between two-body matrix elements calculated with respect to the states (\ref{eqn-chap2-two-particle-states-2}) and two-body matrix elements calculated with respect to the states (\ref{eqn-chap2-two-particle-states-3}) is
\begin{equation}
\langle c d ; J | H | a b ; J \rangle_{\mathrm{NAS}} = (1+\delta_{cd})^{-1/2} (1+\delta_{ab})^{-1/2} \langle c d ; J | H | a b ; J \rangle_{\mathrm{AS}}.
\label{eqn-chap2-nas-two-body-matrix-elements}
\end{equation}
A detailed discussion about the calculation of the two-body matrix elements of the intrinsic Hamiltonian (\ref{eqn-chap2-overview-hamiltonian}) is given in Ref.~\cite{caprio2012:csbasis}. A brief overview is also given in Sec.~\ref{sec-chap2-general-basis}.

\section{The harmonic oscillator single-particle states}
\label{sec-chap2-harmonic-oscillator-radial-functions}

In this section we will review the basic properties of the harmonic oscillator single-particle states $| \alpha \rangle$~\cite{suhonen2007:nucleons-nucleus,moshinsky1996:oscillator,weniger1985:fourier-plane-wave} which are traditionally used in the NCCI approach because their properties facilitate the many-body calculations. The harmonic oscillator single-particle states are also our starting basis for the construction of natural orbitals which are introduced in Chapter~\ref{chap-chap3}.

The first important property of the harmonic oscillator single-particle states (relevant to the NCCI approach) is that they allow for the complete removal of spurious center-of-mass states from the low-lying spectrum of the NCCI Hamiltonian provided that an $\Nmax$ truncation is imposed on the many-body basis. The appearance of spurious center-of-mass states is related to the use of the many-body basis states (\ref{eqn-chap2-many-body-state-harmonic-oscillator}) for the representation of the NCCI Hamiltonian as a square matrix. A detailed explanation for the appearance of spurious center-of-mass states in an NCCI calculation and the procedure followed for their removal from the low-lying spectrum is given in Sec.~\ref{sec-chap2-many-body-basis}.

The second important property of the harmonic oscillator single-particle states (relevant to many-body calculations in general and in particular the NCCI approach) is that they simplify the calculation of two-body matrix elements for operators written in terms of relative coordinates, using the Moshinsky transformation~\citep{moshinsky1996:oscillator}. According to  the transformation, products of harmonic oscillator states expressed in single-particle coordinates $R_{n_{1}l_{1}}(\mathbf{r}_{1}) R_{n_{2}l_{2}}(\mathbf{r}_{2})$ transform into products of harmonic oscillator states expressed in terms of relative and center-of mass coordinates $R_{nl}(\mathbf{r}_{\mathrm{rel}})R_{NL}(\mathbf{R})$, where $2 n_{1} + l_{1} + 2 n_{2} + l_{2}= 2 n + l+ 2 N + L$, $\mathbf{r}_{\mathrm{rel}} = \frac{1}{\sqrt{2}} (\mathbf{r}_{1}-\mathbf{r}_{2})$, and $\mathbf{R} = \frac{1}{\sqrt{2}} (\mathbf{r}_{1}+\mathbf{r}_{2})$. (The factor of $1/\sqrt{2}$ in front of the center-of-mass vector is used instead of the traditional $1/2$ for normalization reasons). The calculation of two-body matrix elements for operators of the form $V(\mathbf{r}_{1}-\mathbf{r}_{2})$ with respect to two-particle states of the form $R_{n_{1}l_{1}} (\mathbf{r}_{1}) R_{n_{2}l_{2}}(\mathbf{r}_{2})$ is then reduced to the calculation of the single-particle matrix element of the operator $V(\mathbf{r}_{\mathrm{rel}})$ with respect to single-particle states of the form $R_{nl}(\mathbf{r}_{\mathrm{rel}})$ and an overlap between single-particle states of the form $R_{NL}(\mathbf{R})$.

Let us now review the harmonic oscillator radial wave functions. If we write the single-particle wave function as $\Psi_{nlm}(\mathbf{r})=r^{-1} R_{nl}(b;r)Y_{lm}(\theta,\phi)$, where $R_{nl}(b;r)$ are radial wave functions and $Y_{lm}(\theta,\phi)$ are spherical harmonics, then $\Psi_{nlm}(\mathbf{r})$ are solutions to the familiar harmonic oscillator central force problem \cite{suhonen2007:nucleons-nucleus} 
\begin{equation}
h(\omega) = \frac{p^{2}}{2m_{\mathrm{N}}} + \frac{m_{\mathrm{N}} \omega^{2} r^{2}}{2},
\label{eqn-chap2-harmonic-oscillator-one-body-hamiltonian}
\end{equation} 
where $m_{\mathrm{N}}c^{2} \approx 938.92$ MeV is the average nucleon mass and $\omega$ is the oscillator frequency. The length scale of the radial functions $R_{nl}(b;r)$ is set by the oscillator length $b$ which depends on the parameters of the oscillator Hamiltonian (\ref{eqn-chap2-harmonic-oscillator-one-body-hamiltonian}) as $b=\sqrt{\hbar / (m_{\mathrm{N}} \omega})$. The eigenvalues corresponding to each single-particle radial wave function are given by $ (N + 3/2) \hw$, where $N = 2n + l$ is the oscillator quantum (notice that the eigenvalues are evenly spaced by one unit of $\hw$). The radial functions $R_{nl}(b;r)$ are given by
\begin{equation}
R_{nl}(b;r) = b N_{nl} (r/b)^{l+1} L_{n}^{l+1/2}[(r/b)^{2}] e^{-(r/b)^{2}/2},
\label{eqn-chap2-harmonic-oscillator-radial-functions}
\end{equation} 
where $L_{n}^{l+1/2}[(r/b)^{2}]$ are generalized Laguerre polynomials, $n$ is the radial quantum number (which gives the number of nodes in the radial function), $l$ is the angular momentum, and $N_{nl}$ is a normalization factor given by
\begin{equation}
N_{nl} = \frac{1}{b^{3/2}} \left[ \frac{2n!}{(l+n+1/2)!} \right]^{1/2}.
\label{eqn-chap2-harmonic-oscillator-functions-normalization}
\end{equation}
For each $l$, the harmonic oscillator radial functions form a complete discrete basis for square integrable functions on $ \mathbb{R}^{+}$
\begin{equation}
\int_{0}^{\infty} R_{nl}(b;r) R_{n'l}(b;r) = \delta_{nn'}.
\label{eqn-chap2-ho-orthonormal}
\end{equation}

\section{Symmetries of the NCCI Hamiltonian and the nuclear many-body \\ basis}
\label{sec-chap2-many-body-basis}

The selection of the nuclear many-body basis used in NCCI calculations is based on the symmetries of the NCCI Hamiltonian~(\ref{eqn-chap2-overview-hamiltonian}). The NCCI Hamiltonian is rotationally invariant therefore it conserves the total angular momentum $J$ and the $z$ projection of total angular momentum $M$. Moreover, the Hamiltonian conserves parity. To diagonalize the Hamiltonian we can choose to use many-body states that have good $J$ (a $J$-scheme basis) or many-body states that have good $M$ (an $M$-scheme basis)~\cite{whitehead1977:shell-methods}. Although the $M$-scheme basis involves a larger number of many-body basis states compared to a $J$-scheme basis, constructing $M$-scheme many-body basis states is straightforward (i.e., we do not have to deal with angular momentum coupling as in the $J$-scheme basis). Therefore, for the NCCI calculations discussed here, an $M$-scheme basis is considered. However, the eigenstates of $H$ still have good $J$ which is recovered after diagonalization by calculating the expectation value of the $J^{2}$ operator with respect to the calculated many-body wave function.

To build many-body basis states we start from single-particle states $| nljm \rangle$ and construct fully antisymmetric states (with good $M$) given by~(\ref{eqn-chap2-many-body-state-harmonic-oscillator}). Calculations can be performed for any possible value of $M$ (which is supported by the single-particle states in the many-body basis). For example, for even nuclei $M=0, 1,\ldots$ and for odd nuclei $M=1/2, 3/2, \ldots$ (we do not consider negative values for $M$ here). However, we usually use many-body states with $M=0$ for even nuclei and $M=1/2$ for odd nuclei. This is because for a given $M$ we can only obtain many-body wave functions with $J \geq | M |$. Thus, choosing the lowest possible $M$ allows us to study the ground and first few excited states of a nucleus.

Commonly, for reasons involving the center-of-mass degree of freedom, as discussed in Sec.~\ref{sec-chap2-com-factorization}, the many-body basis is truncated using the $\Nmax$ truncation scheme~\cite{elliott1955:com-shell}. The scheme dictates that only many-body states with $\Ntot \leq N_{0} + \Nmax$ are permitted in the many-body basis, where $N_{0}$ is the number of quanta in the configuration where all the nucleons occupy the lowest permitted oscillator shells. For example, $N_{0}=2$ for $\isotope[6]{He}$ (in this configuration, the two protons occupy the $N=0$ proton shell, two neutrons occupy the $N=0$ neutron shell, and the last two neutrons occupy the $N=1$ shell) and $N_{0} =3$ for $\isotope[7]{Li}$ (the extra proton, compared to $\isotope[6]{He}$, goes into the $N=1$ proton shell).

Finally, because the nuclear Hamiltonian conserves parity we want our many-body basis to have good parity. Recall that the parity of a harmonic oscillator single-particle state is given by $(-1)^{l}$, or, equivalently, $(-1)^{N}$ since $N=2n+l$. A many-body basis truncated at a given $\Nmax$ contains (many-body) states with $N_{\mathrm{tot}} \leq N_{0} + \Nmax$, where $N_{\mathrm{tot}} = \sum_{i} N_{i} = N_{1} + \ldots + N_{A}$, and $N_{i}$ is the oscillator quantum of a nucleon $i$ in its single-particle state. Hence, many-body basis states which belong to this (many-body) basis have parity which is obtained as the product $(-1)^{N_{1}}\ldots (-1)^{N_{A}} = (-1)^{\Ntot}$. The parity of the many-body states of the lowest allowed configuration is $(-1)^{N_{0}}$, which we call the natural parity of the nucleus. Thus, to obtain the natural parity eigenstates of the nucleus we must build many-body bases with even $\Nmax$ truncations, i.e., $\Nmax = 0, 2, 4, \ldots$. On the other hand, to obtain the unnatural parity eigenstates of the nucleus we must construct many-body bases with odd $\Nmax$ truncations, i.e., $\Nmax = 1, 3, 5, \ldots$. 


\section{Spurious center-of-mass states removal}
\label{sec-chap2-com-factorization}

In this section we describe how spurious center-of-mass states result from the diagonalization of the NCCI Hamiltonian matrix built in terms of the set of the many-body basis states (\ref{eqn-chap2-many-body-state-harmonic-oscillator}). Subsequently, we describe how these spurious states can be removed provided that the $\Nmax$ truncation scheme is used.

The many-body basis states (\ref{eqn-chap2-many-body-state-harmonic-oscillator}) are defined with respect to a fixed point in space. However, physically there is no such point around which all the nucleons are orbiting. Rather we are interested in the internal motion of the nucleons relative to each other within the nucleus, a motion which defines the intrinsic structure of the nucleus.

In principle we can convert to relative coordinates. However, antisymmetrization in Jacobi coordinates (which define the relative motion) is cumbersome. Therefore, we are compelled to work in the full coordinate space which includes both the center of mass and relative degrees of freedom.

Our goal is to describe the relative motion as accurately as possible without unnecessary complications arising from the center of mass motion. In order to accomplish this it helps making certain choices regarding the basis and the Hamiltonian.

We want our eigenfunctions to factorize into center of mass and relative factors. We also want the center of mass factor to be simple and well understood so that it does not interfere with the calculation of energies and observables. In principle, the full Hamiltonian (\ref{eqn-chap2-overview-full-hamiltonian}) is already separated into center of mass and relative coordinates. If we define the center-of-mass momentum as $\mathbf{P} = \sum_{i} \mathbf{p}_{i}$, the one-body kinetic energy operator in (\ref{eqn-chap2-overview-full-hamiltonian}) can be decomposed into a center-of-mass and a relative kinetic energy contributions~\cite{vanhees:1983}
\begin{equation}
T = T_{\mathrm{c.m.}} + T_{\mathrm{rel}} = \frac{\left(\sum\limits_{i=1}^{A}\mathbf{p}_{i}\right)^{2}}{2Am_{\mathrm{N}}}  + \sum_{i \neq j}^{A} \frac{(\mathbf{p}_{i}-\mathbf{p}_{j})^{2}}{4 A m_{\mathrm{N}}}.
\label{eqn-chap2-kinetic-energy-decomposition}
\end{equation}
Therefore, the full Hamiltonian (\ref{eqn-chap2-overview-full-hamiltonian}) also separates into center-of-mass and relative parts
\begin{equation}
H= T_{\mathrm{c.m.}} + (T_{\mathrm{rel}} + V).
\label{eqn-chap2-full-separated-hamiltonian}
\end{equation}
Thus, in the full space our eigenfunctions would factorize into a center-of-mass and a relative wave function.

However, we are confined to work in a truncated space. Factorization can still be exact in an oscillator basis when the $\Nmax$ truncation is used. The truncated space then only contains a limited center-of-mass space spanned by center-of-mass harmonic oscillator states with $N_{\mathrm{c.m.}} = 0$, $2$, $\ldots, \Nmax$. We could attempt to diagonalize the Hamiltonian~(\ref{eqn-chap2-full-separated-hamiltonian}) in the truncated space. However, the $T_{\mathrm{c.m.}}$ operator will mix contributions with different center-of-mass excitations $N_{\mathrm{c.m.}}$ and destroy factorization of the eigenfunctions. The Hamiltonian will force us towards states which attempt to diagonalize the center-of-mass kinetic energy operator $T_{\mathrm{c.m.}}$. Therefore, our NCCI calculations would end up approximating spherical waves in the center of mass coordinates, at the expense of accurately describing the intrinsic motion.

Thus, instead, we choose to work with the intrinsic Hamiltonian
\begin{equation}
H = T_{\mathrm{rel}} + V.
\end{equation} 
The resulting Hamiltonian is block diagonal with respect to the number of quanta $N_{\mathrm{c.m.}}$ for the center-of-mass motion. The resulting eigenfunctions will factorize into a center-of-mass factor of good $N_{\mathrm{c.m.}}$ and an intrinsic factor. The total number of quanta in a many-body state is shared between the center-of-mass motion and the intrinsic motion:
\begin{equation}
\Ntot = \sum_{i} N_{i} = N_{\mathrm{c.m.}} + N_{\mathrm{rel}}.
\label{eqn-chap2-number-of-quanta-com-quanta}
\end{equation}
Therefore, in eigenfunctions with $N_{\mathrm{c.m.}} = 0$, the full $\Nmax$ quanta are available to use in describing the intrinsic motion. The higher $N_{\mathrm{c.m.}}$ eigenfunctions produce spurious copies of the spectrum in which the intrinsic motion is described using fewer quanta $N_{\mathrm{rel}} \leq \Nmax - N_{\mathrm{c.m.}}$.

It is convenient to push the spurious excited center-of-mass states out of the low lying spectrum. We do this by adding a term diagonal in the number of center-of-mass quanta. That is we add a ``Lawson term''~\cite{gloeckner1974:spurious-com} proportional to the $N_{\mathrm{c.m.}}$ operator. This is often equivalently described as adding a center-of-mass harmonic oscillator Hamiltonian. In terms of the center-of-mass coordinate $\mathbf{R} = (\sum_{i} \mathbf{r}_{i})/A$ and momentum $\mathbf{P} = \sum_{i} \mathbf{p}_{i}$, 
\begin{equation}
H_{\mathrm{c.m.}} = \frac{P^{2}}{2Am_{\mathrm{N}}} + \frac{1}{2} A m_{\mathrm{N}} \omega^{2} R^{2}.
\label{eqn-chap2-harmonic-oscillator-many-body-hamiltonian-com}
\end{equation}
Then, $H_{\mathrm{c.m.}} = (N_{\mathrm{c.m.}} + 3/2)\hbar \omega$. Thus we actually diagonalize the Hamiltonian
\begin{equation}
H = T_{\mathrm{rel}} + V + \alpha N_{\mathrm{c.m.}},
\label{eqn-chap2-ncci-hamiltonian-with-lawson}
\end{equation} 
where $\alpha$ is the Lawson term strength (typically a few MeV). This leads to the same eigenstates as the relative Hamiltonian but with the spurious states lifted out of the low-lying spectrum by an amount $\alpha N_{\mathrm{c.m.}}$.

\section{Many-body calculations}
\label{sec-chap2-many-body-calculations}

In this section we will briefly describe the general procedure followed to obtain results using the NCCI approach.

The basic ingredient needed to build the Hamiltonian matrix is the matrix elements of the NCCI Hamiltonian (\ref{eqn-chap2-ncci-hamiltonian-with-lawson}) with respect to the many-body basis states (\ref{eqn-chap2-many-body-state-harmonic-oscillator}). Thus, we need to evaluate matrix elements of the form
\begin{equation}
\langle \Phi_{n} | H | \Phi_{m} \rangle,
\label{eqn-chap2-two-body-matrix-elements}
\end{equation}
where $| \Phi_{n} \rangle = | (n_{1}l_{1}j_{1}m_{1}) \ldots (n_{A}l_{A}j_{A}m_{A}) \rangle$ is a many-body basis state. [The matrix elements (\ref{eqn-chap2-two-body-matrix-elements}) reduce to a sum over two-body matrix elements of the form $\langle c d ; J | H | a b ; J \rangle$ using Wick's theorem]. After calculating the matrix elements (\ref{eqn-chap2-two-body-matrix-elements}) for all the many-body states in our basis we obtain the Hamiltonian matrix
\begin{equation}
  \begin{pmatrix}
    \langle \Phi_{1} | H | \Phi_{1} \rangle & \langle \Phi_{1} | H | \Phi_{2} \rangle & \ldots & \langle \Phi_{1} | H | \Phi_{m} \rangle \\
    \langle \Phi_{2} | H | \Phi_{1} \rangle & \langle \Phi_{2} | H | \Phi_{2} \rangle & \ldots & \langle \Phi_{2} | H | \Phi_{m} \rangle \\
    \vdots & \vdots & \ddots & \vdots \\
   \langle \Phi_{d} | H | \Phi_{1} \rangle & \langle \Phi_{d} | H | \Phi_{2} \rangle & \ldots & \langle \Phi_{d} | H | \Phi_{d} \rangle
  \end{pmatrix}
\label{eqn-chap2-ncci-hamiltonian-matrix}
\end{equation}
The dimension $d$ of the matrix (\ref{eqn-chap2-ncci-hamiltonian-matrix}) is equal to the number of many-body basis states for a given nucleus and $\Nmax$ truncation (see Fig.~\ref{fig-chap1-NNCI-dimension}). Diagonalization of (\ref{eqn-chap2-ncci-hamiltonian-matrix}) yields the nuclear many-body wave function
\begin{equation}
| \Psi \rangle = \sum\limits_{n=1}^{d} c_{n} | \Phi_{n} \rangle.
\label{eqn-chap2-ncci-wave-functions}
\end{equation}

The diagonalization of the Hamiltonian matrix is commonly performed using the Lanczos algorithm~\cite{komzsik2003:lanczos-method,lanczos1950:algorithm}. The algorithm transforms the generally large sparse matrix (\ref{eqn-chap2-ncci-wave-functions}) into a tridiagonal matrix (of lower dimension than the original matrix). The tridiagonal matrix is then diagonalized (using an appropriate algorithm) and the eigenvalues and eigenvectors (of the tridiagonal matrix) are obtained. The eigenvectors of the tridiagonal matrix (Lanczos vectors) must then be converted from the Lanczos vector space to the original many-body basis. The process established by Lanczos proceeds as follows: If $H$ is a Hamiltonian matrix of order $n$ (i.e., the matrix dimension is $n \times n$), then we start with a random initial vector $b_{0}$ which is used to construct $m$ vectors (orthogonal to each other) according to the algorithm
\begin{align}
b_{0} & = \mathrm{random}, \nonumber & \\
b_{1} & = (H - \alpha_{0}) b_{0}, \nonumber \\
b_{2} & = (H - \alpha_{1}) b_{1} - \beta_{0} b_{0}, \nonumber \\
b_{3} & = (H - \alpha_{2}) b_{2} -\beta_{1} b_{1}, \nonumber \\
 \vdots  \nonumber & \\
b_{m} & = (H - \alpha_{m-1}) b_{m-1} - \beta_{m-2} b_{m-2} = 0,
\label{eqn-chap2-lanczos-algorithm}
\end{align}
where $m \leq n$. The equality to zero in the last step means the end of the process. The coefficients $\alpha_{m-1}$ and $\beta_{m-2}$ are determined by requiring that the norm of the vector $\parallel b_{m} \parallel$ is minimized. The coefficients are then used to build the tridiagonal matrix 
\begin{equation}
T =  \begin{pmatrix}
\alpha_{0} & \beta_{0} &   &   & 0\\ 
\beta_{0}  & \alpha_{1} & \beta_{1} & \\
                 & \ddots & \ddots & \ddots & \\
                 &  & \beta_{k-2} & \alpha_{k-1} &\beta_{k-1} \\
           0      & & & \beta_{k-1} & \alpha_{k}
  \end{pmatrix}
\label{eqn-chap2-tridiagonal-matrix}
\end{equation}
The eigenvalues of $T$ converge to the eigenvalues of $H$ provided that a sufficient number of iterations $k$, where $k < n$, is performed.

The generally large sizes of the NCCI Hamiltonian matrices require the use of parallel computing which handles both the storage and the diagonalization of the Hamiltonian matrix. For the calculations presented in this thesis, we use the parallel code Many Fermion Dynamics-Nuclear (MFDn) \cite{maris2010:ncsm-mfdn-iccs10,vary2009:ncsm-mfdn-scidac09} developed by the Iowa State University group handles the construction of the many-body basis, the construction of the Hamiltonian matrix, the storage of the matrix over multiple cores, and the Lanczos diagonalization of the matrix. The input to the code includes the two-body matrix elements of the Hamiltonian, the $\Nmax$ truncation of the many-body basis, the number of protons ($Z$) and neutrons ($N$) of the nucleus we want to study, the number of Lanczos iterations we want to perform, and various other control parameters. The program outputs the eigenvalues (which are the nuclear level energies), the one-body density matrices (described in Sec.~\ref{sec-chap2-one-body-density-matrix}), and other observables.

\section{The one-body density matrix}
\label{sec-chap2-one-body-density-matrix}

The one-body density matrix can be calculated using the many-body wave functions obtained in a many-body calculation. It is more frequently used for the calculation of the matrix elements of one-body operators~\cite{suhonen2007:nucleons-nucleus}. Moreover, the static one-body density matrix contains information about correlations in the many-body wave function as described below. In second quantization, the transition one-body density matrix is given by
\begin{equation}
\rho_{\alpha\beta}^{\mathrm{f\,i}} = \langle \Psi_{\mathrm{f}} | c_{\alpha}^{\dagger} c_{\beta} | \Psi_{\mathrm{i}} \rangle,
\label{eqn-chap2-one-body-density-operator}
\end{equation}
where $| \Psi_{\mathrm{i}} \rangle$ and $| \Psi_{\mathrm{f}} \rangle$ are the many-body wave functions of an initial and a final nuclear state respectively. Using (\ref{eqn-chap2-one-body-density-operator}), the matrix elements of a one-body operator $\mathcal{O}$ are obtained as
\begin{equation}
\langle \Psi_{\mathrm{f}} | \mathcal{O} | \Psi_{\mathrm{i}} \rangle =  \sum_{\alpha\beta} \langle \alpha | \mathcal{O} | \beta \rangle \langle \Psi_{\mathrm{f}} | c_{\alpha}^{\dagger}c_{\beta} | \Psi_{\mathrm{i}} \rangle.
\label{eqn-chap2-one-body-operator-matrix-elements}
\end{equation}

The static one-body density matrix (i.e., the one-body density matrix obtained for the same initial and final many-body wave functions) contains information about correlations in the many-body wave function. By definition, an uncorrelated many-body wave function can be written as a single Slater determinant, i.e., $| \Psi \rangle = | a_{1} \ldots a_{A} \rangle$. For an uncorrelated many-body wave function, it is easy to see that the diagonal matrix elements $\rho_{\alpha\alpha} = \langle \Psi | c_{\alpha}^{\dagger} c_{\alpha} | \Psi \rangle $ are equal to $1$ (which reflects the fact that each single-particle state is occupied by exactly one nucleon), while all the off diagonal matrix elements $\rho_{\alpha\beta}$ are equal to $0$. The sum of the diagonal matrix elements is equal to the total number of nucleons $A$ in the many-body wave function. In the case of a correlated many-body wave function, such as the ones calculated in an NCCI calculation, the one-body density matrix is not (in general) diagonal. The off diagonal matrix elements $\rho_{\alpha\beta}$ of the density matrix provide a measure of how correlated the many-body wave function is, while the diagonal matrix elements $\rho_{\alpha\alpha}$ provide the occupancies of each single-particle state $| a_{i} \rangle$ in the many-body wave function. The sum over the diagonal matrix elements is still equal to the total number of nucleons in the many-body wave function
\begin{equation}
\sum_{\alpha} \langle \Psi | c_{\alpha}^{\dagger} c_{\alpha} | \Psi \rangle = A.
\label{eqn-chap2-density-matrix-number}
\end{equation}
It is important to stress here that a diagonal one-body density matrix does not necessarily indicate that a many-body wave function is uncorrelated unless the diagonal matrix elements $\rho_{\alpha \alpha}$ are exactly equal to $1$. The one-body density matrix is the starting point for the construction of natural orbitals in Chapter~\ref{chap-chap3}.

\section{Example calculations and infrared extrapolations}
\label{sec-chap2-example-calculations}

In this section we perform NCCI calculations to obtain the ground state energy and proton radius of the isotopes $\isotope[3,4]{He}$. Our goal is to revisit the convergence properties of NCCI calculations using the harmonic oscillator basis and to extrapolate the calculated results to the full $\Nmax \rightarrow \infty$ space using the infrared extrapolation method which we present and discuss here.

\begin{figure}
\begin{center}
\includegraphics[width=1.\textwidth]{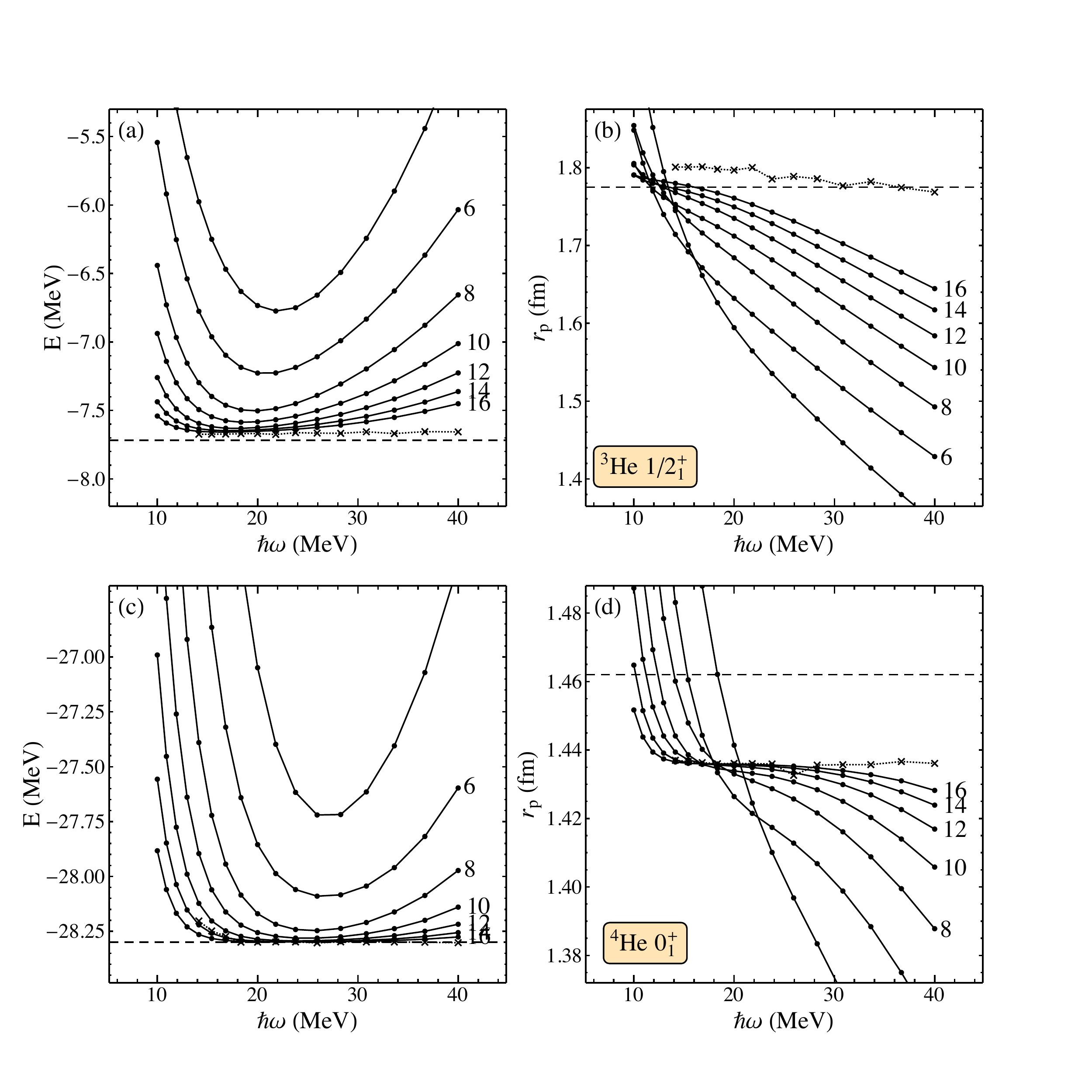}
\caption[Calculated ground state energy and proton radius in the ground state of {\protect $\isotope[3]{He}$} and {\protect $\isotope[4]{He}$}, obtained using the harmonic oscillator basis, the JISP$16$ $NN$ interaction, the Coulomb interaction between protons, truncations of the many-body basis up to $\Nmax = 16$, and $\hw$ parameters in the range $10$-$40$ MeV. The calculated results are extrapolated to the full space using the infrared extrapolation method.]{Calculated ground state energy (left) and proton radius in the ground state (right) of $\isotope[3]{He}$ (top) and $\isotope[4]{He}$ (bottom), obtained using the harmonic oscillator basis, the JISP$16$ $NN$ interaction, the Coulomb interaction between protons, truncations of the many-body basis up to $\Nmax = 16$, and $\hw$ parameters in the range $10$-$40$ MeV. The dashed horizontal lines show the experimental results. The crosses connected by dotted lines are obtained by extrapolating calculated results using the infrared extrapolation method as described in the text.}
\label{fig-chap2-helium3-helium4}
\end{center}
\end{figure}

In Fig.~\ref{fig-chap2-helium3-helium4}, we plot the calculated ground state energy (left) and proton radius (right) of $\isotope[3]{He}$ (top) and $\isotope[4]{He}$ (bottom). Calculations were performed for even truncations of the many-body basis up to $\Nmax = 16$ (in steps of $2$) which scan the $\hw$ range $10$-$40$ MeV. The solid lines correspond to the calculated results, the dashed horizontal lines correspond to the experimental results, and the dotted lines connect extrapolated results obtained using the infrared extrapolation method (discussed below). The experimental binding energies are taken from~\cite{npa-purcell-a-3-data:2010,npa-tilley-a-4-data:1992}, while the proton radii are deduced using the experimentally measured nuclear charge radii reported in~\cite{angeli-marinova-nuclear-charge-radii:2013} and equation ($6$) in Ref.~\cite{bacca2012:6he-hyperspherical}. As we observe, the calculated ground state energy and proton radius of $\isotope[4]{He}$ converge. On the other hand, the calculated ground state energy of $\isotope[3]{He}$ approaches convergence (to the $\sim 0.01$ MeV level), while the proton radius does not converge.

Comparisons of NCCI calculations with experimental results are only meaningful when full convergence is achieved. However, when the calculated results do not converge we can attempt to extrapolate results obtained in truncated spaces to the full $\Nmax \rightarrow \infty$ space. One such example was presented in Chapter~\ref{chap-chap1}, where we used an empirical extrapolation method to extrapolate the calculated ground state energy of $\isotope[6]{He}$ to the full space. Here, we use the infrared extrapolation method to extrapolate the calculated ground state energy and proton radius of $\isotope[3,4]{He}$ to the full space. Extrapolating the converged $\isotope[4]{He}$ results serves as a test of the infrared method as we expect the extrapolated results to match the converged results.

The infrared extrapolation method~\cite{coon2012:nscm-ho-regulator,furnstahl2012:ho-extrapolation, more2013:ir-extrapolation,furnstahl2014:ir-expansion, furnstahl2015:ir-extrapolations,odell-papenbrock-ir-extrapolation-quadruple:2016} starts with the assumption that the truncated many-body basis  induces both a short-range ultraviolet (UV) and a long-range infrared (IR) cutoff on the ability of the many-body basis to describe the many-body wave function. Quantitatively, if the highest occupied single-particle orbital in the many-body basis has $N$ oscillator quanta, then to a first approximation the momentum space (UV) cutoff is~\cite{coon2012:nscm-ho-regulator,furnstahl2012:ho-extrapolation}
\begin{equation}
\Lambda_{\mathrm{UV}} = \sqrt{2(N+3/2)} \,\, \hbar/b,
\label{eqn-chap2-uv-cutoff}
\end{equation}
and the coordinate space (IR) cutoff is
\begin{equation}
L = \sqrt{2(N+3/2)} \,\, b,
\label{eqn-chap2-ir-cutoff}
\end{equation}
where $b$ is the oscillator length.  Assuming that ultraviolet convergence of the calculated results is reached (this typically happens when the ultraviolet cutoff of the internucleon interaction $\lambda$ is smaller than the ultraviolet cutoff $\Lambda_{\mathrm{UV}}$ of the basis), the bound state energies will converge exponentially with respect to $L$
\begin{equation}
E(N_{\mathrm{max}},\hbar\omega) = E_{\infty} +  a_{0} e^{-2 k_{\infty} L(N_{\mathrm{max}},\hbar\omega)},
\label{eqn-chap2-energy-extrapolation-theoretical}
\end{equation}
and the radii will converge as
\begin{equation}
r^{2}(L) = r_{\infty}^{2} [1 - (c_{0}+c_{1}\beta^{-2})\beta^{3}e^{-\beta}],
\label{eqn-chap2-rms-extrapolation}
\end{equation}
where $E_{\infty}$, $a_{0}$, $k_{\infty}$, $c_{0}$, $ c_{1}$, and $r_{\infty}^{2}$ are obtained by fitting (UV converged) calculated results and $\beta = 2k_{\infty} L$ [$k_{\infty}$ is taken from the energy fit in (\ref{eqn-chap2-energy-extrapolation-theoretical})]. The infrared cutoff was later studied in more detail and a more precise expression for $L$ was obtained by finding the lowest eigenvalue of the operator $p^{2}$ in a finite oscillator basis~\cite{more2013:ir-extrapolation} 
\begin{equation}
L(N,\hbar\omega) = \sqrt{2(N+\Delta N + 3/2)} \,\, b(\hbar\omega),
\label{eqn-chap2-ir-length}
\end{equation}
where $\Delta N = 2$ is an offset determined from the diagonalization of $p^{2}$.

The exponential convergence of the bound-state energies with respect to $L$ can be understood as follows~\cite{furnstahl2012:ho-extrapolation}. The finite extent $L$ of the oscillator basis in position space effectively imposes a Dirichlet boundary condition on the bound state wave function at $r = L$. The exponential convergence with respect to $L$ is thus directly related to the exponential fall-off of the bound-state wave functions in position space.

Let us now use the infrared extrapolation method with our calculated results. Here, we perform a three point extrapolation using calculated results obtained at $\Nmax=12$, $14$, and $16$ sharing the same $\hw$ parameter. We use results obtained using $\hw$ parameters $\hw \gtrsim 14$ MeV which have large $\Lambda_{UV}$ somewhat comparable to the ultraviolet cutoff $\Lambda_{\mathrm{UV}}^{\mathrm{JISP}16}$ of the JISP$16$ interaction~\cite{shirokov2007:nn-jisp16}. Specifically, the JISP$16$ interaction is obtained by fitting scattering data in a harmonic oscillator space with $N=8$ and $\hw = 40$ MeV, which means that according to (\ref{eqn-chap2-uv-cutoff}) $\Lambda_{\mathrm{UV}}^{\mathrm{JISP}16} \approx 800$ $\mathrm{MeV}/c$. For the NCCI calculation with $\Nmax = 12$ and $\hw \approx 14$ MeV equation (\ref{eqn-chap2-uv-cutoff}) yields $\Lambda_{\mathrm{UV}} \approx 600$ $\mathrm{MeV}/c$ which is perhaps too low compared to the UV cutoff of JISP$16$ however, the extrapolated result will dictate whether our choice was sensible or not.

In Fig.~\ref{fig-chap2-helium3-helium4}, we show the extrapolated results for the ground state energy [panel (c)] and proton radius [panel (d)] of $\isotope[4]{He}$. As expected, the extrapolated results converge for both observables. A closer inspection reveals that a slight $\hw$ dependence persists which is however very small (at the $0.01$ MeV level and $0.01$ fm level for the extrapolated energy and proton radius respectively). For low $\hw$ parameters, the extrapolated energy does not converge something that we expect since the calculated results used for the extrapolation are not fully UV converged. On the other hand, for high $\hw$ the extrapolated results converge since the calculated results used for the extrapolation are UV converged. The extrapolated ground state energy at $\hw \approx 20$ MeV ($-28.3$ MeV) agrees with the calculated result at the variational minimum of the $\Nmax = 16$ curve ($-28.3$ MeV) and the experimental result ($-28.3$ MeV). On the other hand, the extrapolated proton radius at $\hw \approx 20$ MeV ($1.44$ fm) agrees with the calculated result at $\Nmax = 16$ ($1.44$ fm); however, it is about $\sim 0.02$ fm short of the experimental result [$r_{p} = 1.462(6)$ fm]. Thus, the NCCI calculation using the JISP$16$ interaction correctly predicts the binding energy of $\isotope[4]{He}$, while the prediction for the proton radius is about $\sim 1.5$ $\%$ short of the experimental result.

In Fig.~\ref{fig-chap2-helium3-helium4}, we also show the extrapolated results for the ground state energy [panel (a)] and proton radius [panel (b)] of $\isotope[3]{He}$. We observe that the calculated ground state energy converges (to the $0.01$ MeV level). The extrapolated ground state energy at $\hw \approx 20$ MeV ($-7.67$ MeV) is about $\sim 0.05$ MeV short of the experimental result ($-7.72$ MeV). The extrapolated proton radius staggers around the experimental result [$r_{p} = 1.774(6)$ fm] at the $0.1$ fm level. However, since full convergence (of the extrapolated results) is not achieved we cannot assess whether the calculation correctly predicts the proton radius.

\section{Using a general single-particle basis in NCCI calculations}
\label{sec-chap2-general-basis}

In this section we review the procedure followed to derive the two-body matrix elements of the NCCI Hamiltonian in a general single-particle basis. The discussion follows closely the derivations discussed in Ref.~\cite{caprio2012:csbasis} where the two-body matrix elements of the NCCI Hamiltonian (\ref{eqn-chap2-ncci-hamiltonian-with-lawson}) with respect to the Laguerre basis were derived.

To build the Hamiltonian matrix using a general single-particle basis we need to calculate the two-body matrix elements of the interaction and relative kinetic energy operators in the general basis. In the case where the two-body matrix elements of the interaction with respect to some single-particle basis are known, we can use a two-body transformation to obtain the two-body matrix elements with respect to the general basis. For example, the JISP$16$ interaction is expressed in terms of the harmonic oscillator basis. Therefore, we can use a two-body transformation to obtain the two-body matrix elements in the general basis. Specifically, if $a = (n_{a}l_{a}j_{a})$ is a harmonic oscillator single-particle orbital and $a' = (n_{a'} l_{a'} j_{a'})$ is a general single-particle orbital then for antisymmetrized two-body states such as (\ref{eqn-chap2-two-particle-states-2}) the transformation is given by~\cite{hagen2006:gdm-realistic}
\begin{equation}
\langle c' d'; J | V | a' b'; J \rangle_{\mathrm{AS}} = \sum_{abcd} \langle a | a' \rangle \langle b | b' \rangle \langle c | c' \rangle \langle d | d' \rangle \langle c d;J | V | a b; J \rangle_{\mathrm{AS}},
\label{eqn-chap2-two-body-interaction-transformation}
\end{equation}
where $\langle a | a' \rangle$ is an overlap bracket. For normalized antisymmetrized states such as (\ref{eqn-chap2-two-particle-states-3}) the transformation is given by
\begin{multline}
\langle c' d'; J | V | a' b'; J \rangle_{\mathrm{NAS}} = (1+\delta_{a'b'})^{-1/2} (1+\delta_{c'd'})^{-1/2} \\  \sum_{abcd} (1+\delta_{ab})^{1/2} (1+\delta_{cd})^{1/2} \langle a | a' \rangle \langle b | b' \rangle \langle c | c' \rangle \langle d | d' \rangle \langle c d;J | V | a b; J \rangle_{\mathrm{NAS}}.
\label{eqn-chap2-two-body-interaction-transformation-nas}
\end{multline}
The overlap bracket is given by 
\begin{equation}
\langle a | a' \rangle = \langle R_{n_{a}l_{a}} | R_{n_{a'} l_{a'}} \rangle \delta_{(l_{a}j_{a})(l_{a'}j_{a'})}, 
\label{eqn-chap2-overlap-bracket}
\end{equation}
where 
\begin{equation}
\langle R_{n_{a}l_{a}} | R_{n_{a'}l_{a}} \rangle = \int_{0}^{\infty} dr R_{n_{a}l_{a}}(b_{\mathrm{HO}};r) R_{n_{a'}l_{a}}(b_{a'};r), 
\label{eqn-chap2-overlap-integral}
\end{equation}
and $b_{\mathrm{HO}}$ and $b_{a'}$ are the characteristic lengths of the harmonic oscillator basis and the general basis respectively. The transformation~(\ref{eqn-chap2-two-body-interaction-transformation}) [or (\ref{eqn-chap2-two-body-interaction-transformation-nas})] involves an infinite quadruple sum over orbitals. The sum must be truncated, e.g., according to a one-body shell cutoff $\Ncut$ which must be selected in a way which ensures that NCCI calculations in the general basis are $\Ncut$ independent.

In principle, the two-body transformation~(\ref{eqn-chap2-two-body-interaction-transformation}) can be used to obtain the two-body matrix elements of the relative kinetic energy operator in the general basis (provided that the matrix elements in the harmonic oscillator or some other single-particle basis are known). However, in Ref.~\cite{caprio2012:csbasis} the transformation from the harmonic oscillator to the Laguerre basis was found to yield calculated results which were highly sensitive to the $N_{\mathrm{cut}}$ truncation. Therefore, the two-body matrix elements of the relative kinetic energy operator were calculated directly in the general single-particle basis~\cite{caprio2012:csbasis}. Recall that the relative kinetic energy operator separates into one and two-body parts. Specifically, rearranging (\ref{eqn-chap2-kinetic-energy-decomposition}) we get
\begin{equation}
T_{\mathrm{rel}}  = \frac{1}{4 A m_{\mathrm{N}}} \sum_{i \neq j}^{A} (\mathbf{p}_{i}-\mathbf{p}_{j})^{2} = \frac{1}{2 A m_{\mathrm{N}}} \left[ (A-1) \sum_{i=1}^{A} p_{i}^{2} - \sum_{i \neq j}^{A} \mathbf{p}_{i} \cdot \mathbf{p}_{j} \right].
\label{eqn-chap2-kinetic-energy-decomposition-matrix-elements}
\end{equation}
The first term is a one-body operator the matrix elements of which can be calculated using the momentum space representation of the general single-particle basis. The second term is a two-body operator the matrix elements of which factorize according to Racah's reduction formula~\cite{suhonen2007:nucleons-nucleus}
\begin{equation}
\langle c'd';J | \mathbf{p}_{i} \cdot \mathbf{p}_{j} | a'b'; J \rangle = (-1)^{j_{d'}+j_{a'}+J} \Gj{j_{c'}}{j_{d'}}{J}{j_{b'}}{j_{a'}}{1} \langle c' || \mathbf{p}_{i} || a' \rangle \langle d' || \mathbf{p}_{j} || b' \rangle,
\label{eqn-chap2-kinetic-energy-two-body-matrix-elements}
\end{equation}
where $\langle c' || \mathbf{p} || a' \rangle \propto [\int_{0}^{\infty} dk \tilde{R}_{n_{c'}l_{c'}}(b;k) k \tilde{R}_{n_{a'}l_{a'}}(b;k)] \langle l_{c'} j_{c'} || Y_{1} || l_{a'} j_{a'} \rangle$, $\tilde{R}_{n'l'}(b;k)$ is the momentum space representation of the general single-particle states, and $\mathbf{p} \equiv \hbar \mathbf{k}$.

%% file: chapters/chapter3/chapter3_draft_170402.tex
\chapter{NATURAL ORBITALS FOR NO-CORE CONFIGURATION INTERACTION CALCULATIONS}
\label{chap-chap3}

\section{Overview}
\label{sec-chap3-overview}

Choosing a single-particle basis able to describe the complex multiscale physics of the atomic nucleus is critical for the description of the nuclear many-body wave function. The nuclear wave function must be able to describe both strong short-range correlations between nucleons and long-range asymptotics which are important for the description of halo nuclei for example. In this chapter, we introduce natural orbitals for NCCI calculations in our attempt to efficiently describe the nuclear many-body wave function and accelerate the convergence of observables in truncated spaces.

Natural orbitals were first introduced in atomic physics, where it was shown that they provide a single-particle basis which leads to fast convergence of configuration interaction calculations using a few Slater determinants~\cite{loewdin1955:natural-orbitals-part1,shull1955:natural-orbitals-helium, loewdin1956:natural-orbital, shull1959:natural-orbitals-he, bender1966:natural-orbital-iterative-hydride, davidson1972:natural-orbital}. In nuclear physics, natural orbitals were used to study Jastrow type correlations in closed shell nuclei~\cite{stoitsov1993:natural-orbital-correlation,stoitsov1998:tho-basis} and nuclear charge distributions~\cite{malaguti-nuclear-scharge-distributions-using-natural-orbitals:1982}.

We start by motivating the need for natural orbitals in NCCI calculations and subsequently, we derive natural orbitals by diagonalizing scalar one-body density matrices obtained from initial NCCI calculations in the harmonic oscillator basis (Sec.~\ref{sec-chap3-natural-orbitals}). We then obtain the two-body matrix elements of the NCCI Hamiltonian in the natural orbital basis (Sec.~\ref{sec-chap3-two-body-matrix-elements}) and use them to perform NCCI calculations for the ground state energy and proton radius in the ground state of $\isotope[3]{He}$ and $\isotope[4]{He}$ (Sec.~\ref{sec-chap3-example-calculations}).

\section{Why natural orbitals}
\label{sec-chap3-natural-orbitals}

In the last two chapters we saw that despite the convenient properties afforded by the harmonic oscillator basis, the convergence of observables calculated using the oscillator basis in terms of $\Nmax$ is slow. The problem is severe for long-range operators such as the calculated proton radius which we demonstrated in the example calculations shown in Figs.~\ref{fig-chap1-helium-6-e-rp}(b) and \ref{fig-chap2-helium3-helium4}(b) for $\isotope[6]{He}$ and $\isotope[3]{He}$ respectively. One reason for slow convergence can be attributed to the Gaussian ($\sim e^{-b r^{2}}$) asymptotics carried by the harmonic oscillator basis which do not match the exponential ($\sim e^{-br}$) asymptotics of the nuclear many-body wave function. Another reason might be that using the harmonic oscillator single-particle basis to obtain the nuclear many-body wave function we introduce superficial correlations between nucleons in the many-body wave function. As we saw in Sec~\ref{sec-chap2-one-body-density-matrix}, these correlations can be studied using the one-body density matrix.

Our goal is to accelerate the convergence of NCCI calculations in truncated spaces. Many methods have been proposed to address the problem. In the importance truncated no-core shell model~\cite{kruse-importance-truncation:2013}, the many-body states which have major contributions to the ground state many-body wave function at a given $\Nmax$ are selected using multi-configurational perturbation theory. The diagonalization of the Hamiltonian is then performed in the (reduced) space defined by the selected many-body states and the goal is to reproduce the results obtained using the full $\Nmax$ space as accurately as possible. Another method has already been discussed in Chapter~\ref{chap-chap2}, and it suggests that one can use results calculated in truncated spaces to extrapolate to the full $\Nmax \rightarrow \infty$ space. In the symmetry-adapted no-core shell model (SA-NCSM)~\cite{draayer2012:sa-ncsm-qghn11}, a many-particle basis that exploits the physically relevant SU$(3)$ $\supset$ SO$(3)$ group-subgroup chain is utilized. Using the SU$(3)$ symmetry adapted basis, only a small fraction of the complete model space is needed to model nuclear collective dynamics, deformation, and $\alpha$-particle clustering. In the symplectic no-core configuration interaction scheme (SpNCCI), one uses the Sp$(3,R)$ basis for the expansion of the many-body wave function to take advantage of the Sp$(3,R)$ symmetry which is conserved by the kinetic energy operator. Using the Sp$(3,R)$ basis, the size of the many-body space for a given $\Nmax$ is reduced. In the no-core shell model with continuum~\cite{baroni2013:7he-ncsmc}, the no-core shell model (NCSM), a bound-state technique, is combined with the no-core shell model/resonating group method (NCSM/RGM), a nuclear scattering technique, to describe both bound and scattering states of light nuclei. Finally, one can replace the harmonic oscillator basis, which is traditionally used with the NCCI approach, with another single-particle basis which is the approach we follow here.

We thus seek a physically adapted single-particle basis in which the many-body wave function is efficiently and accurately described in a truncated many-body space. The natural orbital basis minimizes the mean occupancies of single-particle states above the Fermi surface, therefore reducing the contributions from high-lying oscillator orbitals in describing the many-body wave function. Intuitively, the natural orbitals may be understood as an attempt to recover the single-particle basis in terms of which the many-body wave function most resembles a single Slater determinant. However, the many-body wave function is highly correlated, therefore transforming to natural orbitals enhances the role of Slater determinants involving low-lying states, thus leading to faster convergence.

As we saw in Chapter~\ref{chap-chap2} the many-body states used in NCCI calculations are constructed as antisymmetrized products of $| nljm \rangle$ harmonic oscillator single-particle states. We thus want to maintain $l$ and $j$ as good quantum numbers for our natural orbital basis. The scalar one-body density matrix given by 
\begin{equation}
\rho_{ab}^{\,\, (0) } \equiv \langle \Psi | \left[c_{a}^{\dagger} \tilde{c}_{b} \right]_{0\,0} | \Psi \rangle,
\label{eqn-chap3-scalar-one-body-density-matrix}
\end{equation}
where $c_{a}^{\dagger}$ represents the creation operator for a nucleon in orbital $a = (n_{a}l_{a}j_{a})$, $\tilde{c}_{b}$ is the annihilation operator for a nucleon in orbital $b = (n_{b}l_{b}j_{b})$ [the tilde operator means that the operator $\tilde{c}_{b} \equiv (-1)^{j_{b}+m_{b}} c_{b,-m_{b}}$ is a proper spherical tensor of rank $j_{b}$], and $\left[\ldots \right]_{0\,0}$ represents spherical tensor coupling to angular momentum $0$. The diagonal entries give the occupancies of the single-particle orbitals $\mathcal{N}_{a} = (2 j_{a} + 1)^{1/2} \rho_{aa}^{\,\, (0)}$ in the many-body wave function $|\Psi \rangle$. The scalar density matrix only connects orbitals which share the same $l$ and $j$, i.e., they differ only in their radial quantum number $n$. Therefore the natural orbitals obtained by diagonalizing (\ref{eqn-chap3-scalar-one-body-density-matrix}) represent a change of basis on the radial functions separately in each $lj$ space
\begin{equation}
|a' \rangle \equiv |n_{a}'l_{a}j_{a} \rangle = \sum_{n_{a}} \alpha_{n_{a}',n_{a}}^{(l_{a},j_{a})} |n_{a}l_{a}j_{a} \rangle,
\label{eqn-chap3-natural-orbitals-expansion}
\end{equation}
where $\alpha_{n_{a}',n_{a}}^{(l_{a},j_{a})}$ are obtained by the diagonalization of (\ref{eqn-chap3-scalar-one-body-density-matrix}), $n_{a}'$ is a counting index, and the sum over $n_{a}$ runs from $0$ to the radial quantum number of the highest occupied oscillator orbital in the initial $| \Psi \rangle$. That is $0 \leq n_{a} \leq (\Nmax + N_{\mathrm{v}} -l)/2$, where $N_{\mathrm{v}}$ is the nominal shell quantum number of the valence shell in the lowest allowed configuration (i.e., $N_{\mathrm{v}} = 0$ for $s$-shell nuclei and $N_{\mathrm{v}} = 1$ for $p$-shell nuclei). Note that the density matrix does not mix proton and neutron orbitals, therefore the proton and neutron natural orbitals are in general different.

Finally, before using natural orbitals as the single-particle basis for NCCI calculations, we need to make sure that we have a way to truncate our many-body basis built using natural orbitals. The eigenvalues of the scalar one-body density matrix (\ref{eqn-chap3-scalar-one-body-density-matrix}) represent the mean occupancy of each natural orbital in the many-body wave function. We order the natural orbitals by decreasing eigenvalue of the scalar density matrix, i.e., starting with $n=0$ for the natural orbital with the highest eigenvalue $ \left[ \langle \mathcal{N}_{0lj} \rangle \geq \langle \mathcal{N}_{1lj} \rangle \geq \ldots \right]$. Thus, an $n$ quantum number for an $\Nmax$-type truncation scheme is obtained.

\section{Two-body matrix elements in the natural orbital basis}
\label{sec-chap3-two-body-matrix-elements}

To build the many-body Hamiltonian matrix in the natural orbital basis we first need to calculate the two-body matrix elements of the NCCI Hamiltonian (\ref{eqn-chap2-overview-hamiltonian}) in the natural orbital basis. The matrix elements are calculated by taking advantage of the fact that the two-body matrix elements of the Hamiltonian (in the natural orbital basis) can be obtained as linear combinations of the two-body matrix elements of the Hamiltonian in the harmonic oscillator basis (since the natural orbitals are linear combinations of harmonic oscillator orbitals).

Let us start with the two-body matrix elements of the interaction in the natural orbital basis. These can be obtained by transforming two-body matrix elements expressed in the harmonic oscillator basis to the natural orbital basis. Assume that $| a' \rangle$ is a natural orbital obtained by diagonalizing a density matrix which was in turn obtained in an initial NCCI calculation with $\hw \propto b^{-2} $. Moreover, assume that $| \bar{a} \rangle$ is a harmonic oscillator orbital with the same $\hw \propto b^{-2}$ as the natural orbital. Finally, the interaction two-body matrix elements $\langle c d;J | V | a b; J \rangle$ are known and expressed in terms of the harmonic oscillator basis with $\hw_{\mathrm{int}} \propto b_{\mathrm{int}}^{-2}$, where in general $b \neq b_{\mathrm{int}}$. To obtain the matrix elements $\langle c' d';J | V | a' b'; J \rangle$ in the natural orbital basis we can use the two-body transformation (\ref{eqn-chap2-two-body-interaction-transformation})
\begin{equation}
\langle c' d'; J | V | a' b'; J \rangle = \sum_{abcd} \langle a | a' \rangle \langle b | b' \rangle \langle c | c' \rangle \langle d | d' \rangle \langle c d;J | V | a b; J \rangle.
\label{eqn-chap3-two-body-interaction-transformation-natural-orbitals}
\end{equation}
We thus need to evaluate the overlap brackets $\langle a | a' \rangle$. Recall that the natural orbitals are linear combinations of the harmonic oscillator orbitals $| \bar{a} \rangle$. Using~(\ref{eqn-chap3-natural-orbitals-expansion}) we have
\begin{equation}
|a' \rangle = \sum_{\bar{a}} \langle \bar{a} | a' \rangle  | \bar{a} \rangle = \sum_{\bar{n}_{a}} \alpha_{n_{a}',\bar{n}_{a}}^{(\bar{l}_{a},\bar{j}_{a})} | \bar{a} \rangle.
\label{eqn-chap3-natural-orbitals-expansion-bracket-notation}
\end{equation}
Now using (\ref{eqn-chap3-natural-orbitals-expansion-bracket-notation}) the brackets $\langle a | a' \rangle$ are obtained as
\begin{equation}
\langle a | a' \rangle = \sum_{\bar{a}} \langle a | \bar{a} \rangle \langle \bar{a} | a' \rangle,
\label{eqn-chap3-transformation-brackets}
\end{equation}
where $\langle a | \bar{a} \rangle = \langle R_{n_{a}l_{a}} | R_{\bar{n}_{a} \bar{l}_{a}} \rangle \delta_{(l_{a}j_{a})(\bar{l}_{a}\bar{j}_{a})}$. The overlap between radial oscillator functions is given by 
\begin{equation}
\langle R_{n_{a}l_{a}} | R_{\bar{n}_{a} \bar{l}_{a}} \rangle = \int_{0}^{\infty} dr R_{n_{a}l_{a}}(b_{\mathrm{int}};r) R_{\bar{n}_{a}\bar{l}_{a}}(b;r).
\label{eqn-chap3-radial-integral}
\end{equation}
Once the brackets $\langle a | a' \rangle, \ldots, \langle d | d' \rangle$ are obtained, we can plug them into the quadruple sum in (\ref{eqn-chap3-two-body-interaction-transformation-natural-orbitals}). The sum must then be truncated according to a one-body shell cutoff $N_{\mathrm{cut}}$ which ensures that the calculated results are $N_{\mathrm{cut}}$ independent as descibed in Sec.~\ref{sec-chap2-general-basis}.

The two-body matrix elements of the relative kinetic energy operator can be calculated using known single-particle matrix elements in the harmonic oscillator basis. Recall that the relative kinetic energy operator (\ref{eqn-chap2-kinetic-energy-decomposition}) is written as a sum of one-body operator and a separable two-body operator as described in Sec.~\ref{sec-chap2-general-basis}. The two-body matrix elements of the relative kinetic energy operator in the harmonic oscillator basis are obtained using single-particle matrix elements of the form $\langle \bar{a} | \mathcal{O}| \bar{b} \rangle$, where $\mathcal{O}=k$, or $k^{2}$, $p \equiv \hbar k$, and $| \bar{a} \rangle$ is a harmonic oscillator orbital. Now notice that using the matrix elements $\langle \bar{a} | \mathcal{O}| \bar{b} \rangle$ and (\ref{eqn-chap3-natural-orbitals-expansion}), we can obtain the matrix elements $\langle a' | \mathcal{O}| b' \rangle$ in the natural orbital basis
\begin{equation}
\langle a' | \mathcal{O} | b' \rangle = \sum_{\bar{a}\bar{b}} \langle a' | \bar{a} \rangle \langle \bar{a} | \mathcal{O} | \bar{b} \rangle \langle \bar{b} | b' \rangle.
\label{eqn-chap3-matrix-elements-kinetic-energy-operator}
\end{equation}

The diagonalization of the scalar density matrix, the calculation of the overlap brackets (\ref{eqn-chap3-transformation-brackets}), and the calculation of the matrix elements (\ref{eqn-chap3-matrix-elements-kinetic-energy-operator}) is taken care by the suite of computer programs noutils developed for this work. Once the overlap brackets (\ref{eqn-chap3-transformation-brackets}) and the matrix elements (\ref{eqn-chap3-matrix-elements-kinetic-energy-operator}) are obtained, they are given as input to the suite of programs h$2$utils developed for Refs.~\cite{caprio2012:csbasis,caprio2014:cshalo}, which in turn prepares the two-body matrix elements of the Hamiltonian in the natural orbital basis. The two-body matrix elements of the Hamiltonian in the natural orbital basis are then passed as input to MFDn which performs the many-body calculation.

\section{Example calculations using natural orbitals}
\label{sec-chap3-example-calculations}

In this section we test natural orbitals in example NCCI calculations for $\isotope[3,4]{He}$. We begin by studying the properties of the calculated natural orbitals, we then present the calculated ground state properties of $\isotope[3,4]{He}$ in the natural orbital basis, and, finally, we extrapolate the calculated results using infrared extrapolations.

The starting point for the calculations presented here is an NCCI calculation for $\isotope[3,4]{He}$ in the harmonic oscillator basis. The initial calculations are performed using the JISP$16$ internucleon interaction, the Coulomb interaction between protons, $\hw$ parameters in the range $10$-$40$ MeV, and $\Nmax$ truncation of the many-body basis up to $\Nmax = 16$. Subsequently, the natural orbitals are first derived, by diagonalizing the initial scalar one-body density matrices (in the ground state) for each $(\Nmax,\hw)$ pair, and subsequently used as the new single-particle basis for the NCCI calculations.

\begin{figure}[t]
\begin{center}
\includegraphics[width=0.99 \columnwidth]{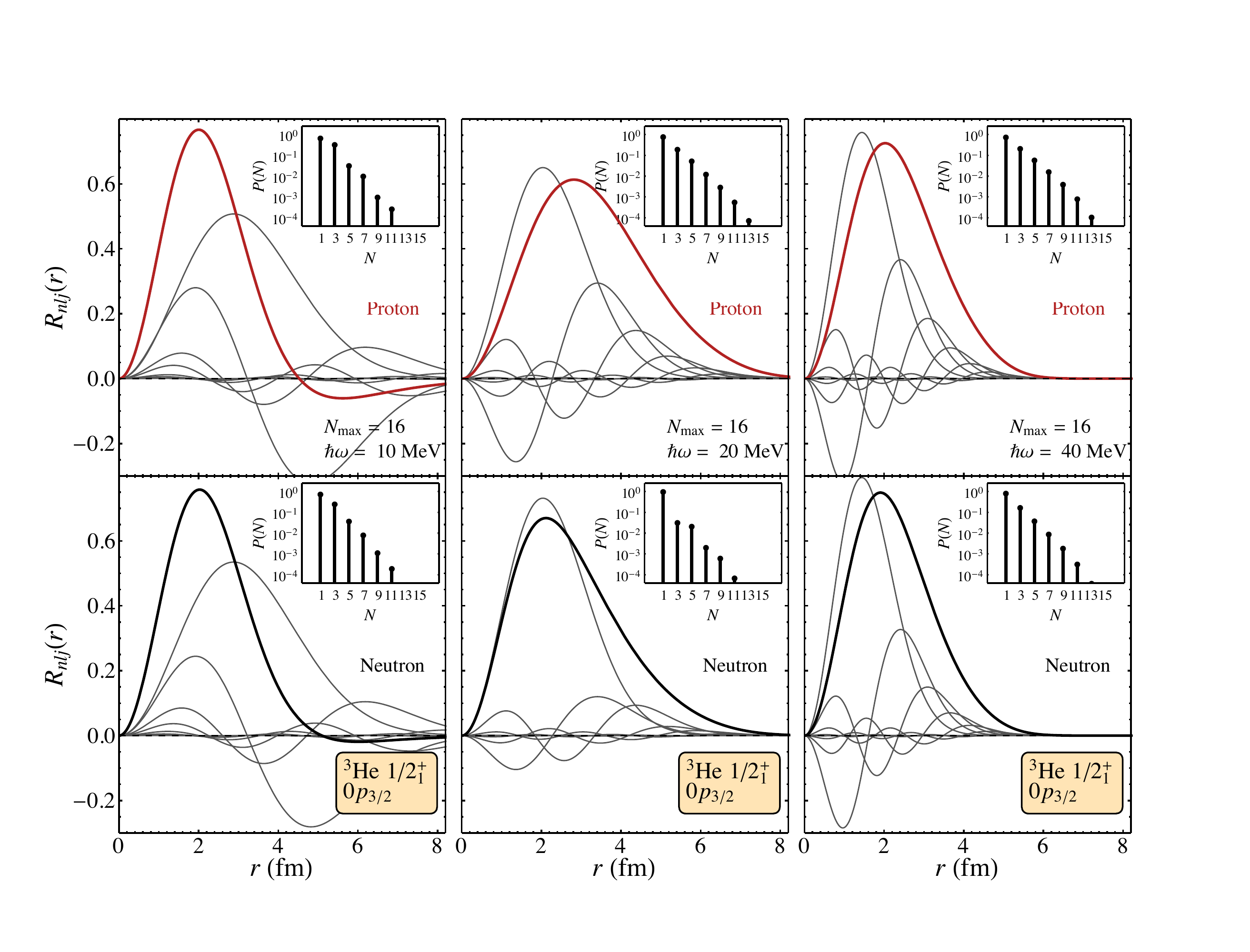}
\caption[Radial wave function for the proton and neutron $0p_{3/2}$ natural orbitals of {\protect $\isotope[3]{He}$} derived from the ground state scalar one-body density in the harmonic oscillator basis. The contributions from individual oscillator basis functions are also shown. The squared amplitude $P(N)$ of these contributions are shown in the inset panel.]{Radial wave function for the proton (top) and neutron (bottom) $0p_{3/2}$ natural orbitals of $\isotope[3]{He}$ derived from the ground state scalar one-body density in the harmonic oscillator basis. The contributions from individual oscillator basis functions are shown with gray curves. The squared amplitude $P(N)$ of these contributions are shown in the inset. The initial calculation was performed at $\Nmax=16$ and $\hw=10$ (left), $20$ (middle), and $40$ MeV (right).}
\label{fig-chap3-z2-n1-0s-natural-orbital-decomposition}
\end{center}
\end{figure}

\begin{figure}
\begin{center}
\includegraphics[width=0.99 \columnwidth]{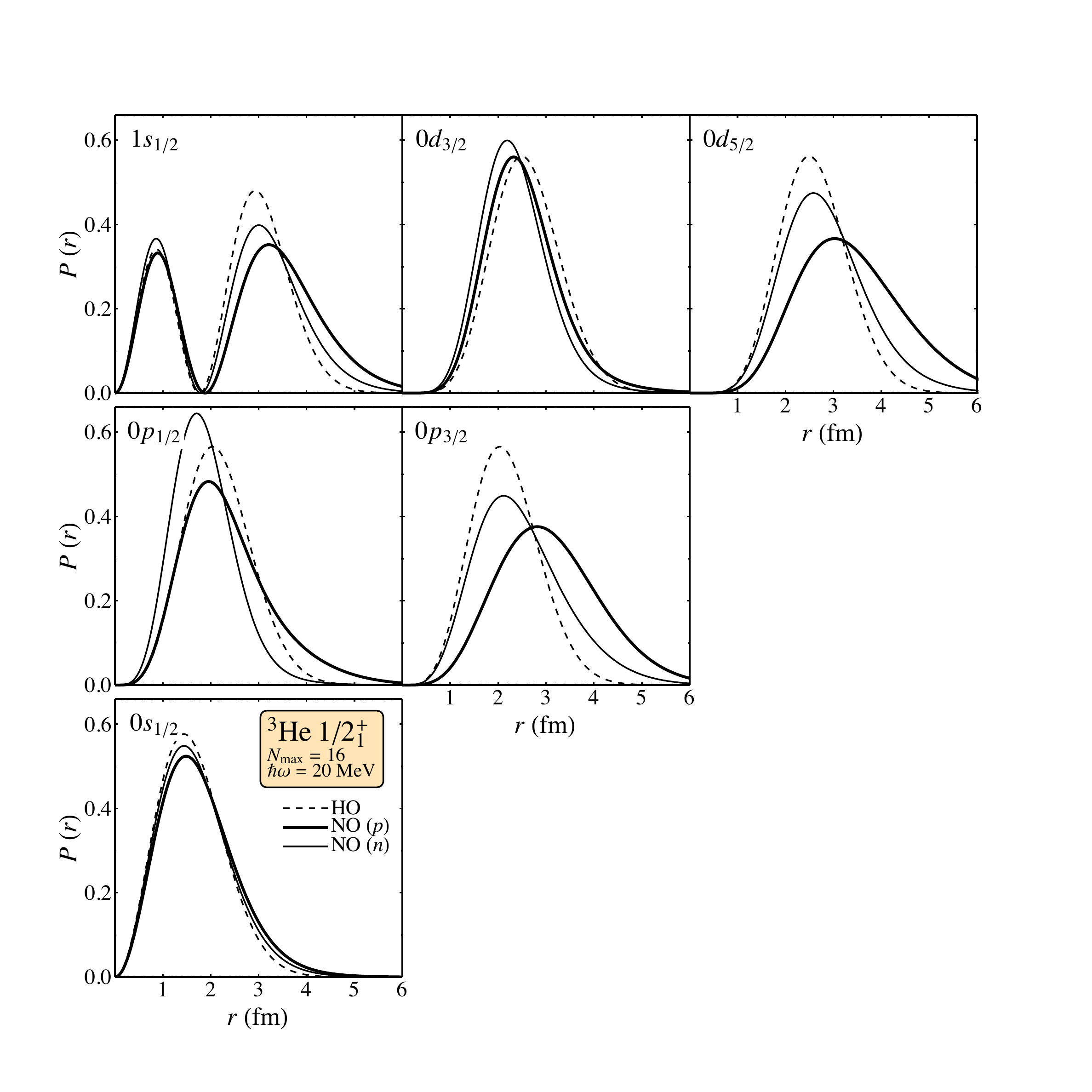}
\caption[Radial probability density functions $P(r) = r^{2} | R_{nlj}(b;r) |^{2}$ for the harmonic oscillator, proton, and neutron orbitals of {\protect $\isotope[3]{He}$} up to the $N=2$ major shell. The natural orbitals were derived from an initial one-body density matrix with $\Nmax = 16$ and $\hw = 20$ MeV.]{Radial probability density functions $P(r) = r^{2} | R_{nlj}(b;r) |^{2}$ for the harmonic oscillator (dashed curves), proton (thick curves), and neutron (dark curves) orbitals of $\isotope[3]{He}$ up to the $N=2$ major shell. The natural orbitals were derived from an initial one-body density matrix with $\Nmax = 16$ and $\hw = 20$ MeV.}
\label{fig-chap3-natural-orbitals-he-3-two-major-shells}
\end{center}
\end{figure}

Let us start by studying how the natural orbitals are built. In Fig.~\ref{fig-chap3-z2-n1-0s-natural-orbital-decomposition}, the proton (top) and neutron (bottom) $0p_{3/2}$ natural orbitals of $\isotope[3]{He}$ derived from an initial scalar one-body density matrix with $\Nmax = 16$ and $\hw = 10$ (left), $20$ (middle), and $40$ MeV (right) are plotted. The contributions from individual oscillator orbitals are shown by the light grey curves. In the inset panel we plot the squared amplitudes of the contributions (to the natural orbital) from each major oscillator shell $N$ [recall that $N=2n+l$ so the sum over $n$ in (\ref{eqn-chap3-natural-orbitals-expansion}) is equivalent to a sum over $N$]. For $\hw = 10$ MeV, the natural orbitals have shorter tails than the initial harmonic oscillator orbitals. The neutron natural orbital has a slightly longer tail than the proton natural orbital, and the main contributions to the natural orbitals come from the $N=1$ and $3$ shells. For $\hw = 20$ MeV, the proton natural orbital receives significant contributions from the $N=1$ and $3$ shells resulting in a significantly more elongated tail than the initial oscillator orbitals. The neutron natural orbital mainly receives contributions from the $N=1$ shell and it also acquires an elongated tail compared to the initial oscillator orbitals. Finally for $\hw = 40$ MeV, both the proton and neutron orbitals acquire elongated tails compared to the initial oscillator orbitals, with the proton orbitals having a longer tail than the neutron orbitals.

It is also interesting to study the behavior of the natural orbitals of the first three major oscillator shells. In Fig.~\ref{fig-chap3-natural-orbitals-he-3-two-major-shells}, we plot the radial probability density $P(r) = r^{2} | R_{nlj}(b;r) |^{2}$ [the $j$-dependence of the natural orbital radial functions comes from the expansion coefficients $\alpha_{n_{a}',n_{a}}^{(l_{a},j_{a})}$ in (\ref{eqn-chap3-natural-orbitals-expansion})] for the oscillator (dashed curves), proton (thick dark curves), and neutron (dark curves) orbitals of $\isotope[3]{He}$ up to the $N=2$ shell. The natural orbitals were obtained by diagonalizing initial density matrices with $\Nmax = 16$ and $\hw = 20$ MeV. We observe that the proton natural orbitals acquire longer tails than both the neutron natural orbitals and the initial oscillator orbitals. The tails of the neutron natural orbitals are also longer than the tails of the initial harmonic oscillator orbitals except for the neutron orbitals $0p_{1/2}$ and $0d_{3/2}$ which have slightly shorter tails than the initial oscillator orbitals.

\begin{figure}
\begin{center}
\includegraphics[width=0.99 \columnwidth]{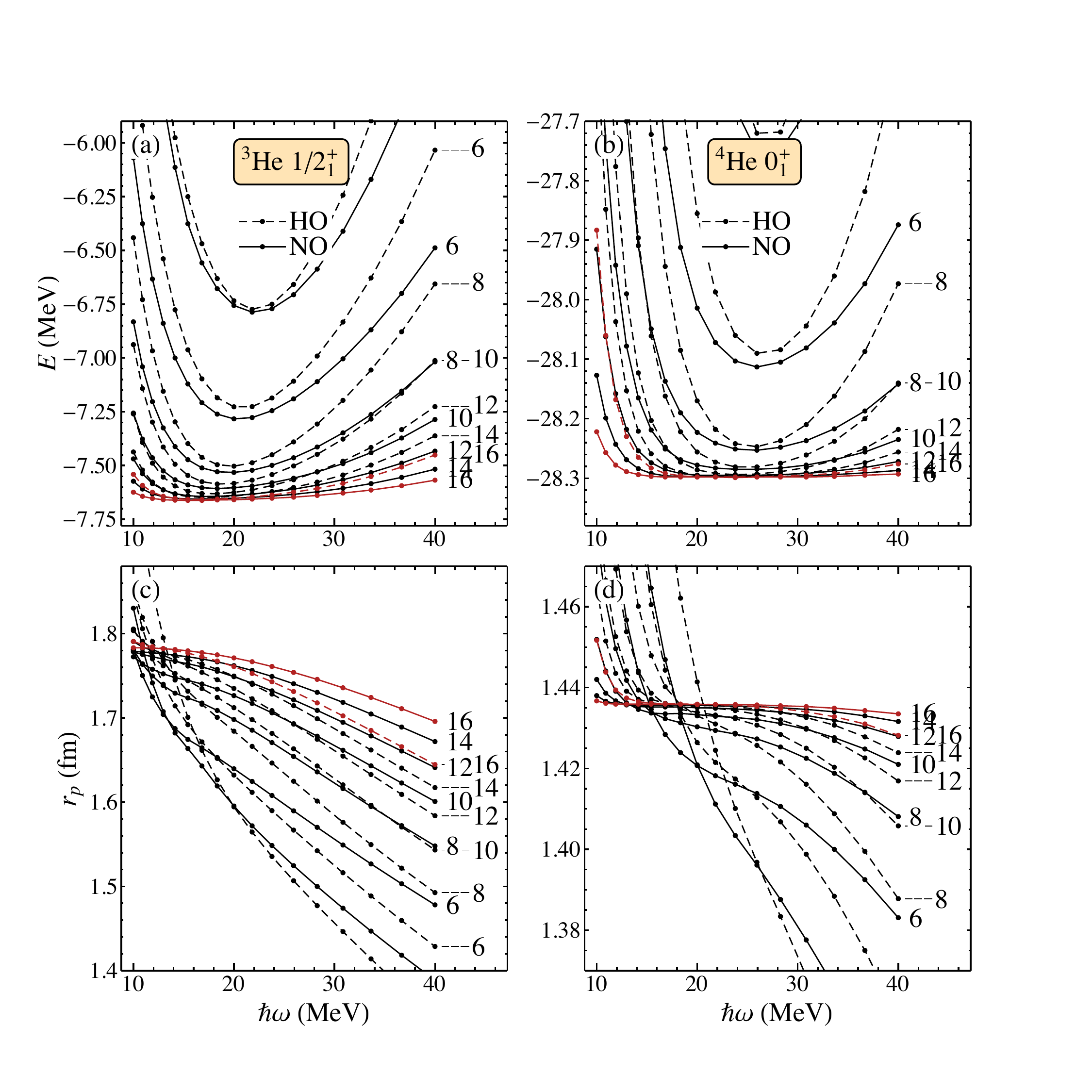}
\caption[The calculated ground state energy and proton radius in the ground state of {\protect $\isotope[3]{He}$}, and {\protect $\isotope[4]{He}$} obtained using harmonic oscillator orbitals and natural orbitals.]{The calculated ground state energy (top) and proton radius in the ground state (bottom) of $\isotope[3]{He}$ (left), and $\isotope[4]{He}$ (right) obtained using harmonic oscillator orbitals (dashed curves) and natural orbitals (solid curves). The red colored curves show results obtained for the highest $\Nmax$ truncation ($\Nmax = 16$).}
\label{fig-chap3-h0-no-energy-rp-comparison}
\end{center}
\end{figure}

We now turn our attention to the calculated ground state energy, shown in Fig.~\ref{fig-chap3-h0-no-energy-rp-comparison}(a) for $\isotope[3]{He}$ and Fig.~\ref{fig-chap3-h0-no-energy-rp-comparison}(b) for $\isotope[4]{He}$. Results obtained using harmonic oscillator orbitals are plotted with dashed curves and results obtained using natural orbitals are plotted with solid curves.

Overall, we observe that convergence in terms of $\Nmax$ is faster for the natural orbital basis compared to the harmonic oscillator basis for both nuclei. Moreover, the natural orbital basis improves convergence in terms of the $\hw$ parameter of the single-particle basis compared to the harmonic oscillator basis. For $\isotope[3]{He}$, the difference between the calculated energy at the variational minimum of the $\Nmax = 14$ and $\Nmax = 16$ natural orbital curves is about $\sim 5$ keV compared to about $\sim 10$ keV for the harmonic oscillator curves (convergence is only approximate). For $\isotope[4]{He}$, the natural orbital basis achieves nearly $\hw$-independent results at $\Nmax = 16$.

Let us now move to the calculated proton radii shown in Fig.~\ref{fig-chap3-h0-no-energy-rp-comparison}(c) for $\isotope[3]{He}$ and Fig.~\ref{fig-chap3-h0-no-energy-rp-comparison}(d) for $\isotope[4]{He}$ respectively. Convergence in terms of $\Nmax$ is also accelerated for both nuclei using natural orbitals (as with the calculated energy case) compared to using harmonic oscillator orbitals. Moreover, the $\hw$ convergence is improved using natural orbitals than using oscillator orbitals. For $\isotope[4]{He}$ the calculated results converge using either basis. Using natural orbitals yields approximately $\hw$ independent results at $\Nmax = 16$. For $\isotope[3]{He}$, full convergence is not reached using the natural orbital basis despite the significant improvement of the convergence in terms of $\Nmax$ afforded by the natural orbitals.

\begin{figure}[t]
\begin{center}
\includegraphics[width=0.7 \columnwidth]{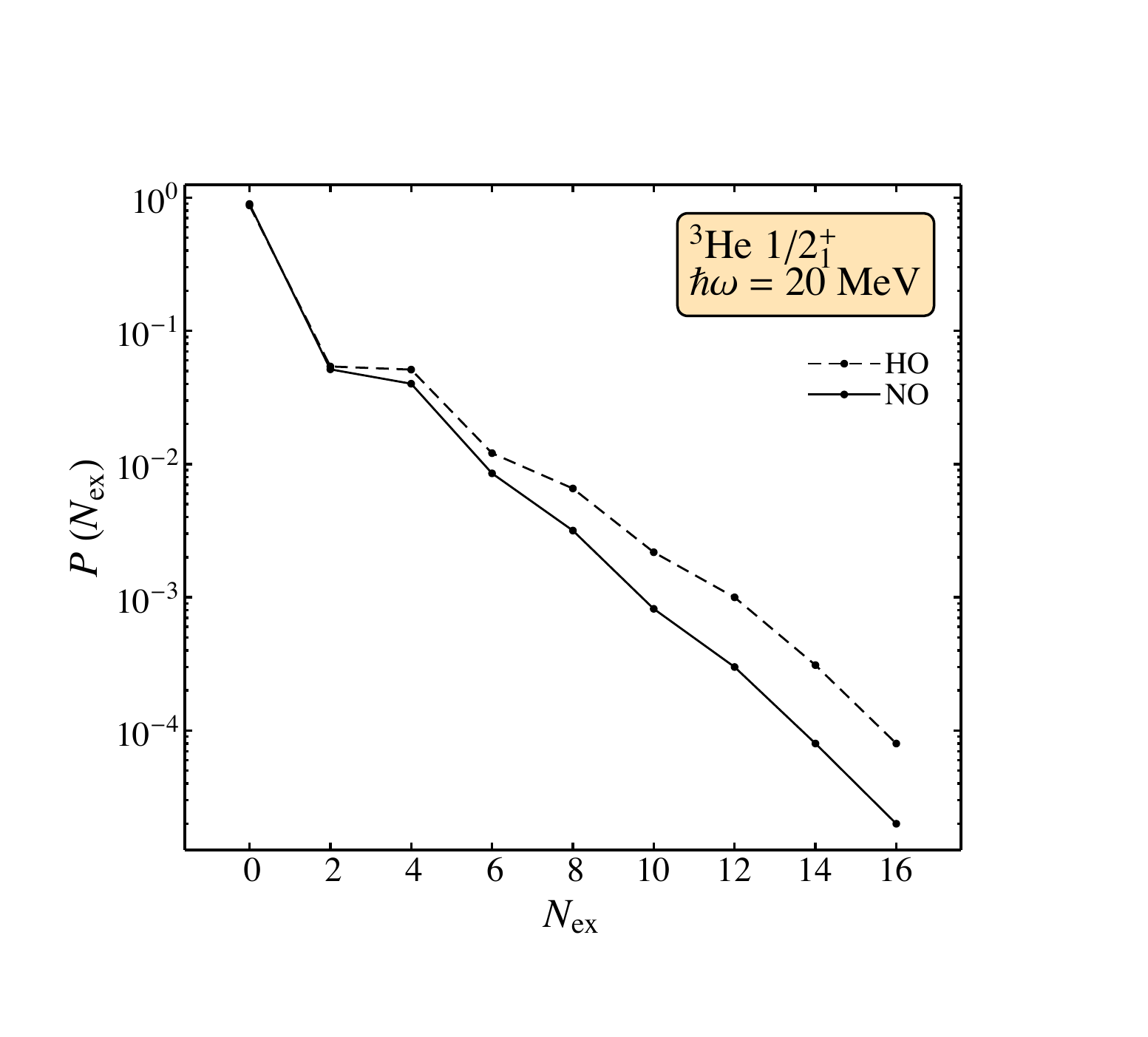}
\caption[Decomposition of the calculated many-body wave function of the $1/2^{+}$ ground state of {\protect $\isotope[3]{He}$} in terms of components with $N_{\mathrm{ex}}$ excitation quanta. The many-body wave function was obtained for $\Nmax = 16$ and $\hw = 20$ MeV using the harmonic oscillator basis and the natural orbital basis.]{Decomposition of the calculated many-body wave function of the $1/2^{+}$ ground state of $\isotope[3]{He}$ in terms of components with $N_{\mathrm{ex}}$ excitation quanta. The many-body wave function was obtained for $\Nmax = 16$ and $\hw = 20$ MeV using the harmonic oscillator basis (dashed curves) and the natural orbital basis (solid curves).}
\label{fig-chap3-h0-no-pnex-comparison}
\end{center}
\end{figure}

To infer whether the natural orbital basis builds in contributions from high-$N$ orbitals of the initial basis, we can plot the decomposition of the many-body wave function in terms of components with $N_{\mathrm{ex}}$ excitation quanta above the minimal configuration (the $N_{0}$ configuration). In Fig.~\ref{fig-chap3-h0-no-pnex-comparison}, we plot this decomposition for the calculated ground state wave-function of $\isotope[3]{He}$ obtained using harmonic oscillator orbitals (dashed curves) and natural orbitals (solid curves) at $\Nmax = 16$ and $\hw = 20$ MeV. We observe, that contributions from high-$N_{\mathrm{ex}}$ components of the oscillator basis are now build into the natural orbital basis. This means that the role of Slater determinants involving (natural) orbitals with low $N$ is enhanced.

\begin{figure}[t]
\begin{center}
\includegraphics[width=0.7 \columnwidth]{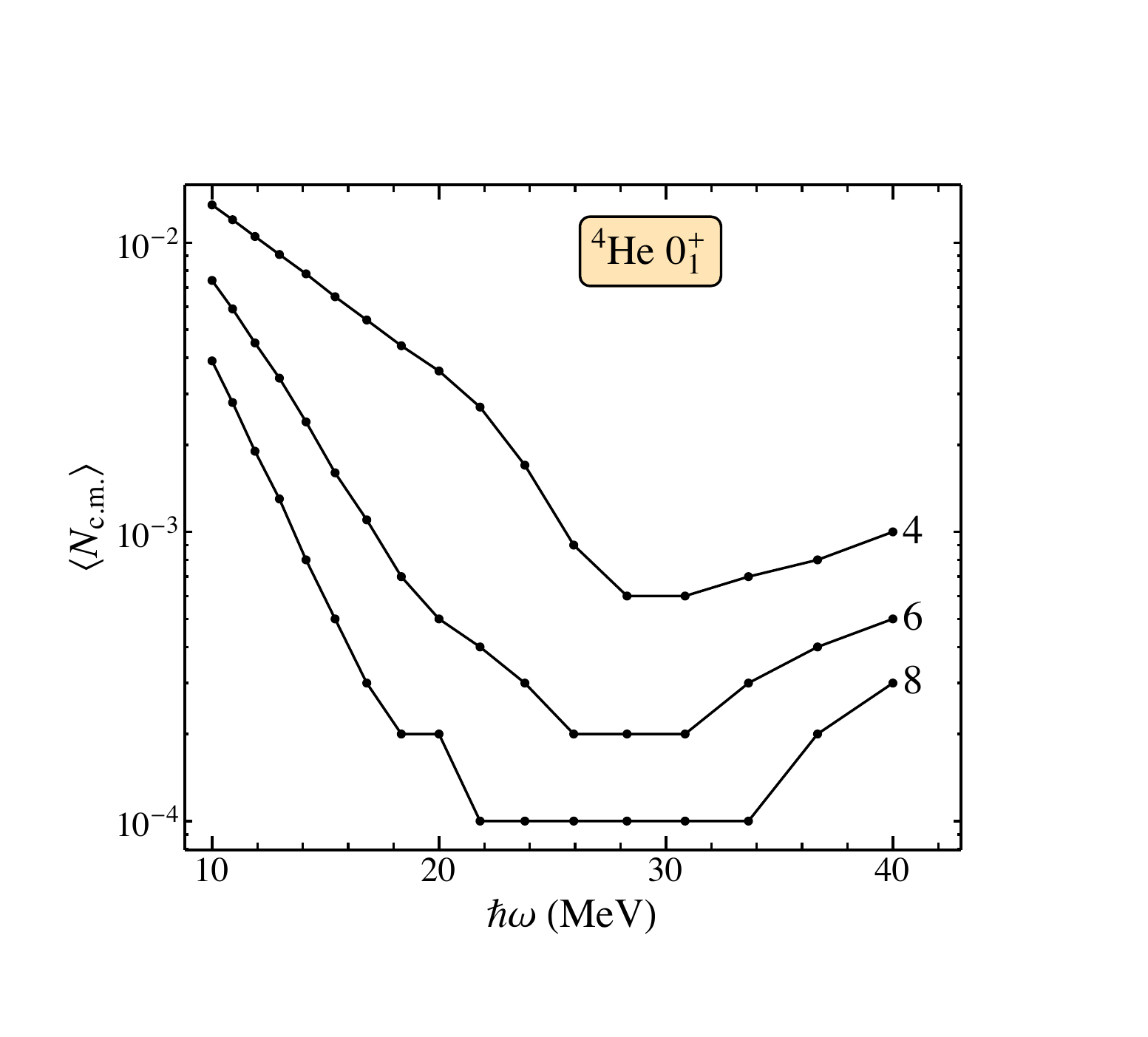}
\caption{The expectation value of the center-of-mass operator $\langle N_{\mathrm{c.m.}} \rangle$ in the calculated ground state wave-function of $\isotope[4]{He}$, obtained using natural orbitals, as a function of $\hw$ at various $\Nmax$ truncations of the many-body basis.}
\label{fig-chap3-z2-n2-cm-vs-hw}
\end{center}
\end{figure}

The removal of spurious center-of-mass states when we move away from the harmonic oscillator basis is very important. To study whether some degree of separability is maintained using natural orbitals, we can study the convergence properties of the expectation value of the operator $N_{\mathrm{c.m.}}$ (which counts the center-of-mass quanta in a many-body wave function) in the ground state of $\isotope[4]{He}$. In Fig.~\ref{fig-chap3-z2-n2-cm-vs-hw}, we plot the expectation value $\langle N_{\mathrm{c.m.}} \rangle$ of the center-of-mass operator against $\hw$ for various $\Nmax$ truncations. (The flat values around the $\hw \approx 20$ MeV region are due to the output precision of MFDn and has no physical meaning). We observe that around the minimum ($\hw \approx 20$ MeV) of the $\Nmax = 4$ curve, the expectation value is approximately $10^{-2}$ and it reduces to $10^{-4}$ by $\Nmax = 8$. Thus, a satisfactory degree of separability is still maintained using the natural orbital basis. Therefore, we can still add a Lawson term to the NCCI Hamiltonian built using natural orbitals to raise the spurious center-of-mass states out of the low-lying spectrum.

\begin{figure}
\begin{center}
\includegraphics[width=0.99 \columnwidth]{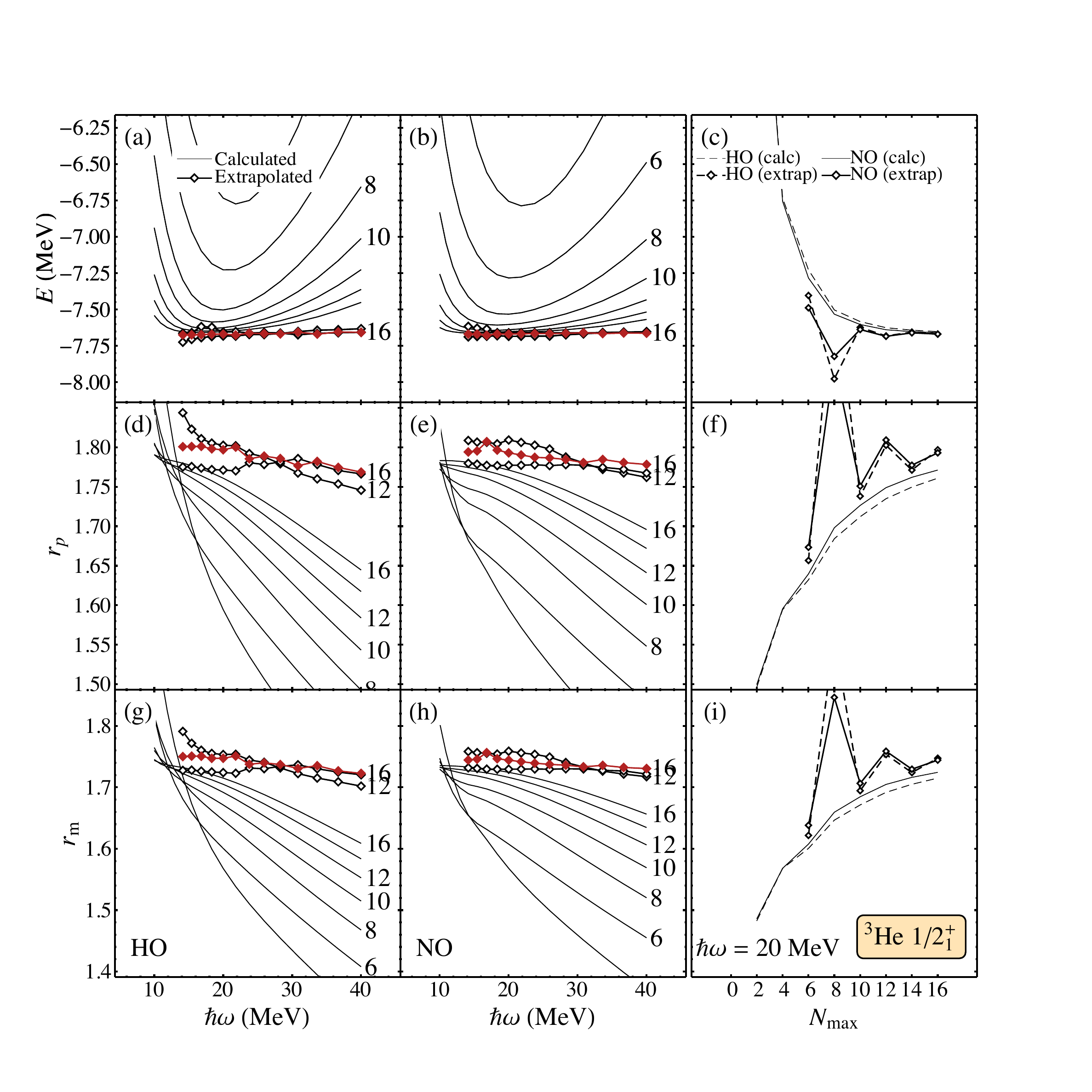}
\caption[Infrared basis extrapolations for the {\protect $\isotope[3]{He}$} ground state energy, proton, and matter radius, based on calculations in the harmonic oscillator basis and natural orbital basis.]{Infrared basis extrapolations for the $\isotope[3]{He}$ ground state energy~(top), proton radius~(middle), and matter radius (bottom), based on calculations in the harmonic oscillator basis~(left) and natural orbital basis~(middle). The extrapolations (diamonds) are shown along with the underlying calculated results (plain curves) as functions of $\hw$ at fixed $\Nmax$ (as indicated). The right column shows the evolution of the calculated and extrapolated results with $\Nmax$ for $\hw = 20$ MeV.}
\label{fig-chap3-z2-n1-extrapolation}
\end{center}
\end{figure}

\begin{figure}[t]
\begin{center}
\includegraphics[width=0.99 \columnwidth]{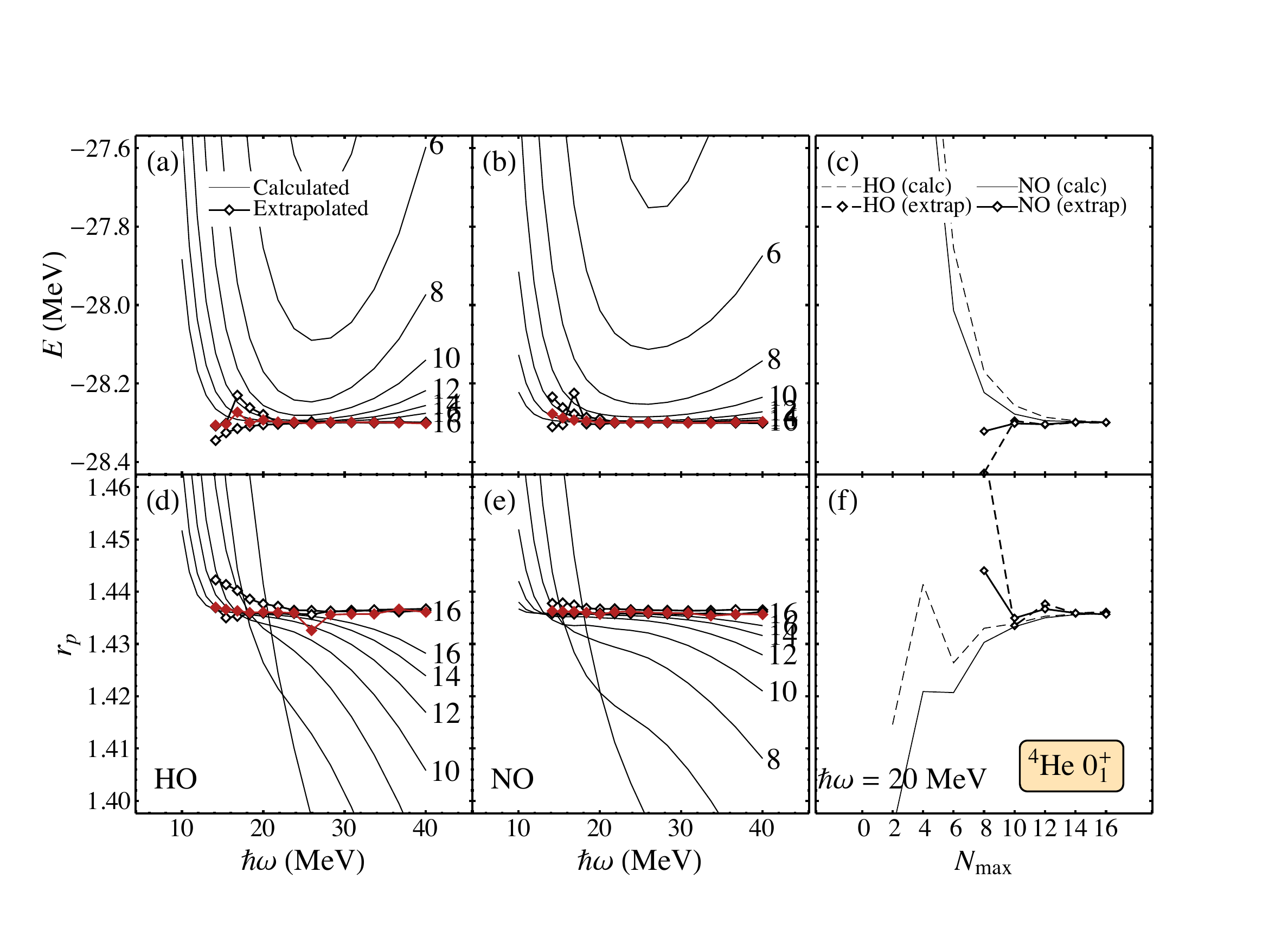}
\caption[Infrared basis extrapolations for the {\protect $\isotope[4]{He}$} ground state energy, and proton radius, based on calculations in the harmonic oscillator basis and natural orbital basis.]{Infrared basis extrapolations for the $\isotope[4]{He}$ ground state energy~(top), proton radius~(middle), and matter radius (bottom), based on calculations in the harmonic oscillator basis~(left) and natural orbital basis~(middle). The extrapolations (diamonds) are shown along with the underlying calculated results (plain curves) as functions of $\hw$ at fixed $\Nmax$ (as indicated). The right column shows the evolution of the calculated and extrapolated results with $\Nmax$ for $\hw = 20$ MeV.}
\label{fig-chap3-z2-n2-extrapolation}
\end{center}
\end{figure}

Although using natural orbitals leads to both faster convergence of calculated observables in terms of $\Nmax$ and improved $\hw$ convergence compared to the harmonic oscillator basis, full convergence is not achieved for the proton radius of $\isotope[3]{He}$. Thus, it is important to test whether we can still use extrapolation methods with results obtained using the natural orbital basis for the cases when convergence is incomplete. The natural orbital basis spans the same single-particle space as the harmonic oscillator basis therefore we can attempt to use the infrared extrapolation method (which was originally developed for the harmonic oscillator basis) with results obtained using natural orbitals. 

In Fig.~\ref{fig-chap3-z2-n1-extrapolation}, we extrapolate the calculated results for $\isotope[3]{He}$, obtained with the harmonic oscillator basis (left), and the natural orbital basis (middle). Moreover, we study the convergence of both the calculated results and the extrapolated results in terms of $\Nmax$ for $\hw = 20$ MeV (right). We perform a three-point infrared extrapolation for results sharing the same $\hw$ and obtained at three different consecutive $\Nmax$ truncations of the many-body basis ($\Nmax =12$, $14$, $16$) which are deemed UV converged as described in Chapter~\ref{chap-chap2}.

In the top row [panels (a), (b), (c)], we observe that the extrapolated ground state energy converges in terms of $\hw$, both for the harmonic oscillator basis [panel (a)] and the natural orbital basis [panel (b)] (however, a slight $\hw$ dependence persists for the harmonic oscillator extrapolated results). The evolution of the extrapolated results in terms of $\Nmax$ [panel (c)] confirms that the extrapolated results are approximately identical for the two bases. Moreover, the extrapolated results are somewhat stable with respect to $\Nmax$ (i.e., extrapolating the $\Nmax = 10$, $12$, and $14$ calculated results yields approximately the same extrapolated energy as the $\Nmax = 12$, $14$, and $16$ calculated results). The extrapolated result at $\hw = 20$ MeV is $-7.67$ MeV for both bases, $5$ keV short of the experimental result.

In the middle row [panels (d), (e), (f)], we observe that the harmonic oscillator extrapolated proton radius results stagger with respect to $\hw$, while the natural orbital results have a smoother dependence on $\hw$. Finally, in the bottom row, we observe (similarly to the proton extrapolations) that the harmonic oscillator extrapolated matter radius staggers with $\hw$, while the natural orbital extrapolated matter radius is more stable with respect to $\hw$. For the proton radius, the extrapolated results (across the range of $\hw$ parameters shown) are found in the range $1.77$-$1.80$ fm and $1.78$-$1.81$ fm for the harmonic oscillator and natural orbital basis respectively. These results are consistent with the experimental result [$r_{p} = 1.774(6)$ fm] however, they are not fully reliable as they depend on the $\hw$ parameter.

In Fig.~\ref{fig-chap3-z2-n2-extrapolation}, we extrapolate the calculated results for $\isotope[4]{He}$ obtained using the harmonic oscillator basis (left), and the natural orbital basis (middle). We also plot the evolution of both the calculated and the extrapolated results with $\Nmax$ for constant $\hw = 20$ MeV. In the top row, we observe that the extrapolated ground state energy results converge for both bases. For the natural orbital basis, the extrapolated ground state energy results are nearly $\hw$ independent. In the bottom row, the extrapolated proton radii converge in terms of $\hw$ for both bases. For the natural orbital basis, the extrapolated proton radii are $\hw$ independent. Quantitatively, the extrapolated ground state energy obtained at $\hw = 20$ MeV using either basis ($-28.3$ MeV) is consistent with the experimental result ($-28.3$ MeV). The extrapolated proton radius obtained for either basis at $\hw = 20$ MeV ($r_{p} = 1.44$ fm) is $\sim 0.02$ fm short of the experimental result [$r_{p} = 1.462(6)$ fm].

The overall conclusion is that we can still use the infrared extrapolation method with results calculated using natural orbitals. Moreover, extrapolating results calculated using the natural orbital basis yields improved convergence (of the extrapolated results) in terms of $\Nmax$ (see the right column in Figs.~\ref{fig-chap3-z2-n1-extrapolation} and~\ref{fig-chap3-z2-n2-extrapolation}) and $\hw$, compared to extrapolating results calculated using the harmonic oscillator basis.

%% file: chapters/chapter4/chapter4_draft_170402.tex
\chapter{HALO NUCLEI $\isotope[6]{He}$ AND $\isotope[8]{He}$ IN A NATURAL ORBITAL BASIS}
\label{chap-chap4}

\section{Overview}
\label{sec-chap4-overview}

Halo nuclei~\cite{jonson2004:light-dripline,tanihata1985:radii-he,tanihata1988:radii-be-b-halo,tanihata2013:halo-expt} are nuclei which can be be described as an inert core nucleus surrounded by (valence) nucleons orbiting around the core at large distances, forming a halo. The separation energy of the halo (valence) nucleons is small compared to the energy required to separate nucleons from the core, and the bound halo nuclear states are close to the continuum. For example, the halo nuclei $\isotope[6]{He}$ and $\isotope[8]{He}$ consist of a (tightly bound) $\isotope[4]{He}$ core surrounded by two and four weakly bound halo neutrons, respectively. The combination of weak binding and short-range nuclear force means that the halo nucleons can tunnel out into a volume well beyond the nuclear core and into the classically forbidden region~\cite{halo-nuclei-al-khalili:2004}. To understand this, consider the example of a simple one-dimensional square well. The deeply bound states of the square well are confined within the potential, and have very little extension beyond the walls of the potential however, the weakly bound states near the surface of the potential can penetrate well outside the walls of the well.

The accurate description of the structure of a halo nucleus depends on the correct description of the long-range part of the many-body wave function~\cite{quaglioni2009:ncsm-rgm}. In the example of the $\isotope[6]{He}$ system, the stability of the nucleus results from the pairing of the two valence neutrons and the effects of the three-body nuclear force (note that the $\isotope[5]{He}$ system is unbound). Because the core and valence nucleons can be separated, the nucleus is often described as a cluster system~\cite{zhukov-halo-nuclei-he6-be11-cluster-review:1993,jensen-few-body-effects-halo-structure:2001}. In a recent study, $\isotope[6]{He}$ was studied as a cluster system consisting of an alpha particle core with two valence neutrons orbiting around the core ($\isotope[4]{He}+n+n$ system)~\cite{redondo6hecluster:2016}. This approach uses the NCCI model space supplemented with cluster degrees of freedom to describe the ground and resonant states of $\isotope[6]{He}$. The study concluded that the convergence of the matter rms radius improves compared to using the NCCI model space alone, while the approach allows for the description of the resonant states of $\isotope[6]{He}$. Within the basic NCCI approach, the nuclei $\isotope[6]{He}$ and $\isotope[8]{He}$ were previously studied using the traditional harmonic oscillator basis and the Laguerre basis~\cite{caprio2014:cshalo}. For the Laguerre basis the effect of using different length parameters for the neutron and proton radial functions was investigated. However, slow convergence of the calculated radii and energies with respect to the truncation of the model space persists even when different lengths for protons and neutrons are used. Here we consider natural orbitals for the challenging many-body calculation of $\isotope[6,8]{He}$. These results were reported in part in Ref.~\cite{constantinouxxxx:natorb}.

We start by presenting the results from the many-body calculations performed for $\isotope[6]{He}$ and $\isotope[8]{He}$ using the harmonic oscillator basis and the natural orbital basis (Sec.~\ref{sec-chap4-results}). The calculated results obtained using both bases are then extrapolated to the full space using the infrared extrapolation method (Sec.~\ref{sec-chap4-extrapolation}). Finally, an estimation of the converged radii based on the crossover point is also made (Sec.~\ref{sec-chap4-crossover-point-analysis}).

\section{Results} 
\label{sec-chap4-results}

As in the previous chapter, we start with an initial many-body calculation using the harmonic oscillator basis, the JISP$16$ internucleon interaction, and the Coulomb interaction between protons. For the initial calculations, we used truncations of the many-body basis up to $\Nmax = 16$ for $\isotope[6]{He}$ and $\Nmax = 14$ for $\isotope[8]{He}$, respectively, and $\hw$ parameters in the range $10$-$40$ MeV.

Subsequently, the calculations were repeated using natural orbitals obtained by diagonalizing the initial one-body density matrices for each $(\Nmax,\hw)$ pair. To obtain the two-body matrix elements of the JISP$16$ interaction in the natural orbital basis, we started with the JISP$16$ two-body matrix elements expressed in the harmonic oscillator basis with $\hw_{\mathrm{int}} = 40$ MeV, and performed the two-body transformation~(\ref{eqn-chap3-two-body-interaction-transformation-natural-orbitals}) to get the two-body matrix elements in the natural orbital basis. 

\begin{figure}
\begin{center}
\includegraphics[width=0.65 \textwidth]{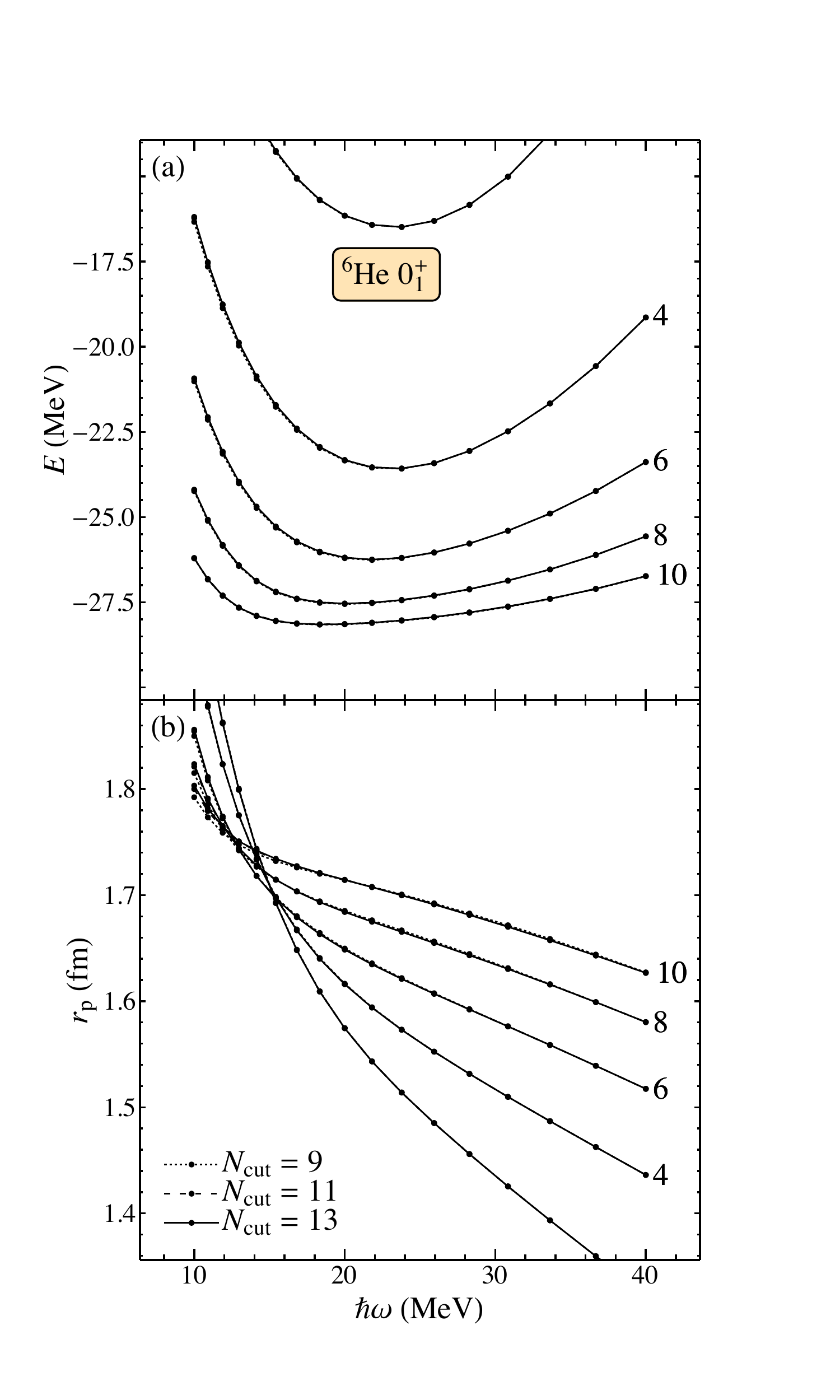}
\caption[The calculated ground state energy and proton radius in the ground state of {\protect $\isotope[6]{He}$}, obtained using the natural orbital basis and one-body shell truncations $N_{\mathrm{cut}}=9$, $11$, and $13$. The initial JISP$16$ interaction is expressed in the harmonic oscillator basis with $\hw_{\mathrm{int}} = 40$ MeV.]{The calculated ground state energy (a) and proton radius in the ground state (b) of $\isotope[6]{He}$, obtained using the natural orbital basis and one-body shell truncations of the quadruple sum in (\ref{eqn-chap3-two-body-interaction-transformation-natural-orbitals}) $N_{\mathrm{cut}}=9$ (dotted curves), $11$ (dashed curves), and $13$ (solid curves). The initial JISP$16$ interaction is expressed in the harmonic oscillator basis with $\hw_{\mathrm{int}} = 40$ MeV.}
\label{fig-chap4-ncut-convergence-he6}
\end{center}
\end{figure}

To assess which shell truncation $N_{\mathrm{cut}}$ in the quadruple sum in (\ref{eqn-chap3-two-body-interaction-transformation-natural-orbitals}) yields $N_{\mathrm{cut}}$-independent results we first performed NCCI calculations in the natural orbital basis (up to $\Nmax = 10$) with $N_{\mathrm{cut}} = 9, 11, 13$. In Fig.~\ref{fig-chap4-ncut-convergence-he6}, we plot the calculated ground state energy (a) and proton radius in the ground state (b) obtained using the natural orbital basis and one-body shell truncations $N_{\mathrm{cut}}=9$ (dotted curves), $11$ (dashed curves), and $13$ (solid curves). For the calculated ground state energy [panel (a)], we observe that there is some slight $N_{\mathrm{cut}}$ dependence of the calculated results for low $\Nmax$ truncations (and low $\hw$ parameters) which diminishes as $\Nmax$ increases. For the calculated proton radius [panel (b)], there is a slight dependence on $N_{\mathrm{cut}}$ at the highest $\Nmax = 10$ truncation (and low $\hw$ parameters) which is however very small ($\sim 10^{-3}$ fm). Thus, we conclude that a shell truncation of $N_{\mathrm{cut}} = 13$ provides sufficiently $N_{\mathrm{cut}}$ independent results. Throughout this thesis, we start from the JISP$16$ interaction expressed in the harmonic oscillator basis with $\hw_{\mathrm{int}} = 40$ MeV and transform to the natural orbital basis using a one-body shell truncation $N_{\mathrm{cut}} = 13$.

\subsection{Natural orbitals}
\label{sec-chap4-natural-orbitals}

\begin{figure}
\begin{center}
\includegraphics[width=1.\textwidth]{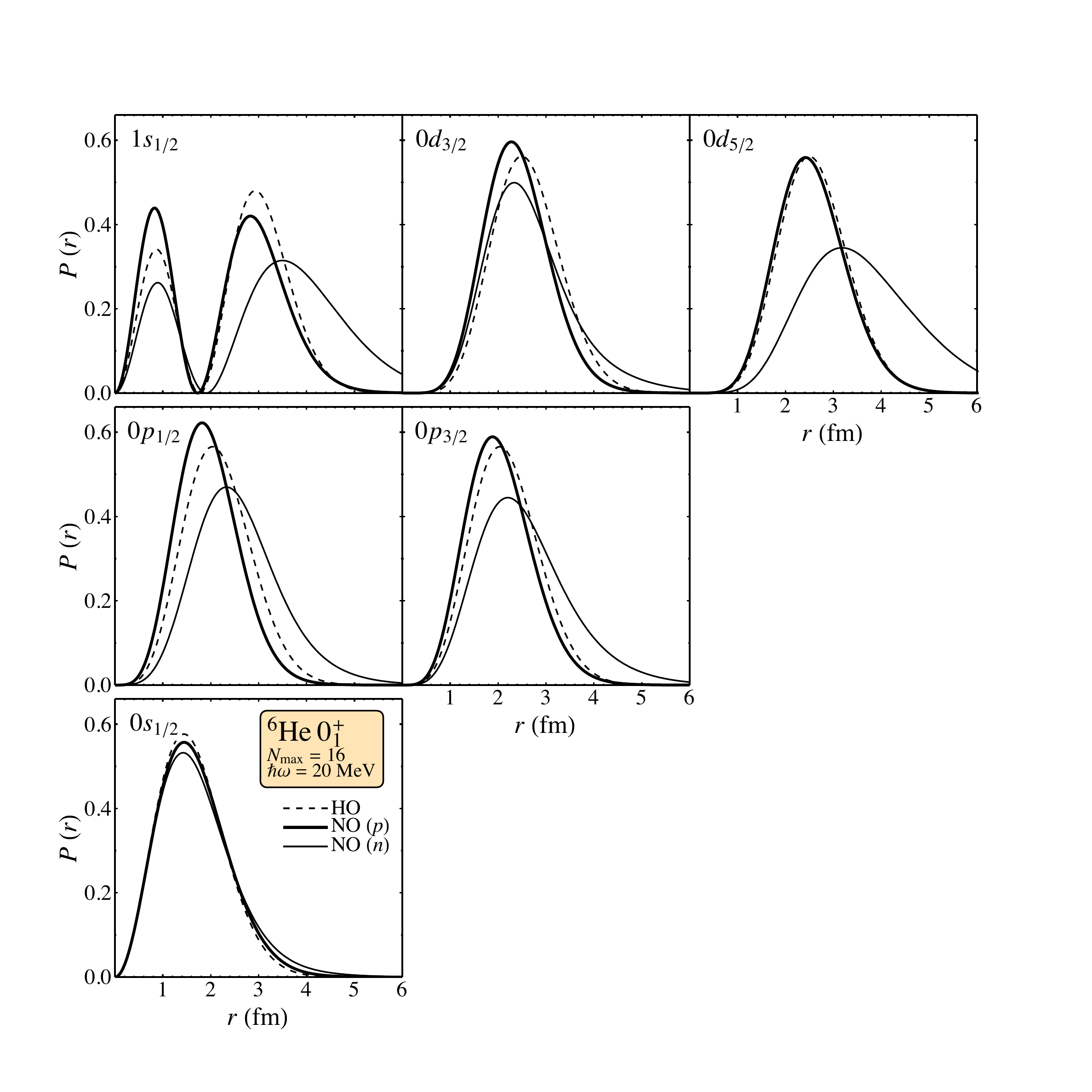}
\caption[Radial probability density functions $P(r) = r^{2} | R_{nlj}(b;r) |^{2}$ for harmonic oscillator, proton, and neutron orbitals of {\protect $\isotope[6]{He}$} up to the $N=2$ major shell.]{Radial probability density functions $P(r) = r^{2} | R_{nlj}(b;r) |^{2}$ for harmonic oscillator (dashed curves), proton (thick dark curves), and neutron (dark curves) orbitals of $\isotope[6]{He}$ up to the $N=2$ major shell. The natural orbitals were derived from an initial scalar one-body density matrix obtained in the harmonic oscillator basis at $\Nmax = 16$ and $\hw = 20$ MeV.}
\label{fig-chap4-natural-orbitals-N-2-he-6}
\end{center}
\end{figure}
\begin{figure}
\begin{center}
\includegraphics[width=1.\textwidth]{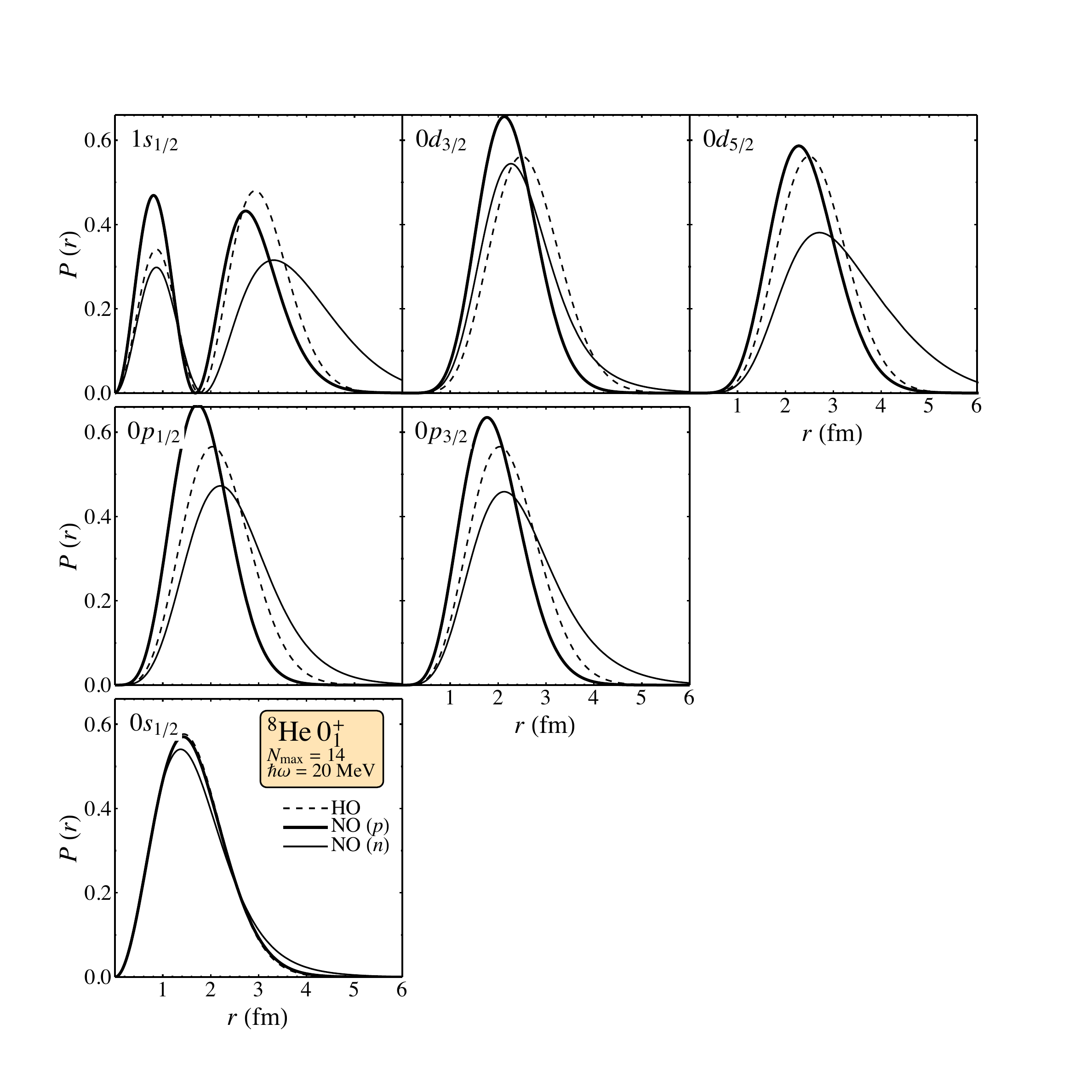}
\caption[Radial probability density functions $P(r) = r^{2} | R_{nlj}(b;r) |^{2}$ for harmonic oscillator, proton, and neutron orbitals of {\protect $\isotope[8]{He}$} up to the $N=2$ major shell.]{Radial probability density functions $P(r) = r^{2} | R_{nlj}(b;r) |^{2}$ for harmonic oscillator (dashed curves), proton (thick dark curves), and neutron (dark curves) orbitals of $\isotope[8]{He}$ up to the $N=2$ major shell. The natural orbitals were derived from an initial scalar one-body density matrix obtained in the harmonic oscillator basis at $\Nmax = 16$ and $\hw = 20$ MeV.}
\label{fig-chap4-natural-orbitals-N-2-he-8}
\end{center}
\end{figure}

Before presenting the many-body calculations, it is instructive to study the properties of the natural orbitals of $\isotope[6,8]{He}$. In Figs.~\ref{fig-chap4-natural-orbitals-N-2-he-6} and~\ref{fig-chap4-natural-orbitals-N-2-he-8}, we plot the radial probability density $P(r) = r^{2} | R_{nlj}(b;r) |^{2}$ for the harmonic oscillator (dashed curves), proton (thick dark curves), and neutron (dark curves) orbitals of $\isotope[6]{He}$ and $\isotope[8]{He}$ up to the $N=2$ major shell, obtained from initial harmonic oscillator scalar one-body densities with $\hw = 20$ MeV and $\Nmax=16$ and $\Nmax=14$ for $\isotope[6]{He}$ and $\isotope[8]{He}$ respectively. We observe that the natural orbitals of $\isotope[6]{He}$ and $\isotope[8]{He}$ are comparable. Specifically, the tails of the neutron natural orbitals are longer than the tails of both the proton natural orbitals and the initial harmonic oscillator orbitals. Moreover, the tails of the proton natural orbitals are shorter than the tails of the initial harmonic oscillator orbitals.

\begin{figure}
\begin{center}
\includegraphics[width=1.\textwidth]{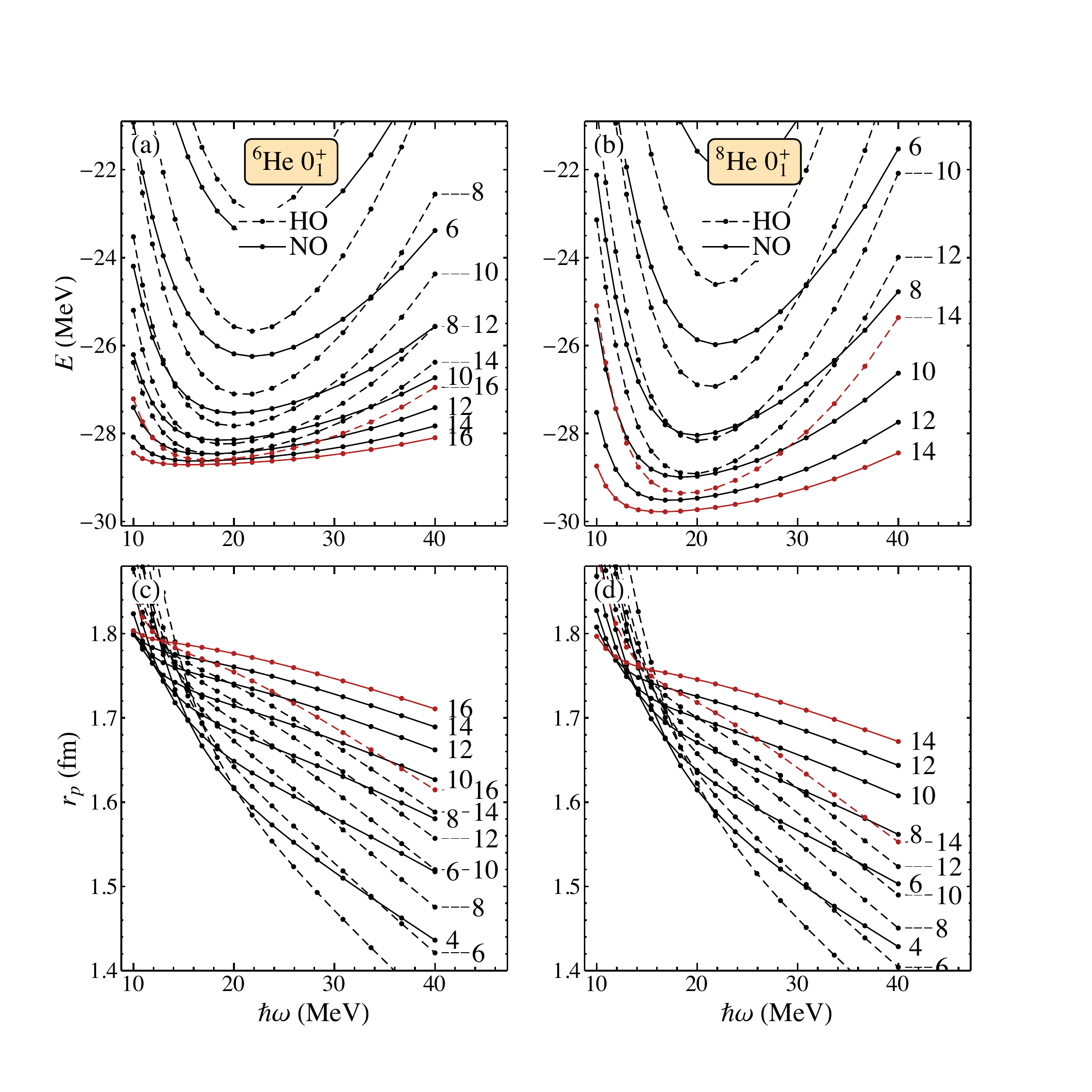}
\caption[The calculated ground state energy and point-proton rms radius in the ground state of {\protect $\isotope[6]{He}$}, and {\protect $\isotope[8]{He}$} obtained using harmonic oscillator orbitals and natural orbitals. The JISP$16$ internucleon interaction and the Coulomb interaction between protons were used.]{The calculated ground state energy (top) and point-proton rms radius in the ground state (bottom) of $\isotope[6]{He}$ (left), and $\isotope[8]{He}$ (right) obtained using harmonic oscillator orbitals (dashed curves) and natural orbitals (solid curves). The JISP$16$ internucleon interaction and the Coulomb interaction between protons were used. The red colored curves show results obtained at the highest $\Nmax$ truncation.}
\label{fig-chap4-he6-he8-gs-ho-no-comparison}
\end{center}
\end{figure}

\subsection{Calculated energies and radii}
\label{sec-chap4-calculated-energies}

Several experimental properties of the ground state of $\isotope[6]{He}$ support the interpretation that it consists of a weakly-bound two-neutron halo surrounding a tightly-bound $\alpha$ core~\cite{jonson2004:light-dripline,tanihata1985:radii-he}. Similarly, the $\isotope[8]{He}$ nucleus is interpreted as an $\alpha$ core surrounded by four halo neutrons. The two-neutron separation energy for $\isotope[6]{He}$ is only $0.97$ MeV, out of a total binding energy of $29.27$ MeV, while the two-neutron separation energy of $\isotope[8]{He}$ is $2.13$ MeV out of a total binding energy of $31.40$ MeV~\cite{tilley-a6-8:2002}. Experimentally, the onset of halo structure along the $\isotope[]{He}$ isotopic chain is indicated by a jump in the measured charge and matter radii, from $\isotope[4]{He}$ to $\isotope[6]{He}$ (the charge and matter radii of $\isotope[8]{He}$ are comparable to those of $\isotope[6]{He}$). The root mean square (rms) point-proton distribution radius $r_{p}$, which may be deduced from the measured charge radius $r_{c}$~\cite{lu2013:laser-neutron-rich}, increases by $32 \%$ from $\isotope[4]{He}$ [$r_{p} = 1.462(6)$ fm] to $\isotope[6]{He}$ [$r_{p} = 1.934(9)$ fm]~\cite{lu2013:laser-neutron-rich,wang2004:6he-radius-laser,brodeur-he6-radius:2012}. [The point-proton rms radius of $\isotope[8]{He}$ is $r_{p} = 1.881(17)$ fm~\cite{lu2013:laser-neutron-rich}]. This increase may be understood as a consequence of halo structure, arising from the recoil of the charged $\alpha$ core against the halo neutrons [as well as possible contributions from swelling of the $\alpha$ core~\cite{lu2013:laser-neutron-rich}]. 

The matter radii are obtained with considerably greater uncertainties, from either nuclear interaction cross sections~\cite{tanihata1985:radii-he} or proton-nucleus elastic scattering data~\cite{alkhazov2002:elastic-halo-radii}. These methods yield model-dependent and often contradictory results along the $\isotope[]{He}$ isotopic chain. Specifically, the reported values are in the range $1.46$-$1.66$ fm for $\isotope[4]{He}$, $2.23$-$2.75$ fm for $\isotope[6]{He}$, and $2.38$-$2.61$ fm for $\isotope[8]{He}$~\cite{tanihata1988:radii-be-b-halo,tanihata1992:neutron-skins,alkhalili2003:inelastic-halo-radii,alkhazov2002:elastic-halo-radii}.

Theoretically, the point-nucleon rms radii (derived by assuming that the nucleon is a point particle) are derived by evaluating the expectation value of the point-nucleon operators with respect to the calculated many-body wave function. Formally, the point-nucleon rms radii are two-body operators determined with respect to the center-of-mass~\cite{bacca2012:6he-hyperspherical}
\begin{align}
r_{p}^{2} &= \frac{1}{Z} \sum_{i=1}^{Z}  (\mathbf{r}_{i} - \mathbf{R})^{2} \\
r_{n}^{2} &= \frac{1}{N} \sum_{i=1}^{N} (\mathbf{r}_{i} - \mathbf{R})^{2} \\
r_{\mathrm{rel}}^{2} &= \sum_{i=1}^{A} (\mathbf{r}_{i} - \mathbf{R})^{2},
\label{eqn-chap4-rms-radii-definitions}
\end{align} 
where $\mathbf{r}_{i}$ is the nucleon's position vector, $\mathbf{R}$ is the center-of-mass vector, and $Z$, $N$ is the number of protons and neutrons respectively. The point-proton ($r_{p}$), point-neutron ($r_{n}$), and point-matter ($r_{\mathrm{m}}$) rms radii are related via $A r_{\mathrm{m}}^{2} = Z r_{p}^{2} + N r_{n}^{2}$.

Let us now discuss the calculated ground and excited state energies and radii of $\isotope[6]{He}$ and $\isotope[8]{He}$. In Fig.~\ref{fig-chap4-he6-he8-gs-ho-no-comparison}, we plot the calculated ground state energy (top) and point-proton rms radius in the $0^{+}$ ground state (bottom) of $\isotope[6]{He}$ (left) and $\isotope[8]{He}$ (right) obtained using the harmonic oscillator basis (dashed curves) and the natural orbital basis (solid curves).

We observe that the energies calculated using natural orbitals are lower than the energies calculated using harmonic oscillator orbitals. This means that using natural orbitals we come closer to the true value due to the variational principle. Quantitatively, the natural orbital curves converge faster than the oscillator curves by (roughly) one step in $\Nmax$ in the vicinity of the variational minimum and two (or more) steps in $\Nmax$ at high or low $\hw$ parameters.

For the calculated point-proton rms radii, results obtained using natural orbitals converge faster than results obtained using harmonic oscillator orbitals. Specifically, at $\hw \approx 12$ MeV the natural orbital curves are about a step in $\Nmax$ ahead of the oscillator curves, while at high $\hw$ the natural orbital curves are several steps in $\Nmax$ ahead of the oscillator curves. Finally, radii obtained using natural orbitals are less $\hw$ dependent than radii obtained using oscillator orbitals.

\section{Infrared extrapolations}
\label{sec-chap4-extrapolation}

\begin{figure}
\begin{center}
\includegraphics[width=1.\textwidth]{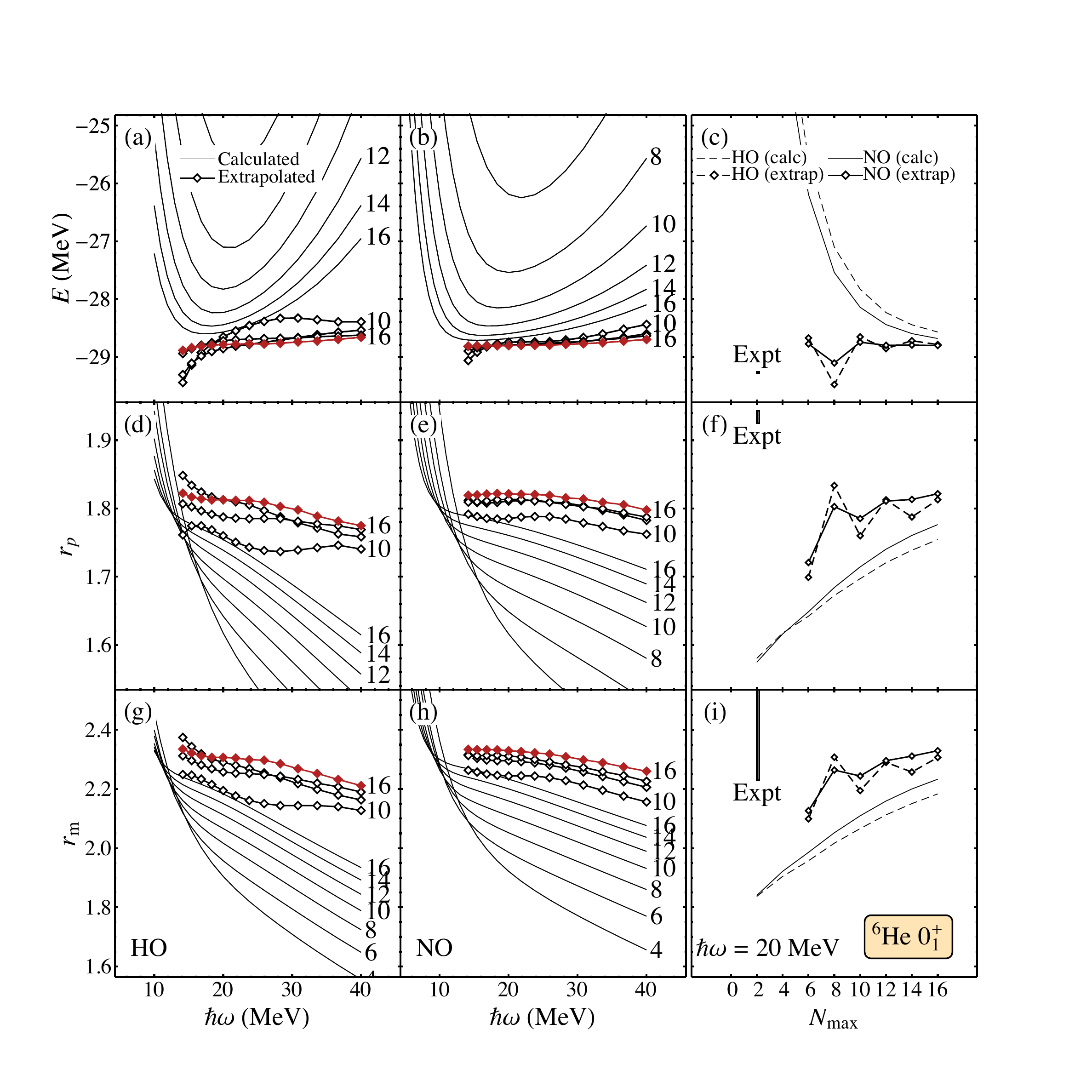}
\caption[Infrared extrapolations of the {\protect $\isotope[6]{He}$} calculated ground state energy, point-proton rms radius, and point-matter rms radius.]{Infrared basis extrapolations for the $\isotope[6]{He}$ ground state energy~(top), point-proton rms radius~(middle), and point-matter rms radius (bottom), based on calculations in the harmonic oscillator basis~(left) and natural orbital basis~(middle). The evolution of the calculated and extrapolated results with $\Nmax$ at $\hw = 20$ MeV and the experimental values (rectangles) are shown in the right column. The extrapolations (diamonds) are shown along with the underlying calculated results (plain curves) as functions of $\hw$ at fixed $\Nmax$ (as indicated).}
\label{fig-chap4-he6-ground-state-extrapolation}
\end{center}
\end{figure}

In this section we use the infrared extrapolation method (see Chapter~\ref{chap-chap2}) with our calculated results for $\isotope[6,8]{He}$. In Fig.~\ref{fig-chap4-he6-ground-state-extrapolation}, we perform a three point extrapolation of the calculated results of $\isotope[6]{He}$ obtained using the harmonic oscillator basis (left column) and the natural orbital basis (middle column). Moreover, we plot the evolution of the calculated and extrapolated results with $\Nmax$ at $\hw = 20$ MeV (right column). In the same column (right), we also show the experimental results (plotted as rectangles, where the center of the rectangle is the experimental result and the height of the rectangle indicates the uncertainty in the experimental result). In the top row we show the ground state energy results, in the middle row we show the point-proton rms radius results, and in the bottom row we show the point-matter rms radius results. The originally calculated results are shown as light curves.

We start with the extrapolated ground state energies of $\isotope[6]{He}$, shown in Fig.~\ref{fig-chap4-he6-ground-state-extrapolation}(a) for the harmonic oscillator basis and Fig.~\ref{fig-chap4-he6-ground-state-extrapolation}(b) for the natural orbital basis. We observe that the extrapolated natural orbital results are considerably less $\hw$-dependent than the extrapolated harmonic oscillator results. Moreover, by looking in Fig.~\ref{fig-chap4-he6-ground-state-extrapolation}(c), we can infer that results obtained by extrapolating calculated natural orbital results are less $\Nmax$ dependent than results obtained by extrapolating harmonic oscillator results. Going back to Fig.~\ref{fig-chap4-he6-ground-state-extrapolation}(b), we can infer that the extrapolated ground state energy at $\hw \approx 20$ MeV is approximately converged (at the $30$ keV level). The extrapolated ground state energy result at $\hw = 20$ MeV is $-28.79$ MeV (for comparison the extrapolated ground state energy result for the harmonic oscillator basis at the same $\hw$ is $-28.80$ MeV, which is consistent with the natural orbital result). This means that using the JISP$16$ internucleon interaction our calculation underbinds $\isotope[6]{He}$ by about $\sim 0.5$ MeV.

Let us now move to the extrapolated point-proton rms radii in the ground state of $\isotope[6]{He}$, shown in Fig.~\ref{fig-chap4-he6-ground-state-extrapolation}(d) for the harmonic oscillator basis and Fig.~\ref{fig-chap4-he6-ground-state-extrapolation}(e) for the natural orbital basis. We observe that the extrapolated harmonic oscillator results [Fig.~\ref{fig-chap4-he6-ground-state-extrapolation}(d)] are considerably more $\hw$ dependent than the extrapolated natural orbital results [Fig.~\ref{fig-chap4-he6-ground-state-extrapolation}(e)]. Moreover, in Fig.~\ref{fig-chap4-he6-ground-state-extrapolation}(f) we observe that the natural orbital extrapolations are less $\Nmax$ dependent than the harmonic oscillator extrapolated results. Overall, it is not clear whether we get $\Nmax$ converegnce of the extrapolated results in either basis. However, going back to Fig.~\ref{fig-chap4-he6-ground-state-extrapolation}(e), notice that for $\Nmax = 16$ and across all the $\hw$ values shown the extrapolated $r_{p}$ varies by only $\sim 0.02$ fm. Taking the extrapolated natural orbital proton radius at $\hw = 20$ MeV and $\Nmax = 16$ as representative gives $r_{p} \approx 1.82$ fm, which is about $0.1$ fm short of the experimental result [$r_{p} = 1.462(6)$ fm].

The extrapolated point-matter rms radii of $\isotope[6]{He}$ are shown in Fig.~\ref{fig-chap4-he6-ground-state-extrapolation}(g) for the harmonic oscillator basis, and Fig.~\ref{fig-chap4-he6-ground-state-extrapolation}(h) for the natural orbital basis. Similarly to the extrapolated proton radii, the extrapolated matter rms radii obtained from the natural orbital results [Fig.~\ref{fig-chap4-he6-ground-state-extrapolation}(h)] are less $\hw$ and $\Nmax$ dependent than the extrapolated matter radii obtained from the oscillator basis results [Fig.~\ref{fig-chap4-he6-ground-state-extrapolation}(g)]. At $\Nmax = 16$, the extrapolated matter radius varies by $\sim 0.07$ fm across the $\hw$ values shown, which means that a less reliable estimation (compared to the proton radius) of the matter radius can be made. If we again take the extrapolated natural orbital radius at $\hw = 20$ MeV and $\Nmax = 16$ as representative we get $r_{\mathrm{m}} \approx 2.33$ fm, which is within the range of experimentally reported values [$r_{m} = 2.23$-$2.75$ fm].

\begin{figure}
\begin{center}
\includegraphics[width=1.\textwidth]{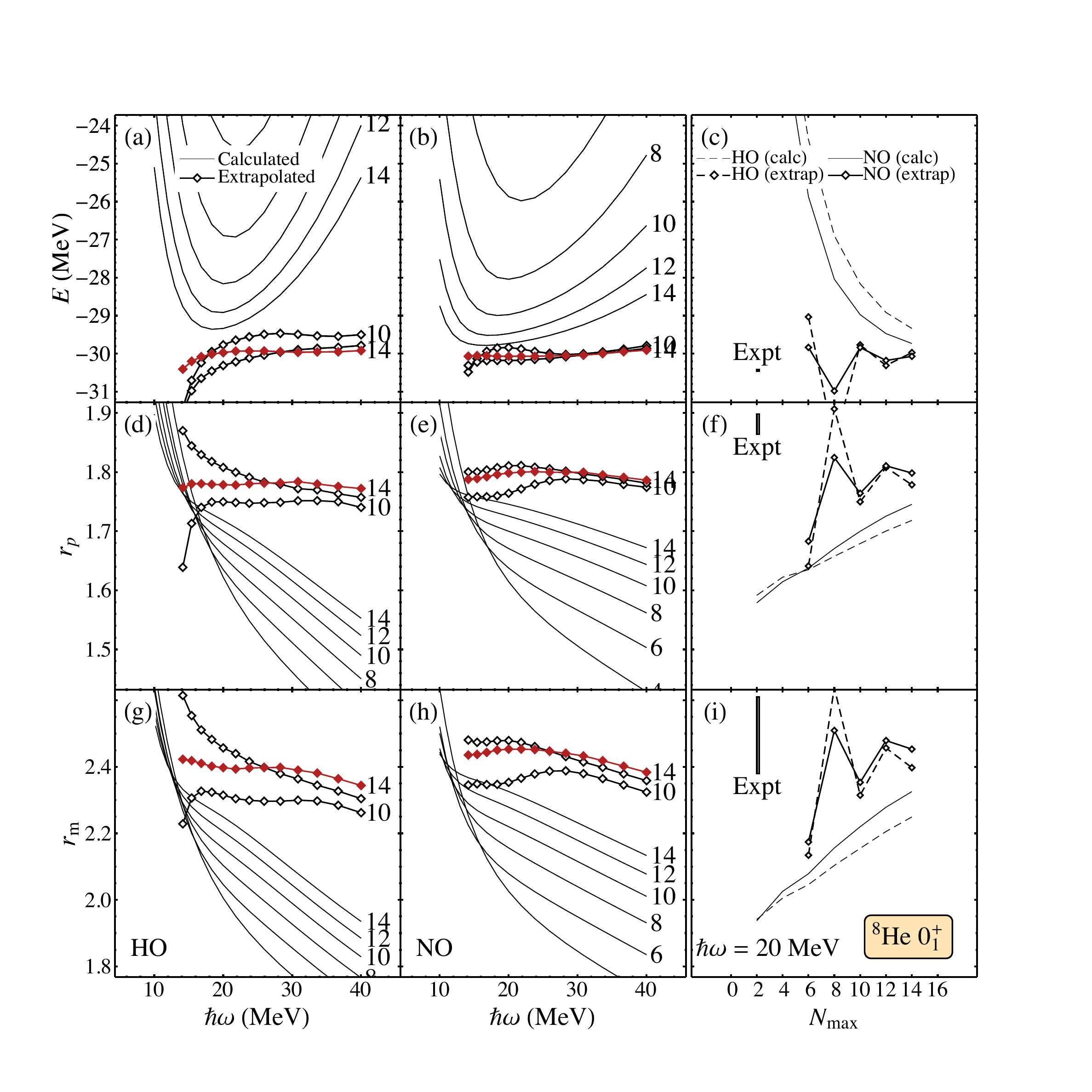}
\caption[Infrared extrapolations of the {\protect $\isotope[8]{He}$} calculated ground state energy, point-proton rms radius, and point-matter rms radius.]{Infrared basis extrapolations for the $\isotope[8]{He}$ ground state energy~(top), point-proton rms radius~(middle), and point-matter rms radius (bottom), based on calculations in the harmonic oscillator basis~(left) and natural orbital basis~(middle). The evolution of the calculated and extrapolated results with $\Nmax$ at $\hw = 20$ MeV and the experimental values (rectangles) are shown in the right column. The extrapolations (diamonds) are shown along with the underlying calculated results (plain curves) as functions of $\hw$ at fixed $\Nmax$ (as indicated).}
\label{fig-chap4-he8-ground-state-extrapolation}
\end{center}
\end{figure}

Let us now extrapolate the calculated results of $\isotope[8]{He}$. The ground state energy extrapolations are shown in Figs.~\ref{fig-chap4-he8-ground-state-extrapolation}(a) for the harmonic oscillator basis and \ref{fig-chap4-he8-ground-state-extrapolation}(b) for the natural orbital basis. As with the case of $\isotope[6]{He}$, the natural orbital extrapolations are significantly less $\hw$ and $\Nmax$ dependent than the harmonic oscillator extrapolations. However, because overall the calculated results are less UV converged than the $\isotope[6]{He}$ results (recall that for $\isotope[8]{He}$, calculations were only performed only up to $\Nmax = 14$) we observe that the extrapolated results in Fig.~\ref{fig-chap4-he8-ground-state-extrapolation}(b) do not approximately converge with respect to $\Nmax$ (or $\hw$) like the $\isotope[6]{He}$ results. If we nevertheless consider the extrapolated result for $\hw = 20$ MeV and $\Nmax = 14$ as representative, the extrapolated ground state energy (from the natural orbital extrapolations) is $-30.07$ MeV (for comparison, the extrapolated ground state energy is $-29.97$ MeV for the harmonic oscillator basis), which is $\sim 0.4$ MeV short of the experimental result ($-31.40$ MeV).

The extrapolated point-proton (middle in Fig.~\ref{fig-chap4-he8-ground-state-extrapolation}) and point-matter (bottom in Fig.~\ref{fig-chap4-he8-ground-state-extrapolation}) rms radii of $\isotope[8]{He}$ depend significantly on $\hw$ and $\Nmax$ in either basis. The dependence on $\hw$ and $\Nmax$ is ``smoother'' for the natural orbital extrapolations however, notice that in Fig.~\ref{fig-chap4-he8-ground-state-extrapolation}(e) the $\Nmax = 12$ extrapolated curve crosses the $\Nmax = 14$ extrapolated curve in the low $\hw$ region. Similar conclusions apply to the extrapolated matter radii in Fig.~\ref{fig-chap4-he8-ground-state-extrapolation}(h). For $\Nmax = 14$, the natural orbital extrapolated results are in the range $1.79$-$1.80$ fm for the proton radius ($1.77$-$1.78$ fm for the harmonic oscillator basis), which is about $\sim 0.13$ fm short of the experimental result [$r_{p} = 1.934(9)$ fm]. Similarly for $\Nmax = 14$, the natural orbital extrapolated matter radii are in the range $2.38$-$2.45$ fm ($2.34$-$2.42$ fm for the harmonic oscillator basis), consistent with the experimentally reported results $2.38$-$2.61$ fm.

\section{Crossover point analysis}
\label{sec-chap4-crossover-point-analysis}

If we take a close look at the dependence of the calculated point-nucleon radii as functions of $\hw$ (Fig.~\ref{fig-chap4-he6-he8-gs-ho-no-comparison}), we will notice that there is a qualitative similarity between the result calculated using the harmonic oscillator or the natural orbital basis. Specifically, in the low $\hw$ region (and below the variational minimum) results obtained for a given $\Nmax$ truncation and the immediately higher $\Nmax + 2$ cross at the so called ``crossover point''~\cite{caprio2014:cshalo}. To the left of the crossover point the calculated results decrease with $\Nmax$, and to the right of the crossover point the calculated results increase with $\Nmax$. Thus, at the crossover point the calculated results are approximately $\Nmax$ independent and the calculated results at the crossover point can be used as a reasonable estimate of the converged radius.

\begin{figure}[t]
\begin{center}
\includegraphics[width=0.7 \textwidth]{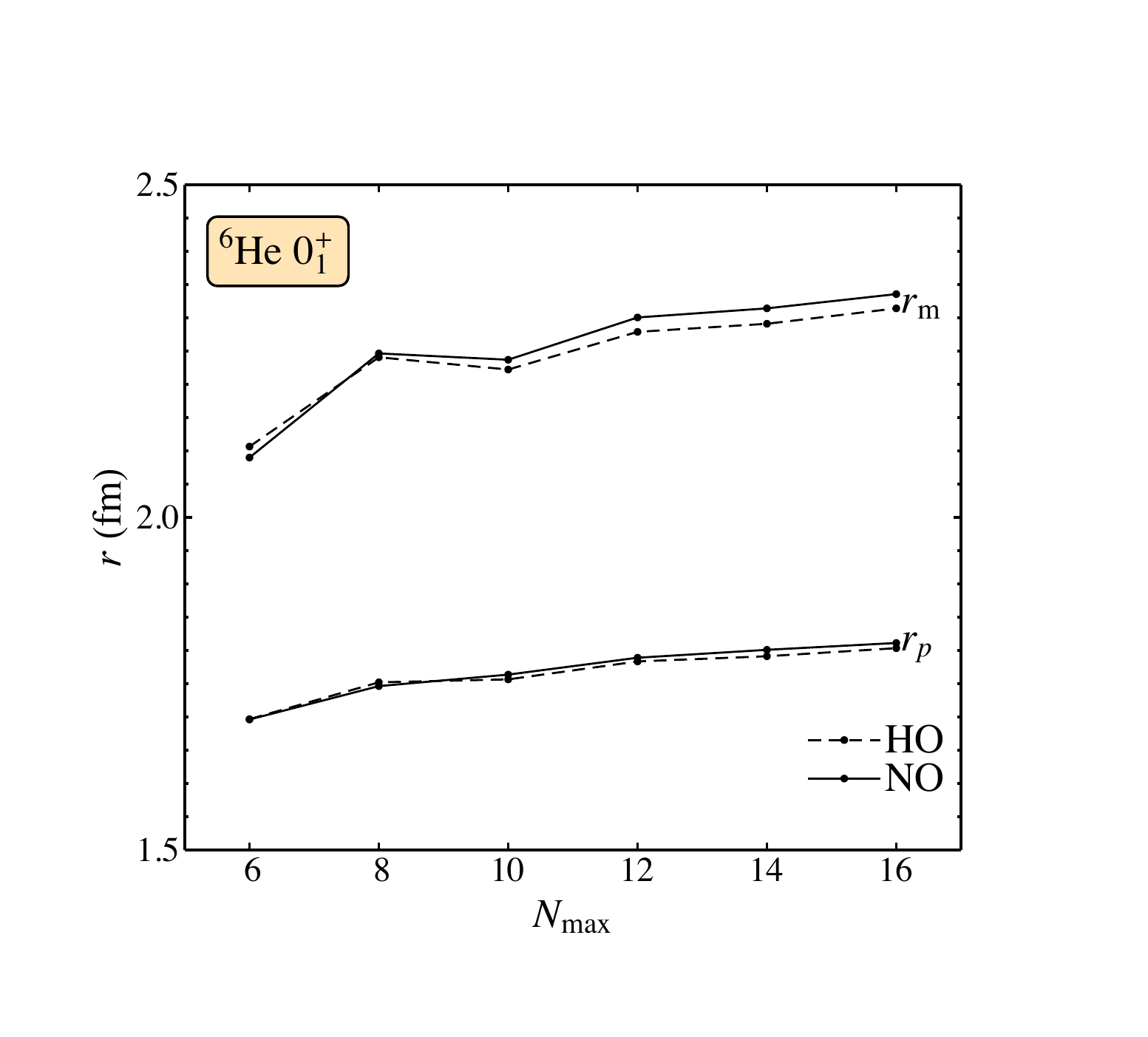}
\caption[The point-proton and point-matter rms radii in the $0^{+}$ ground state of {\protect $\isotope[6]{He}$}, deduced from the crossover point of two consecutive $\Nmax$ curves.]{The point-proton and point-matter rms radii in the ground state of $\isotope[6]{He}$, deduced from the crossover point of two consecutive $\Nmax$ curves (see text). The radii deduced from the harmonic oscillator curves are plotted using dashed curves and radii deduced from the natural orbital basis curves are plotted using solid curves.}
\label{fig-chap4-he6-crossover-point-analysis}
\end{center}
\end{figure}

\begin{figure}[t]
\begin{center}
\includegraphics[width=0.7 \textwidth]{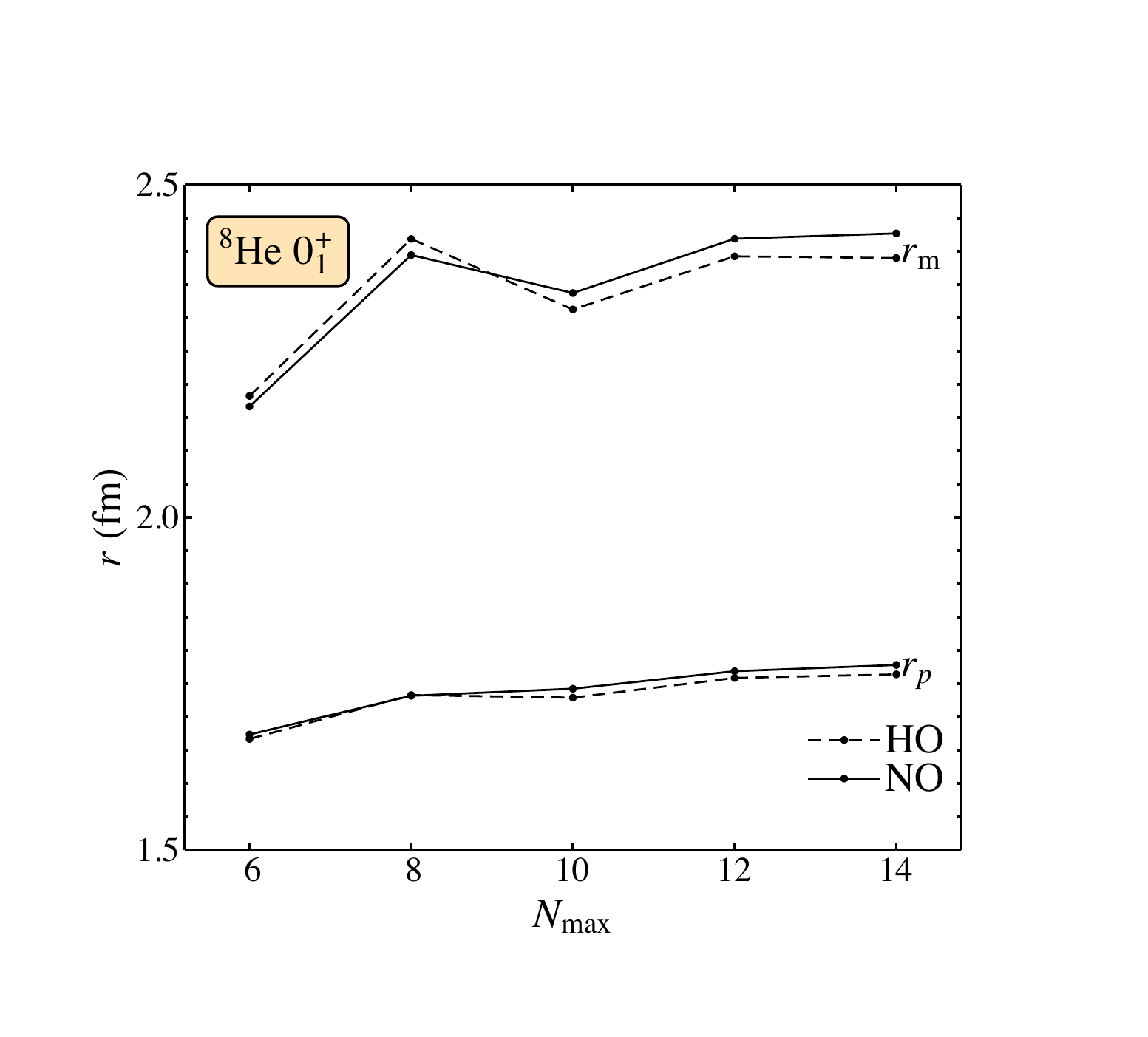}
\caption[The point-proton and point-matter rms radii in the $0^{+}$ ground state of {\protect $\isotope[8]{He}$}, deduced from the crossover point of two consecutive $\Nmax$ curves.]{The point-proton and point-matter rms radii in the $0^{+}$ ground state of $\isotope[8]{He}$, deduced from the crossover point of two consecutive $\Nmax$ curves (see text). The radii deduced from the harmonic oscillator curves are plotted using dashed curves and radii deduced from the natural orbital basis curves are plotted using solid curves.}
\label{fig-chap4-he8-crossover-point-analysis}
\end{center}
\end{figure}

To deduce the radius at the crossover point we first use a cubic interpolation to fit the calculated results as a function of $\hw$ and consequently, we find the radius at the intersection between curves obtained at two consecutive truncations of the many-body basis. In Fig.~\ref{fig-chap4-he6-crossover-point-analysis}, we plot the calculated radii at the crossover point as a function of the $\Nmax$ truncation for $\isotope[6]{He}$, and in Fig.~\ref{fig-chap4-he8-crossover-point-analysis}, we plot the corresponding radii for $\isotope[8]{He}$. Results obtained using the harmonic oscillator basis are plotted using dashed curves and results obtained using natural orbitals are plotted using solid curves.

For $\isotope[6]{He}$, the point-proton rms radius at the crossover point converges slowly with respect to $\Nmax$ for both bases. Specifically, for the harmonic oscillator basis the proton radius varies from $r_{p} = 1.70$ fm for $\Nmax = 6$ to $r_{p} = 1.80$ fm for $\Nmax = 16$ and for the natural orbitals it varies from $r_{p} = 1.70$ fm ($\Nmax = 6$) to $r_{p} = 1.81$ fm ($\Nmax = 16$). These results are consistent with our ``best estimate'' for the point-proton rms radius taken by extrapolating natural orbital results at $\Nmax = 16$ and $\hw = 20$ MeV ($r_{p} = 1.82$ fm). The point-matter rms radius at the crossover point varies more significantly with respect to $\Nmax$ for both bases, reflecting the fact that the calculated point-matter rms radius (Fig.~\ref{fig-chap4-he6-he8-gs-ho-no-comparison}) converges slower with respect to $\Nmax$ than the calculated point-proton rms radius. At the highest $\Nmax = 16$ truncation, we get $r_{m} = 2.31$ fm for the harmonic oscillator basis and $r_{m} = 2.34$ fm for the natural orbital basis (the natural orbital result reflects the faster convergence of the matter radius with respect to $\Nmax$ obtained using the natural orbital basis instead of using the harmonic oscillator basis). Both results are consistent with the natural orbital extrapolated result ($r_{m} = 2.33$ fm) within $0.02$ fm.

For $\isotope[8]{He}$, the convergence of the deduced proton and matter radii at the crossover point with respect to $\Nmax$ is qualitatively similar to the convergence with respect to $\Nmax$ of the $\isotope[6]{He}$ radii. At the highest $\Nmax = 14$ truncation, the deduced point-proton rms radius is $1.76$ fm for the harmonic oscillator basis and $r_{p} = 1.78$ fm for the natural orbital basis, consistent with the extrapolated results ($r_{p} = 1.77$-$1.78$ fm and $r_{p}=1.79$-$1.8$ fm for the harmonic oscillator and natural orbital bases respectively). The deduced matter radii at $\Nmax = 14$ are $2.39$ fm and $2.43$ fm for the harmonic oscillator and natural orbital bases respectively. These results are within the range of extrapolated results $2.34$-$2.42$ fm and $2.38$-$2.45$ fm, obtained for the harmonic oscillator and natural orbital bases respectively.

%% file: chapters/chapter5/chapter5_draft_170402.tex
\chapter{THE MIRROR NUCLEI $\isotope[7]{Li}$ AND $\isotope[7]{Be}$ IN A NATURAL ORBITAL BASIS}
\label{chap-chap5}

\section{Overview}
\label{sec-chap5-overview}

In this chapter we use natural orbitals to study the convergence properties of calculated observables for the mirror nuclei $\isotope[7]{Li}$ and $\isotope[7]{Be}$. Both nuclei have a bound first excited state, while some of their higher excited states are narrow resonances. The first excited $1/2^{-}$ state of both $\isotope[7]{Li}$ or $\isotope[7]{Be}$ decays to the ground $3/2^{-}$ state via an $E2$ or an $M1$ transition. The electromagnetic transition probability for this decay can be calculated using the calculated one-body transition density (\ref{eqn-chap2-one-body-density-operator}). Here we will study the convergence properties of the $B(E2)$ and $B(M1)$ values for this transition.

In this chapter we start by presenting the calculated ground state energy and point-proton rms radius in the ground state of $\isotope[7]{Li}$ and $\isotope[7]{Be}$ (Sec.~\ref{sec-chap5-many-body-calculations}). Subsequently, we present the calculated $B(M1;1/2^{-} \rightarrow 3/2^{-})$ and $B(E2;1/2^{-} \rightarrow 3/2^{-})$ values for the decay of the first $1/2^{-}$ excited state of $\isotope[7]{Li}$ and $\isotope[7]{Be}$ to the $3/2^{-}$ ground state (Sec.~\ref{sec-chap5-electromagnetic-transitions}). Finally, we extrapolate the calculated ground state energy and point-proton and matter rms radii in the ground state to the full space using the infrared extrapolation method (Sec.~\ref{sec-chap5-extrapolation}).
 
\section{Results}
\label{sec-chap5-many-body-calculations}

We perform an initial many-body calculation in the harmonic oscillator basis using the JISP$16$ interaction, the Coulomb interaction between protons, truncations of the many-body basis up to $\Nmax = 14$, and $\hw$ parameters in the range $10$-$40$ MeV. Note that $N_{0} = 3$ for these nuclei therefore, the natural parity spectrum is negative.

As in the calculations described in Chapter~\ref{chap-chap4}, after performing the initial calculation we use the calculated scalar one-body density matrices to deduce the natural orbitals for each $(\Nmax,\hw)$ pair, which we then use in the subsequent NCCI calculations in the natural orbital basis. The JISP$16$ internucleon interaction expressed in the harmonic oscillator basis at $\hw_{\mathrm{int}} = 40$ MeV is transformed to the natural orbitals basis using an $N_{\mathrm{cut}} = 13$ one-body shell truncation.

\begin{figure}
\begin{center}
\includegraphics[width=0.999 \columnwidth]{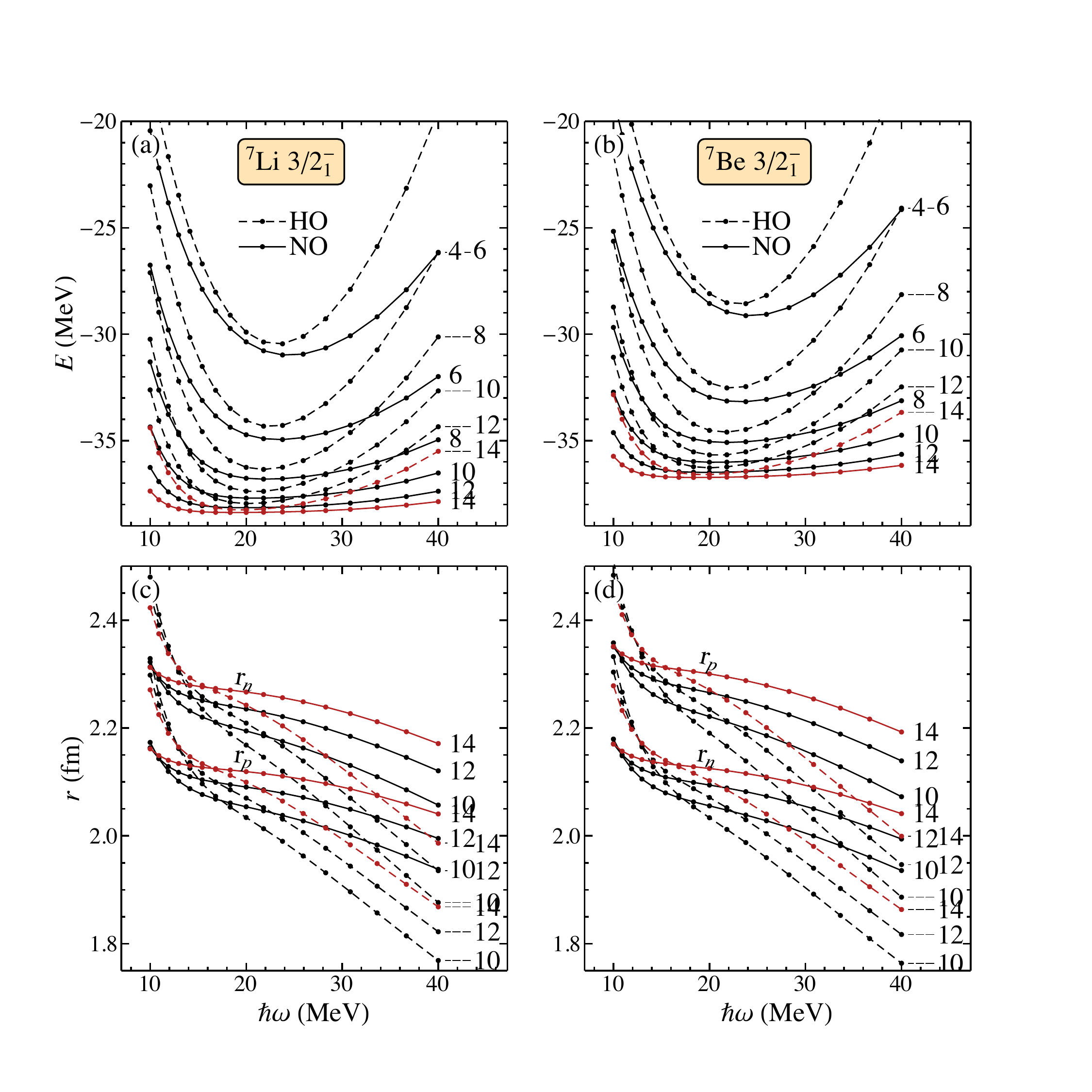}
\caption[Calculated ground state energy and point-proton (neutron) rms radius in the ground state of {\protect $\isotope[7]{Li}$} and {\protect $\isotope[7]{Be}$}. The results are obtained using the harmonic oscillator basis and the natural orbital basis, truncations of the many-body basis up to $\Nmax = 14$, $\hw$ parameters in the range $10$-$40$ MeV, the JISP$16$ NN interaction, and the Coulomb interaction between protons.]{Calculated ground state energy (top) and point-proton (neutron) rms radius in the ground state (bottom) of $\isotope[7]{Li}$ (left) and $\isotope[7]{Be}$ (right). The results are obtained using the harmonic oscillator basis (dashed curves) and the natural orbital basis (solid curves), truncations of the many-body basis up to $\Nmax = 14$, $\hw$ parameters in the range $10$-$40$ MeV, the JISP$16$ NN interaction, and the Coulomb interaction between protons. Results obtained at the highest $\Nmax$ truncation are plotted using red color.}
\label{fig-chap5-comparison-ho-no-li7-be7}
\end{center}
\end{figure}

\begin{figure}[t]
\begin{center}
\includegraphics[width=0.7 \columnwidth]{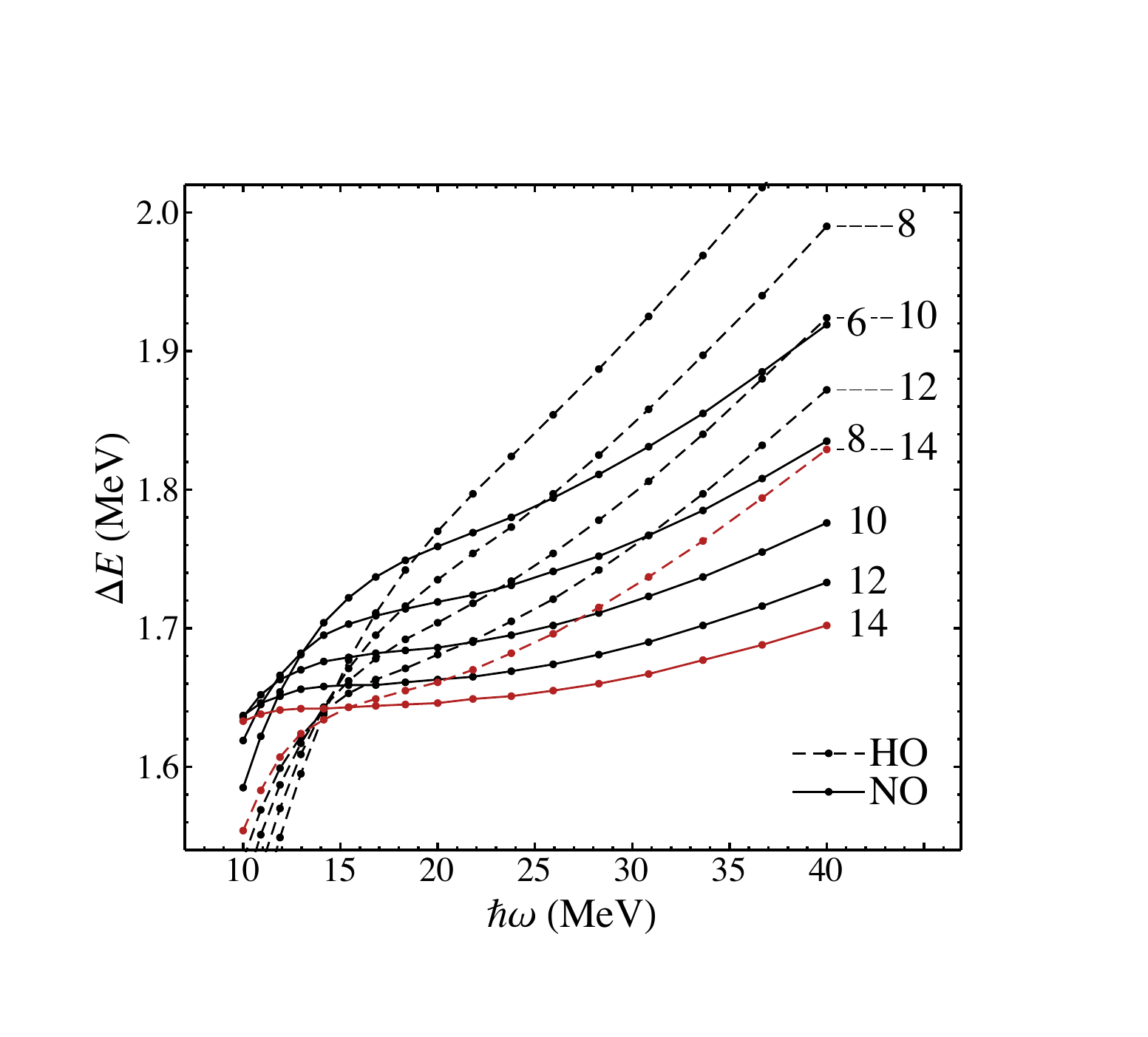}
\caption[The energy difference $\Delta E$ between the calculated ground state energy of {\protect $\isotope[7]{Li}$} and {\protect $\isotope[7]{Be}$} obtained using the harmonic oscillator basis and the natural orbital basis.]{The energy difference $\Delta E$ between the calculated ground state energy of $\isotope[7]{Li}$ and $\isotope[7]{Be}$ obtained using the harmonic oscillator basis (dashed curves) and the natural orbital basis (solid curves). The red colored curves show results obtained at the maximum $\Nmax = 14$ truncation.}
\label{fig-chap5-mirror-nuclei-coulomb-ho-no-li7-be7}
\end{center}
\end{figure}

In Figs.~\ref{fig-chap5-comparison-ho-no-li7-be7}(a)-(b), we plot the calculated ground state energy of $\isotope[7]{Li}$ and $\isotope[7]{Be}$ respectively. Results obtained using harmonic oscillator orbitals are shown with dashed curves, while results obtained using natural orbitals are shown with solid curves. We observe that the calculations performed using natural orbitals accelerate convergence in terms of $\Nmax$ and improve convergence in terms of $\hw$ compared to the harmonic oscillator basis for both nuclei (something we have already seen in all the other nuclei studied in this thesis). Let us now focus on one of the two nuclei namely $\isotope[7]{Li}$. Quantitatively, at the variational minimum ($\hw \approx 20$ MeV), the step from $\Nmax = 10$ to $\Nmax = 12$ brings us closer to convergence by $0.58$ MeV for the harmonic oscillator basis and $0.45$ MeV for the natural orbital basis. The step from $\Nmax = 12$ to $\Nmax = 14$ brings us closer to convergence by $0.3$ MeV for the harmonic oscillator basis and $0.22$ MeV for the natural orbital basis. Moreover, the calculated energy at the variational minimum of the natural orbital curves is $\sim 0.12$ MeV lower than the calculated energy at the variational minimum of the oscillator curves. These results indicate that using natural orbitals substantially accelerates convergence in terms of $\Nmax$ thus, due to the variational principle, they bring us closer to the true ground state energy. Similar conclusions apply to the calculated results of $\isotope[7]{Be}$.


In Figs.~\ref{fig-chap5-comparison-ho-no-li7-be7}(c)-(d), we plot the calculated point-proton and point-neutron rms radii of $\isotope[7]{Li}$ and $\isotope[7]{Be}$ respectively (to avoid cluttering in the figure we only plot results calculated for $\Nmax = 10, 12, 14$). For both nuclei, we observe that using natural orbitals results in a significant improvement of the convergence of the calculated radii with respect to $\Nmax$. For low $\hw$, a narrow shoulder begins to form for the natural orbital results. Finally, because of the isospin invariance of the nuclear Hamiltonian we expect that the proton radius of $\isotope[7]{Li}$ will be approximately identical to the neutron radius of $\isotope[7]{Be}$ (except for small differences due to the Coulomb interaction) and vice versa which is indeed what we see in Figs.~\ref{fig-chap5-comparison-ho-no-li7-be7}(c) and (d).

It is also interesting to study the convergence properties of the difference between the calculated ground state energies of $\isotope[7]{Li}$ and $\isotope[7]{Be}$, i.e., $\Delta E = E (\isotope[7]{Li}) - E (\isotope[7]{Be})$ (see Sec.~\ref{sec-chap5-extrapolation} for experimental information). Since $\isotope[7]{Li}$ and $\isotope[7]{Be}$ are mirror nuclei and the JISP$16$ interaction is isospin invariant, this difference is solely due to the Coulomb interaction between protons.

In Fig.~\ref{fig-chap5-mirror-nuclei-coulomb-ho-no-li7-be7}, we plot the calculated $\Delta E$ as a function of $\hw$ at various $\Nmax$ truncations obtained using harmonic oscillator (dashed curves) and natural orbitals (solid curves). We observe that the calculated $\Delta E$ obtained using natural orbitals converges faster with respect to $\Nmax$ than the $\Delta E$ obtained using harmonic oscillator orbitals.

\section{Electromagnetic transition probabilities}
\label{sec-chap5-electromagnetic-transitions}

Electromagnetic transitions between nuclear states probe the structure of nuclei and have been traditionally used to check the validity of nuclear models. Here we are interested in the calculation of the reduced transition probability between two nuclear states with total angular momenta $J_{i}$ and $J_{f}$. (From the reduced transition probability one can deduce the expected lifetime of a nuclear state). The reduced transition probabilty is given by~\cite{suhonen2007:nucleons-nucleus}
\begin{equation}
B(\sigma\lambda;J_{i} \rightarrow J_{f}) \equiv \frac{1}{2J_{i}+1} | \langle \Psi_{f} || \mathcal{M}_{\sigma\lambda} || J_{i} \rangle |^{2},
\label{eqn-chap5-matrix-elements-be2-bm1}
\end{equation}
where $\mathcal{M}_{\sigma\lambda}$ is the one-body spherical tensor operator responsible for the transition. For electric transitions, the tensor operator is written as $\mathcal{M}_{E\lambda} = Q_{\lambda}$. For magnetic transitions, the tensor operator is written as $\mathcal{M}_{M\lambda} = M_{\lambda}$. The reduced matrix element in (\ref{eqn-chap5-matrix-elements-be2-bm1}) is calculated using the reduced one-body transition density matrix
\begin{equation}
\langle \Psi_{\mathrm{i}} || \mathcal{M}_{\sigma\lambda} || \Psi_{\mathrm{f}} \rangle = \hat{\lambda}^{-1}  \sum_{ab} \langle a || \mathcal{M}_{\sigma\lambda} || b \rangle \langle \Psi_{\mathrm{i}} || \left[ c_{a}^{\dagger}\tilde{c}_{b} \right]_{\lambda} || \Psi_{\mathrm{f}} \rangle.
\label{eqn-chap5-matrix-elements-q2-m1}
\end{equation}
The electric tensor operator is given by~\cite{wong-introductory-nuclear-physics:2007, suhonen2007:nucleons-nucleus} 
\begin{equation}
Q_{\lambda \mu} = \sum_{i = 1}^{A} e(i) r_{i}^{\lambda} Y_{\lambda \mu}(\theta_{i}, \phi_{i}), 
\label{eqn-chap5-e-lambda-pole-operators}
\end{equation} 
and the magnetic tensor operator by
\begin{equation}
M_{\lambda \mu} = \sum_{i=1}^{A} \left[ g_{s}(i) \mathbf{s}_{i} + g_{l}(i) \frac{2 \mathbf{l}_{i}}{\lambda + 1} \right] \cdot \mathbf{\nabla}   \left[ r_{i}^{\lambda} Y_{\lambda \mu} (\theta_{i},\phi_{i}) \right],
\label{eqn-chap5-m-lambda-pole-operators}
\end{equation} 
where $e(i)$ is the electric charge, $\mathbf{s}_{i}$ and $\mathbf{l}_{i}$ are the spin and orbital angular momenta respectively, and $g_{s}(i)$ and $g_{l}(i)$ are the spin and orbital gyromagnetic ratios of nucleon $i$ respectively. Here we take the electric charge to be equal to $e(i)=e$ for a proton and $e(i)=0$ for a neutron, the spin gyromagnetic ratio to be equal to $g_{s}(i) = g_{p}$ for a proton and $g_{s}(i) = g_{n}$ for a neutron (where $g_{p} = 5.586$ $\mu_{\mathrm{N}}$ and $g_{n} = -3.826$ $\mu_{\mathrm{N}}$) and the orbital gyromagnetic ratio to be equal to $g_{l}(i) = \mu_{\mathrm{N}}$ for a proton and $g_{l}(i) = 0$ for a neutron.

\begin{figure}[t]
\begin{center}
\includegraphics[width=0.7 \columnwidth]{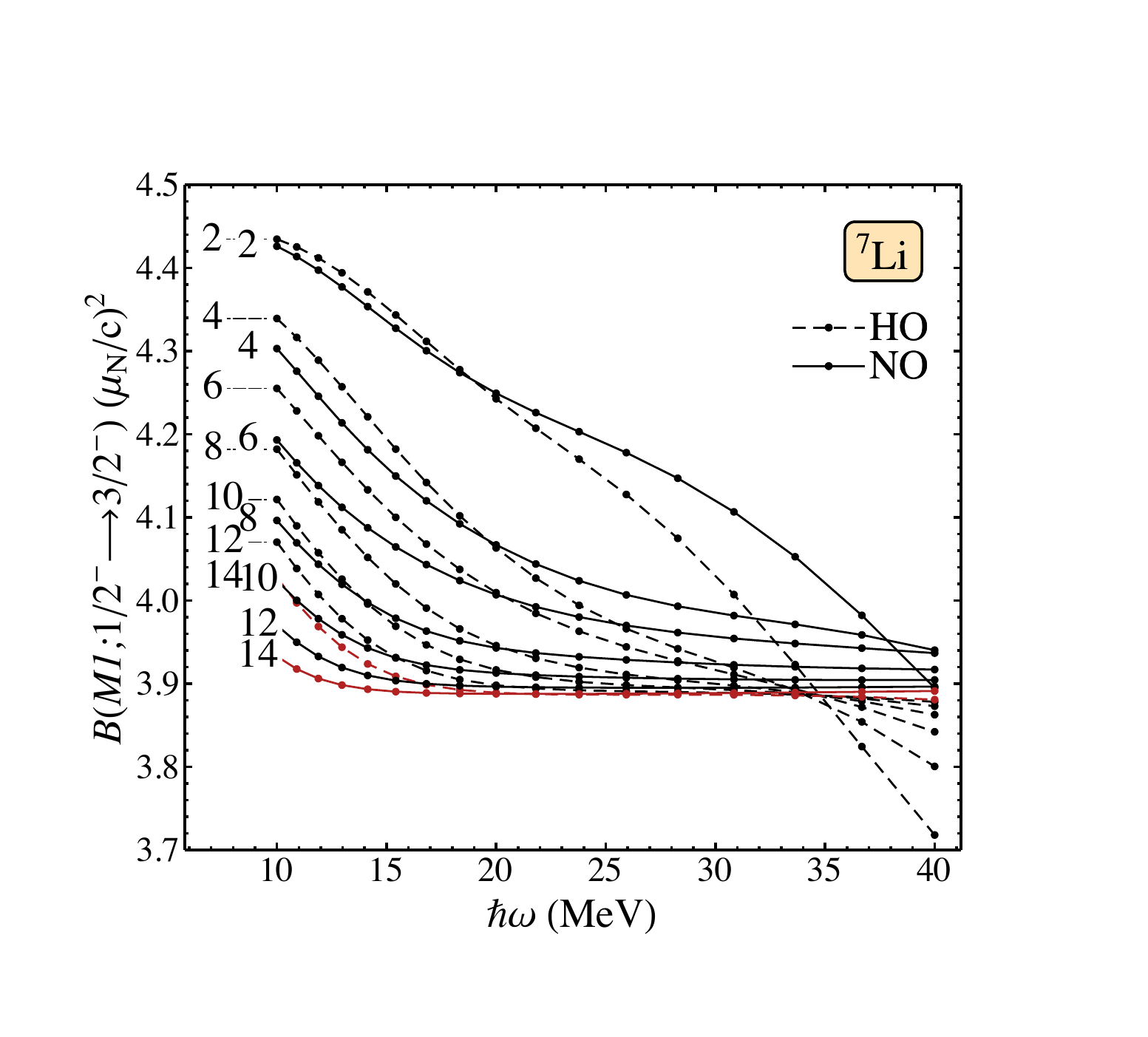}
\caption[The calculated $B(M1)$ values for the transition from the first $1/2^{-}$ excited state of {\protect $\isotope[7]{Li}$} to the ground $3/2^{-}$ state.]{The calculated $B(M1)$ values for the transition from the first $1/2^{-}$ excited state of $\isotope[7]{Li}$ to the $3/2^{-}$ ground state, obtained using harmonic oscillator orbitals (dashed curves) and natural orbitals (solid curves), the JISP$16$ internucleon interaction, and the Coulomb interaction between protons.}
\label{fig-chap5-comparison-ho-no-li7-bm1}
\end{center}
\end{figure}
\begin{figure}[t]
\begin{center}
\includegraphics[width=0.7 \columnwidth]{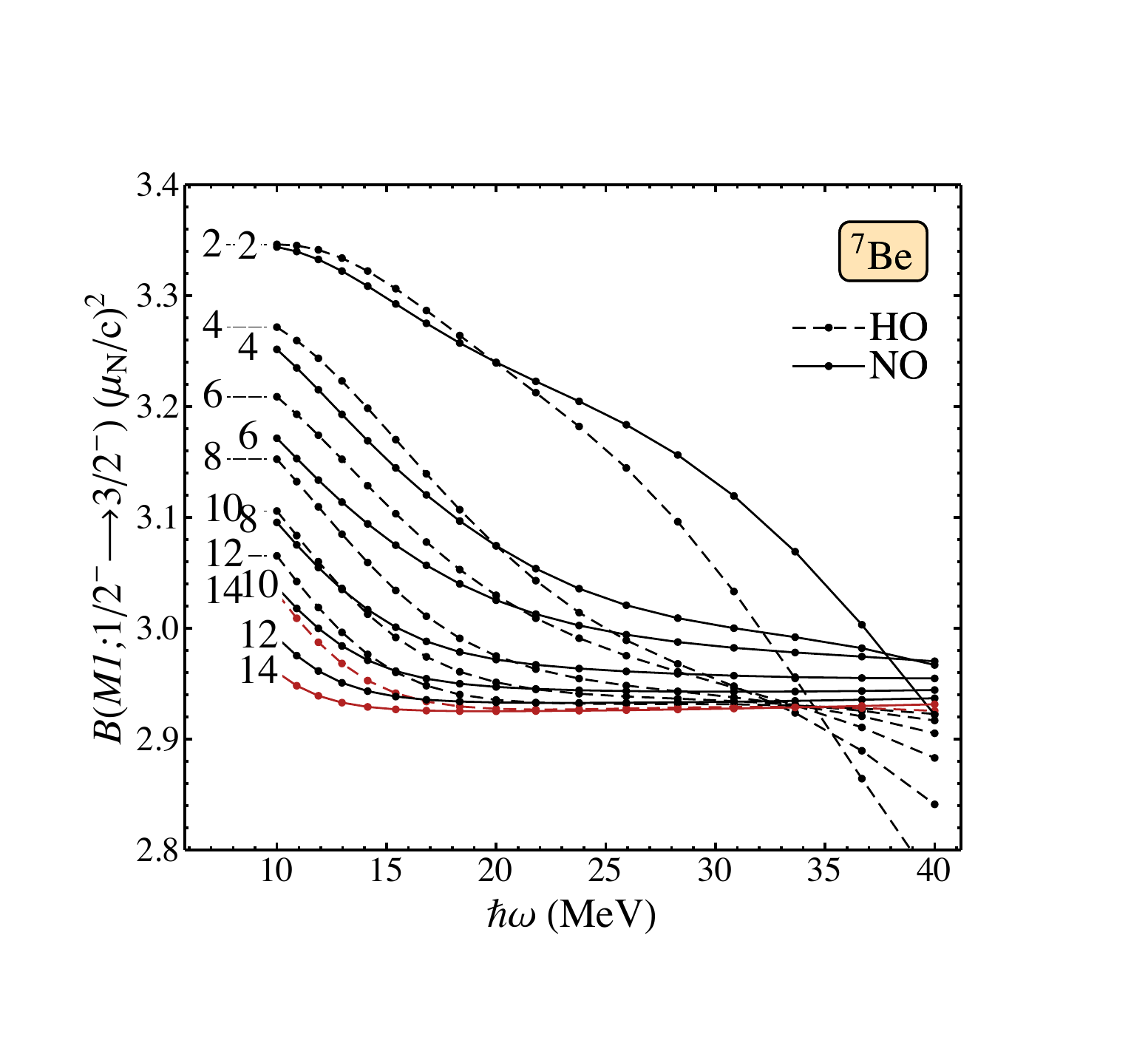}
\caption[The calculated $B(M1)$ values for the transition from the first $1/2^{-}$ excited state of {\protect $\isotope[7]{Be}$} to the ground $3/2^{-}$ state.]{The calculated $B(M1)$ values for the transition from the first $1/2^{-}$ excited state of $\isotope[7]{Be}$ to the $3/2^{-}$ ground state, obtained using harmonic oscillator orbitals (dashed curves) and natural orbitals (solid curves), the JISP$16$ internucleon interaction, and the Coulomb interaction between protons.}
\label{fig-chap5-comparison-ho-no-be7-bm1}
\end{center}
\end{figure}

We will focus on the electromagnetic transition probability between the (bound) $1/2^{-}$ first excited state to the $3/2^{-}$ ground state of the nuclei $\isotope[7]{Li}$ and $\isotope[7]{Be}$. According to the selection rules~\cite{suhonen2007:nucleons-nucleus}, this transition can either be an $M1$ or an $E2$ transition. The $B(M1)$ and $B(E2)$ values are obtained using the reduced one-body transition density matrix and equation (\ref{eqn-chap5-matrix-elements-be2-bm1}).

Let us start with the calculated $B(M1)$ values for this decay. In Fig.~\ref{fig-chap5-comparison-ho-no-li7-bm1}, we plot the reduced transition probability $B(M1;1/2^{-} \rightarrow 3/2^{-})$ for $\isotope[7]{Li}$, and in Fig.~\ref{fig-chap5-comparison-ho-no-be7-bm1} the reduced transition probability for $\isotope[7]{Be}$. Results obtained using the harmonic oscillator basis are shown with dashed curves and results obtained using natural orbitals are shown with solid curves. We observe that full convergence is achieved using wither basis. Notice that according to (\ref{eqn-chap5-m-lambda-pole-operators}) the magnetic dipole operator $M1$ is not a long-range observable (it does not depend on $r$), hence convergence does not depend on the long-range asymptotics of the many-body wave function. Compared to the harmonic oscillator basis, the natural orbital basis improves the convergence of the $B(M1)$ values in terms of the $\hw$ parameter and it also accelerates convergence in terms of $\Nmax$.

\begin{figure}[t]
\begin{center}
\includegraphics[width=0.7 \columnwidth]{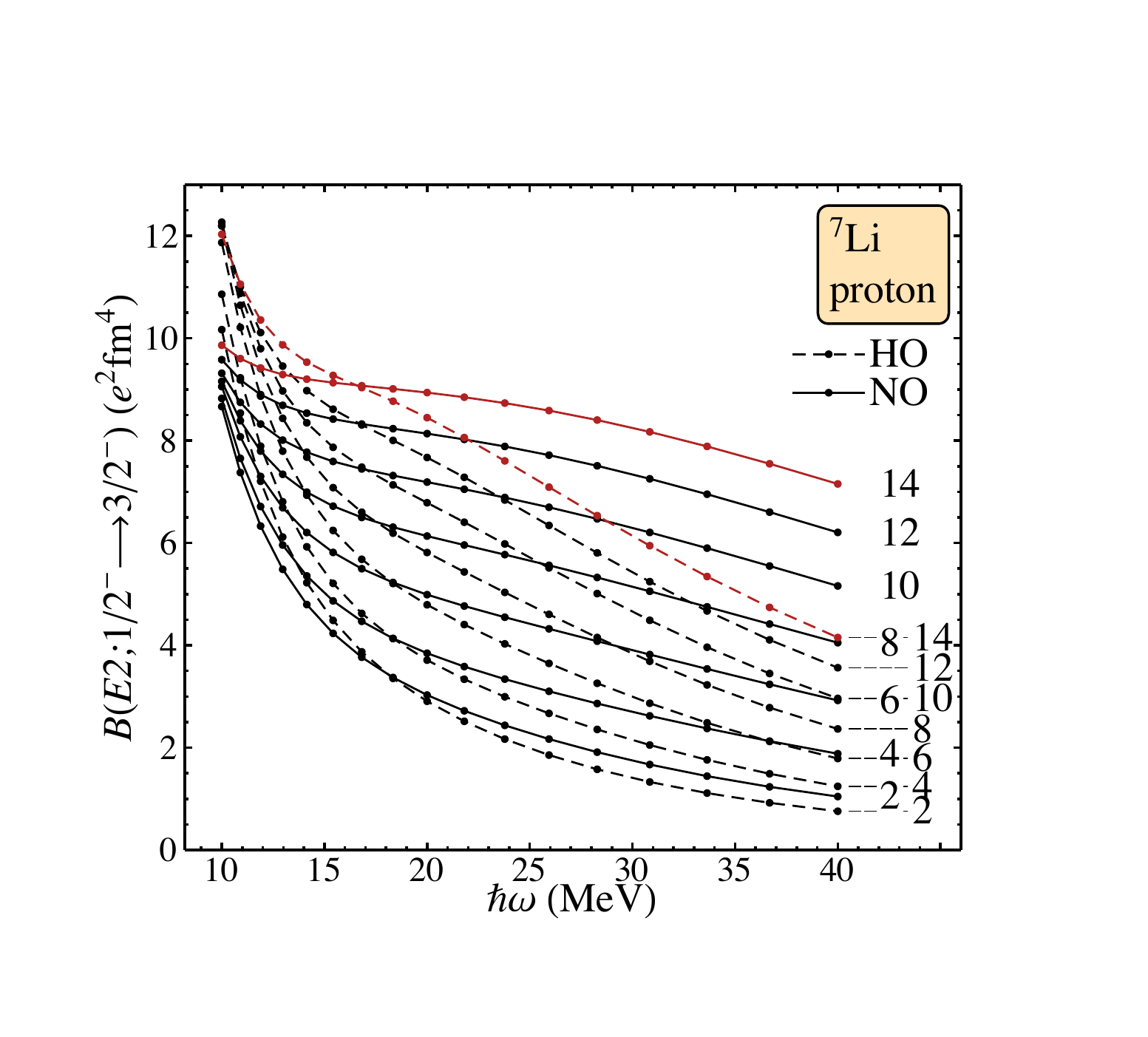}
\caption[The calculated $B(E2)$ values for the transition from the first $1/2^{-}$ excited state of {\protect $\isotope[7]{Li}$} to the ground $3/2^{-}$ state.]{The calculated $B(E2)$ values for the transition from the first $1/2^{-}$ excited state of $\isotope[7]{Li}$ to the $3/2^{-}$ ground state. The results are obtained using harmonic oscillator orbitals (dashed curves) and natural orbitals (solid curves), the JISP$16$ internucleon interaction, and the Coulomb interaction between protons.}
\label{fig-chap5-comparison-ho-no-li7-be2}
\end{center}
\end{figure}

\begin{figure}[t]
\begin{center}
\includegraphics[width=0.7 \columnwidth]{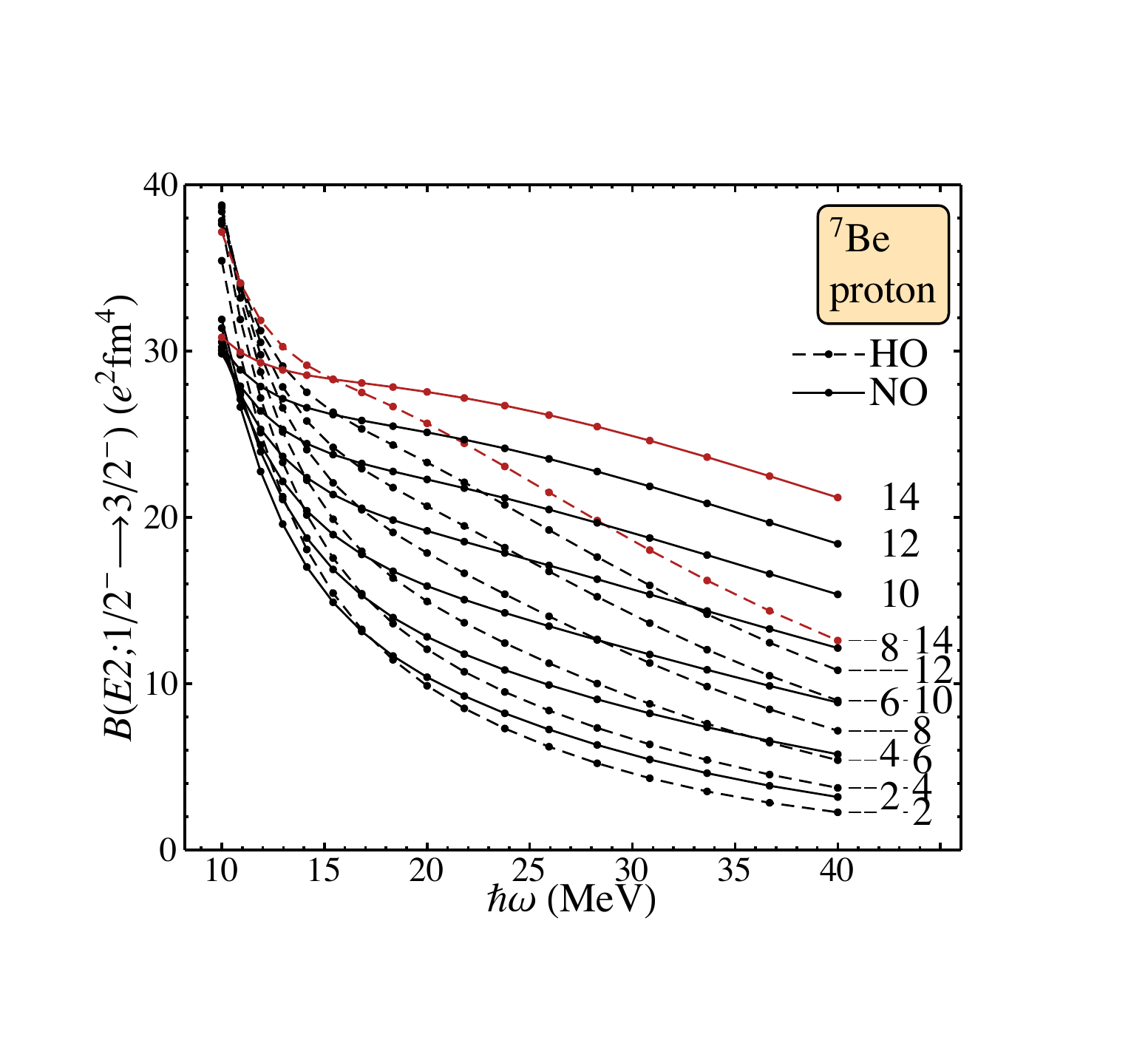}
\caption[The calculated $B(E2)$ values for the transition from the first $1/2^{-}$ excited state of {\protect $\isotope[7]{Be}$} to the ground $3/2^{-}$ state.]{The calculated $B(E2)$ values for the transition from the first $1/2^{-}$ excited state of $\isotope[7]{Be}$ to the $3/2^{-}$ ground state. The results are obtained using harmonic oscillator orbitals (dashed curves) and natural orbitals (solid curves), the JISP$16$ internucleon interaction, and the Coulomb interaction between protons.}
\label{fig-chap5-comparison-ho-no-be7-be2}
\end{center}
\end{figure}

Because the calculated $B(M1)$ values converge we can compare our calculated results against experimental data. The nuclei $\isotope[7]{Li}$ and $\isotope[7]{Be}$ are isobars; therefore, their Weisskopf estimates~\cite{suhonen2007:nucleons-nucleus} are identical and equal to $B_{\mathrm{W}}(M1) = 1.790$ $(\mu_{\mathrm{N}}/c)^{2}$. In Ref.~\cite{npa2002:005-007}, the experimentally reported values are $2.75 \pm 0.14$ Wu and $2.07 \pm 0.27$ Wu for the $B(M1)$ values of $\isotope[7]{Li}$ and $\isotope[7]{Be}$ respectively. Our calculated results from the $\Nmax = 14$ and $\hw > 15$ MeV calculations using natural orbitals suggest that the reduced transition probabilities are $ 2.17$ Wu and $1.64$ Wu for $\isotope[7]{Li}$ and $\isotope[7]{Be}$ respectively. This suggests that the NCCI calculation using the JISP$16$ interaction underestimates the $B(M1)$ values for both nuclei even when we account for the experimental error.

We now turn our attention to the calculated $B(E2)$ values. In Figs.~\ref{fig-chap5-comparison-ho-no-li7-be2} and \ref{fig-chap5-comparison-ho-no-li7-be2}, we show the calculated $B(E2;1/2^{-} \rightarrow 3/2^{-})$ values for $\isotope[7]{Li}$ and $\isotope[7]{Be}$ respectively, obtained using harmonic oscillator orbitals (dashed curves) and natural orbitals (solid curves). We observe that the natural orbital basis improves convergence in terms of $\Nmax$ compared to the harmonic oscillator basis; however, full convergence is not achieved. A narrow shoulder forms at $\Nmax = 14$ and low $\hw$ parameters for results obtained using natural orbitals.


\section{Infrared extrapolations}
\label{sec-chap5-extrapolation}

We close this chapter by extrapolating the calculated ground state energy, point proton, and point matter rms radii in the ground state of $\isotope[7]{Li}$ using the infrared extrapolation method. Due to the qualitative similarity between the calculated results of $\isotope[7]{Li}$ and $\isotope[7]{Be}$ (see Fig.~\ref{fig-chap5-comparison-ho-no-li7-be7}), we are discussing the extrapolation of the $\isotope[7]{Li}$ results in detail and we only provide final results for $\isotope[7]{Be}$.

\begin{figure}
\begin{center}
\includegraphics[width=0.9 \columnwidth]{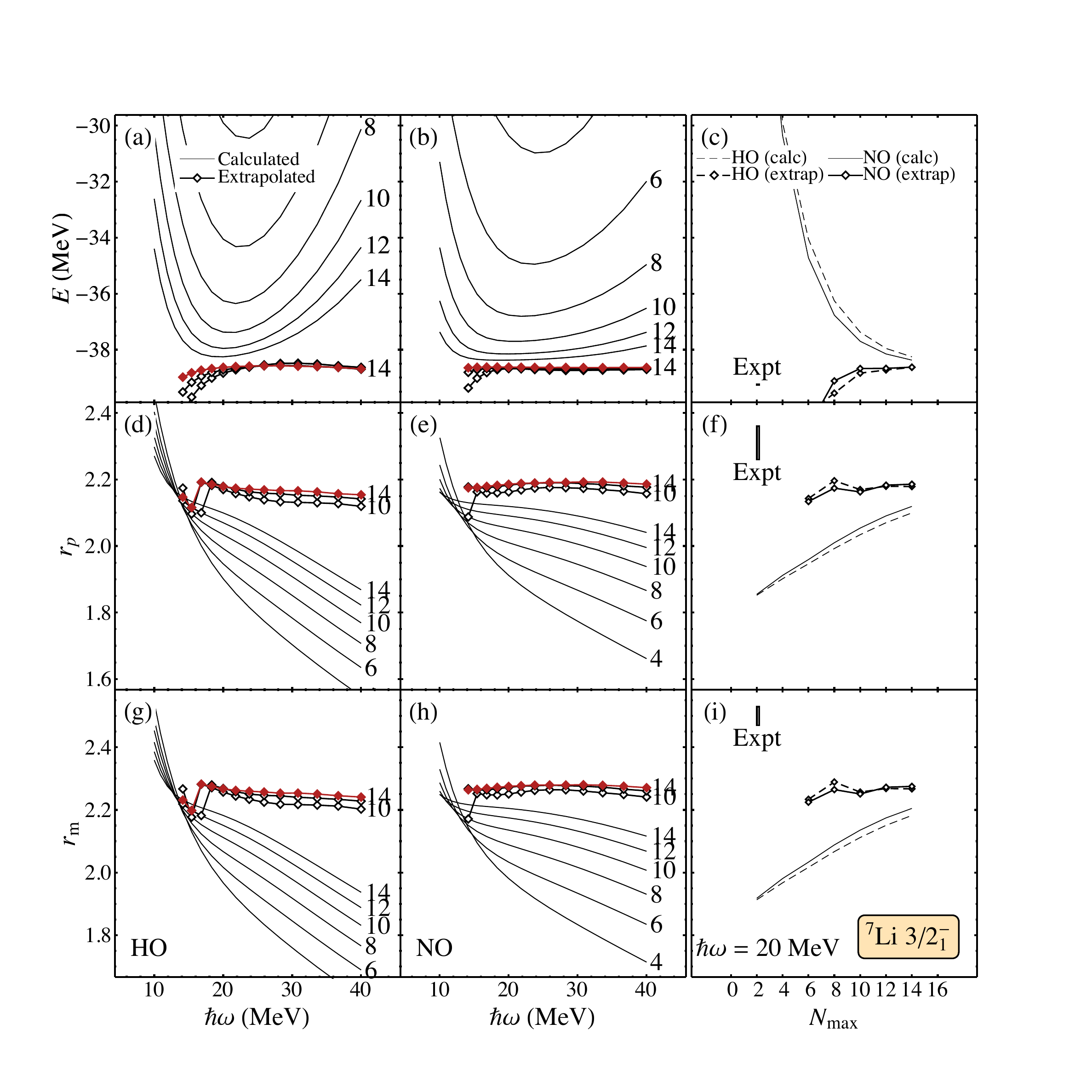}
\caption[Infrared basis extrapolations for {\protect $\isotope[7]{Li}$} calculated results obtained using the harmonic oscillator basis and the natural orbital basis.]{Infrared basis extrapolations for the $\isotope[7]{Li}$ ground state energy~(top), point-proton rms radius~(middle), point-matter rms radius (bottom), based on calculations in the harmonic oscillator basis~(left) and natural orbital basis~(middle). The extrapolations (diamonds) are shown along with the underlying calculated results (plain curves) as functions of $\hw$ at fixed $\Nmax$ (as indicated). The evolution of the calculated and extrapolated results with $\Nmax$ at $\hw = 20$ MeV and the experimental values (rectangles) are shown in the right column.}
\label{fig-chap5-li7-extrapolation-ground-state}
\end{center}
\end{figure}

In Fig.~\ref{fig-chap5-li7-extrapolation-ground-state}, we extrapolate the calculated results of $\isotope[7]{Li}$, obtained using the harmonic oscillator basis (left), and the natural orbital basis (middle). Moreover, we study the convergence of both the calculated and extrapolated results with respect to $\Nmax$ for $\hw = 20$ MeV (right). In the same column (right), we also show the experimental results (plotted as rectangles, where the center of the rectangle is the experimental result and the height of the rectangle indicates the uncertainty in the experimental result). We extrapolate three calculated data points that share the same $\hw$ and come from three different consecutive truncations of the many-body basis (as we did in the previous chapters), which are deemed to be approximately UV converged as described in Chapter~\ref{chap-chap2}.

In the top row, we observe that the extrapolated ground state energy results of $\isotope[7]{Li}$ have an $\hw$ dependence for the harmonic oscillator basis [panel (a)], and they converge in terms of both $\Nmax$ and $\hw$ for the natural orbital basis [panel (b)]. For the natural orbital basis, extrapolated results with $\hw \lesssim 20$ MeV do not fully converge due to perhaps an imperfect UV convergence. 

In the middle row, we observe that point-proton rms radii obtained by extrapolating the harmonic oscillator results cross in the low $\hw$ region. That is extrapolations performed using results obtained at different $\Nmax$ truncations cross. Moreover, the extrapolated results depend on $\hw$ (specifically they decrease as $\hw$ increases). On the other hand, the natural orbital extrapolations have a smoother dependence on both $\Nmax$ and $\hw$. As $\hw$ increases the extrapolated results increase, however the increase is very small (at the $\sim 0.01$ fm).

In the bottom row, the extrapolated point-matter rms radii depend on $\Nmax$ and $\hw$ for the harmonic oscillator basis. The situation is (similarly to the proton radius extrapolations) better for the natural orbital extrapolations. Specifically, the extrapolated matter radius changes by only $\sim 0.02$ fm across the range of $\hw$ parameters shown.

Let us now take the natural orbital extrapolated results at $\Nmax=14$ and $\hw \approx 20$ MeV as our best estimates of the converged results. For $\isotope[7]{Li}$, the extrapolated ground state energy is $-38.63$ MeV, the extrapolated point-proton rms radius is $2.19$ fm, and the extrapolated point-matter rms radius is $2.28$ fm. For $\isotope[7]{Be}$, the extrapolated ground state energy is $-37.01$ MeV, the extrapolated point-proton rms radius is $2.39$ fm, and the extrapolated point-matter rms radius is $2.31$ fm. 

Finally, we will attempt to compare our ``best'' extrapolated results against experimental results starting with the calculated energies. Experimentally, the ground state energy of $\isotope[7]{Li}$ is reported to be~\cite{npa2002:005-007} $-39.25$ MeV and that of $\isotope[7]{Be}$ $-37.60$ MeV. This means that our calculation using the JISP$16$ interaction underbinds both nuclei by about $\sim 0.6$ MeV. However, our calculation estimates that the energy difference $\Delta E$ (see Sec.~\ref{sec-chap5-many-body-calculations}) between the ground state energy of $\isotope[7]{Li}$ and $\isotope[7]{Be}$ is $\sim 1.62$ MeV which is very close to the experimental result ($1.65$ MeV).

The point-proton rms radii are deduced from the experimentally measured charge radii. Charge radii are based on isotope shift measurements in $\isotope[]{Li}$ atoms~\cite{nortershauser-lithium-charge-radii:2011} and $\isotope[]{Be}^{+}$ ions~\cite{krieger-beryllium-charge-radii:2012} and are referenced to the values of the stable $\isotope[6]{Li}$ and $\isotope[9]{Be}$, respectively, which are independently determined from electron scattering experiments. The reported values are~\cite{lu2013:laser-neutron-rich} $r_{p} = 2.31(5)$ fm and $r_{p} = 2.507(17)$ fm for $\isotope[7]{Li}$ and $\isotope[7]{Be}$ respectively. Our extrapolated results are about $\sim 0.1$ fm short of the experimental result for both nuclei.

The point-matter rms radii are obtained using interaction cross sections and are model dependent as discussed in Chapter~\ref{chap-chap4} for the $\isotope[]{He}$ isotopes. In Ref.~\cite{tanihata-matter-radii-lithium-beryllium:1985}, the reported point-matter rms radius is $r_{m} = 2.50(3)$ fm and $r_{m} = 2.48(3)$ fm for $\isotope[7]{Li}$ and $\isotope[7]{Be}$ respectively. Our ``best'' extrapolated results are $r_{m} = 2.28$ fm and $r_{m} = 2.31$ fm for $\isotope[7]{Li}$ and $\isotope[7]{Be}$, which underestimate $r_{m}$ by $\sim 0.2$ fm. However, the extrapolated results are approximately equal to each other for these two nuclei, like the experimental results.


%% file: chapters/chapter6/chapter6_draft_170402.tex
\chapter{CONCLUSION}
\label{chap-chap6}

The no-core configuration interaction (NCCI) approach strives to describe the structure of nuclei from first principles, i.e., starting from the internucleon interaction between protons and neutrons. The approach uses a many-body basis expansion to cast the problem of finding the eigenvalues and eigenvectors of the Hamiltonian into a matrix eigenvalue problem. The many-body basis is truncated according to the $\Nmax$ truncation scheme (as described in Chapters~\ref{chap-chap1} and \ref{chap-chap2}), and the many-body basis states are built using antisymmetrized products of single-particle states. Hence, the calculated observables (obtained by diagonalizing the Hamiltonian matrix) depend on the $\Nmax$ truncation of the many-body basis and the characteristic length of the single-particle basis states used (here the $\hw$ parameter). Convergence (of a calculated observable) is signaled by an independence of the calculated observable from both $\Nmax$ and $\hw$. The predictive power of the NCCI approach is compromised when we are unable to obtain results which are independent of the two parameters $\Nmax$ and $\hw$ of the basis. Calculations performed using the traditional harmonic oscillator basis rarely provide fully converged results as discussed in this thesis.

In this work we introduced natural orbitals for NCCI calculations. The natural orbitals are obtained by diagonalizing a scalar one-body density matrix obtained from an initial calculation using harmonic oscillator orbitals as described in Chapter~\ref{chap-chap3}. Subsequently, we used natural orbitals as the single-particle basis for \textit{ab initio} NCCI calculations for the nuclei $\isotope[3,4,6,8]{He}$, $\isotope[7]{Li}$, and $\isotope[7]{Be}$. Specifically, we calculate the ground state energy ($\isotope[3,4,6,8]{He}$, $\isotope[7]{Li}$, $\isotope[7]{Be}$), the point-proton (matter) rms radii in the ground state ($\isotope[3,4,6,8]{He}$, $\isotope[7]{Li}$, $\isotope[7]{Be}$), and the reduced transition probabilities $B(M1)$ and $B(E2)$ for the decay of the first excited state to the ground state ($\isotope[7]{Li}$, $\isotope[7]{Be}$). Upon comparison of the calculated results (using natural orbitals) against results obtained using harmonic oscillator orbitals, we found that convergence in terms of both $\Nmax$ and $\hw$ is improved using natural orbitals. Specifically, for long range observables such as the calculated rms radii and $B(E2)$ values, we found that results obtained using natural orbitals converge faster than results obtained using harmonic oscillator orbitals by one to two steps in $\Nmax$ (see Chapters~\ref{chap-chap3}, \ref{chap-chap4}, and \ref{chap-chap5}). For the calculated energies we found that convergence is accelerated by about one step in $\Nmax$, and for the reduced transition probabilities $B(M1)$ we find that calculated results fully converge using either the harmonic oscillator or the natural orbital basis. Whenever full convergence is not obtained, we used the infrared extrapolation method with the calculated results in both bases. Because of the overall improvement in the convergence of the calculated observables afforded by the natural orbital basis, the extrapolated results obtained from the natural orbital calculated results are more stable than the extrapolated results obtained from the harmonic oscillator calculated results (see Chapters~\ref{chap-chap3}, \ref{chap-chap4}, and \ref{chap-chap5}).

Beyond the basic implementation of natural orbitals developed in this work there are natural extensions which can potentially provide substantial additional improvement in convergence. Let us briefly discuss some of these directions for future development.

The natural orbitals are obtained using the one-body density matrix from an initial calculation in the harmonic oscillator basis. As we saw in this thesis, they accelerate convergence by mixing in contributions from high-$N$ orbitals in the initial basis into the final basis, thus providing a physically adapted basis for the nucleus under study. Using these natural orbitals in a subsequent NCCI calculation yields a new one-body density matrix, which can in turn be diagonalized to provide new natural orbitals. The new natural orbitals will (potentially) further mix in contributions form high-$N$ orbitals of the initial basis (into the final basis), thus (potentially) leading to faster convergence of the subsequent NCCI calculation. This procedure can be continued in an iterative fashion~\cite{bender1966:natural-orbital-iterative-hydride}.

In this work we derived natural orbitals from initial scalar density matrices in the traditional harmonic oscillator basis. However, as discussed in Chapter~\ref{chap-chap1}, the initial harmonic oscillator basis carries Gaussian ($\sim e^{-b r^{2}}$) asymptotics which are inadequate for the description of the exponential ($\sim e^{-br}$) asymptotics of the nuclear many-body wave function. Therefore, we can build natural orbitals starting from single-particle bases which are physically adequate for the description of the nuclear many-body wave function such as the Laguerre basis (discussed in Chapter~\ref{chap-chap2}). The derivation of natural orbitals from initial single-particle bases other than the harmonic oscillator basis will be straightforward provided that the initial single-particle basis forms a discrete, complete, orthonormal basis over $\mathbb{R}^{+}$. 

The $\Nmax$ truncation scheme provides a convenient way to remove the spurious center-of-mass states from the calculated low-lying spectrum when we use the harmonic oscillator basis. However, when we move away from the harmonic oscillator basis there is no need to maintain the $\Nmax$ truncation scheme. For instance, we can select single-particle orbitals according to their importance in the many-body wave function. When we use natural orbitals, the criterion which yields which natural orbitals are the most important is the eigenvalue (of the scalar one-body density matrix) corresponding to a given natural orbital. This eigenvalue, expresses the expected occupation number of a given natural orbital in the many-body wave function. Using these occupations, we can make a sensible choice of the natural orbitals which will be included in the many-body basis.

Finally, in this work we derived natural orbitals by diagonalizing the ground state scalar density matrices. These natural orbitals provide a physically adapted single-particle basis for the description of the ground state many-body wave function. Similarly, by diagonalizing the scalar density matrices of excited states we can potentially obtain natural orbitals which can provide a better description of excited states.